\documentclass[aps,prb,nolongbibliography,notitlepage,twocolumn,superscriptaddress]{revtex4-2}

\usepackage{amsmath}
\usepackage{amssymb}
\usepackage{bm}
\usepackage{bbm}

\usepackage[dvipsnames]{xcolor}
\usepackage{braket}
\usepackage{graphicx}
\usepackage{CJKutf8}
\usepackage{mathrsfs}
\usepackage{empheq}
\usepackage{etoolbox}
\usepackage{color}
\usepackage{epstopdf}
\usepackage{upgreek} 
\usepackage{mathtools}
\usepackage{soul}
\usepackage{dsfont}
\usepackage[toc,page,title,titletoc,header]{appendix}
\usepackage[colorlinks=true]{hyperref}
\hypersetup{
    colorlinks=true,
    linkcolor=blue,
    citecolor=blue,
    urlcolor=blue
} 

\usepackage[normalem]{ulem}

\def\L{{\mathcal L}}
\def\K{{\mathcal K}}

\def\A{{\mathcal A}}

\def\bA{{\bm A}}

\def\bxi{{\bm \xi}}

\def\br{{\bm r}}

\def\bp{{\bm p}}

\renewcommand{\v}[1]{{\boldsymbol{#1}}}

\graphicspath{ {./figs/} }

\begin{document}

\title{Bound states of anyons: a geometric quantization approach
}

\author{Qingchen Li}
\affiliation{Department of Physics, Harvard University, Cambridge, MA 02138, USA}
\author{Pavel A. Nosov}
\affiliation{Department of Physics, Harvard University, Cambridge, MA 02138, USA}
\author{Taige Wang}
\affiliation{Department of Physics, Harvard University, Cambridge, MA 02138, USA}
\author{Eslam Khalaf}
\affiliation{Department of Physics, Harvard University, Cambridge, MA 02138, USA}

\date{\today}
\begin{abstract}
The question of anyon interactions and their possible binding plays a key role in the physics of quantum Hall states and the more recently discovered fractional quantum anomalous Hall states. Here, we introduce a controlled and scalable approach to study anyon binding by working entirely within the Hilbert space of anyons. The resulting theory is characterized by an effective potential, which captures the electrostatic energy of classical anyon configurations, and a K\"ahler potential, which simultaneously encodes the anyon Berry phase and the structure of their Hilbert space; both quantities are readily computed using Monte Carlo methods for large systems, enabling reliable extrapolation to the thermodynamic limit. By applying the formalism of geometric quantization on K\"ahler manifolds, we construct the anyon Hamiltonian, which can be exactly diagonalized in the few-anyon Hilbert space. The approach is fully controlled for short-range interactions for any anyon excitations that are zero modes of a Trugman-Kivelson-type potential. Applying this formalism to the quasiholes of the $\nu=1/3$ Laughlin state with screened Coulomb interaction, we find that Laughlin quasiholes form bound states for screening lengths comparable or smaller than the magnetic length. Remarkably, binding occurs despite both the bare electron-electron interaction and the quasihole electrostatic potential being purely repulsive. The bound-state formation is a Berry phase effect, driven by oscillations in the quasihole density profile on the $\ell_B$ scale that are invisible in the quasihole electrostatic potential alone. For multiple quasiholes, we identify a sequence of phases as the screening length is reduced: free $e/3$ anyons, paired $2e/3$ bound states, three-anyon charge-$e$ clusters, and larger composite objects. Finally, we discuss possible signatures in charge imaging experiments on quantum Hall systems and the relevance to the phase diagram of itinerant anyon phases in fractional quantum anomalous Hall materials.

\end{abstract}
\maketitle

\addtocontents{toc}{\protect\setcounter{tocdepth}{-10}}

\emph{Introduction}--- Anyons, emergent quasiparticles with fractional quantum numbers, are among the most remarkable phenomena associated with topological phases of matter. Although the properties of single anyons in fractional quantum Hall
(FQH) states are relatively well understood, the question of whether anyons can form bound states is
comparatively less explored \cite{Tafelmayer1993topological, AnyonMoleculesJain, AnyonClustersYang, bernevig2025ED}. This is surprising given that the existence of such bound states has direct consequences for
the low-energy charge excitation spectrum and thus for transport and thermodynamic properties.
Interest in this question was recently revived due to two developments. First, advances in local probes
such as scanning tunneling microscopy (STM) now enable direct imaging of charge profiles at
sub-magnetic-length resolution, making the charge structure of anyon clusters directly
accessible in experiments \cite{Yazdani2025_STM,Yazdani2025_PNAS}. Moreover, recent shot-noise measurements revealed elementary fractional charge excitations at elevated temperatures, whereas higher multiples of the elementary charge were observed at lower temperatures \cite{Ghosh2025_anyon_bunching,Bid2009_shot_noise_2/3,Chung2003_Bunched_anyons}, consistent with the formation of anyon bound states. Second, the discovery of fractional quantum anomalous Hall (FQAH)
states in twisted MoTe${}_2$ \cite{xu2023observation, zeng2023thermodynamic, cai2023signatures, park2023observation} and rhombohedral pentalayer graphene on hexagonal
boron nitride \cite{Lu2024} has motivated interest in itinerant anyon phases --- in particular,
anyon superconductors \cite{LaughlinAnyonSC, FetterHannaLaughlin, 
laughlin1988, WilczekWittenHalperinAnyonSC,PhysRevLett.63.903,PhysRevB.39.11413,PhysRevB.42.342,PhysRevB.41.240,doi:10.1142/S0217979291001607,PhysRevB.41.11101,shi2025doping, divic2024anyon, NosovAnyonSCPRL, ShiSenthilAnyonDelocalization,PICHLERanyonSC,han2025anyon,Pichlerspectral} --- where the identity of the lowest charged excitations is
central to the theoretical description. 

Existing numerical methods for studying multi-anyon physics face significant limitations. Exact
diagonalization (ED) is restricted to small system sizes \cite{bernevig2025ED, AnyonClustersYang} that may not capture the thermodynamic
behavior, particularly for larger multi-anyon bound states. Density matrix renormalization group (DMRG) methods are largely
confined to cylinder geometries with exponential scaling in the cylinder radius.
Furthermore, both approaches are, in some sense, black boxes: they can reproduce spectra but do not easily
reveal the physical mechanism underlying the results. Composite fermion (CF) variational wavefunctions
offer more physical transparency and can reproduce certain numerics \cite{AnyonMoleculesJain, CFED1997, CFED2013}, but are
generally uncontrolled, and require benchmarking against ED when applied to new interaction
potentials or geometries.

\begin{figure}
    \centering
    \includegraphics[width=0.98\linewidth]{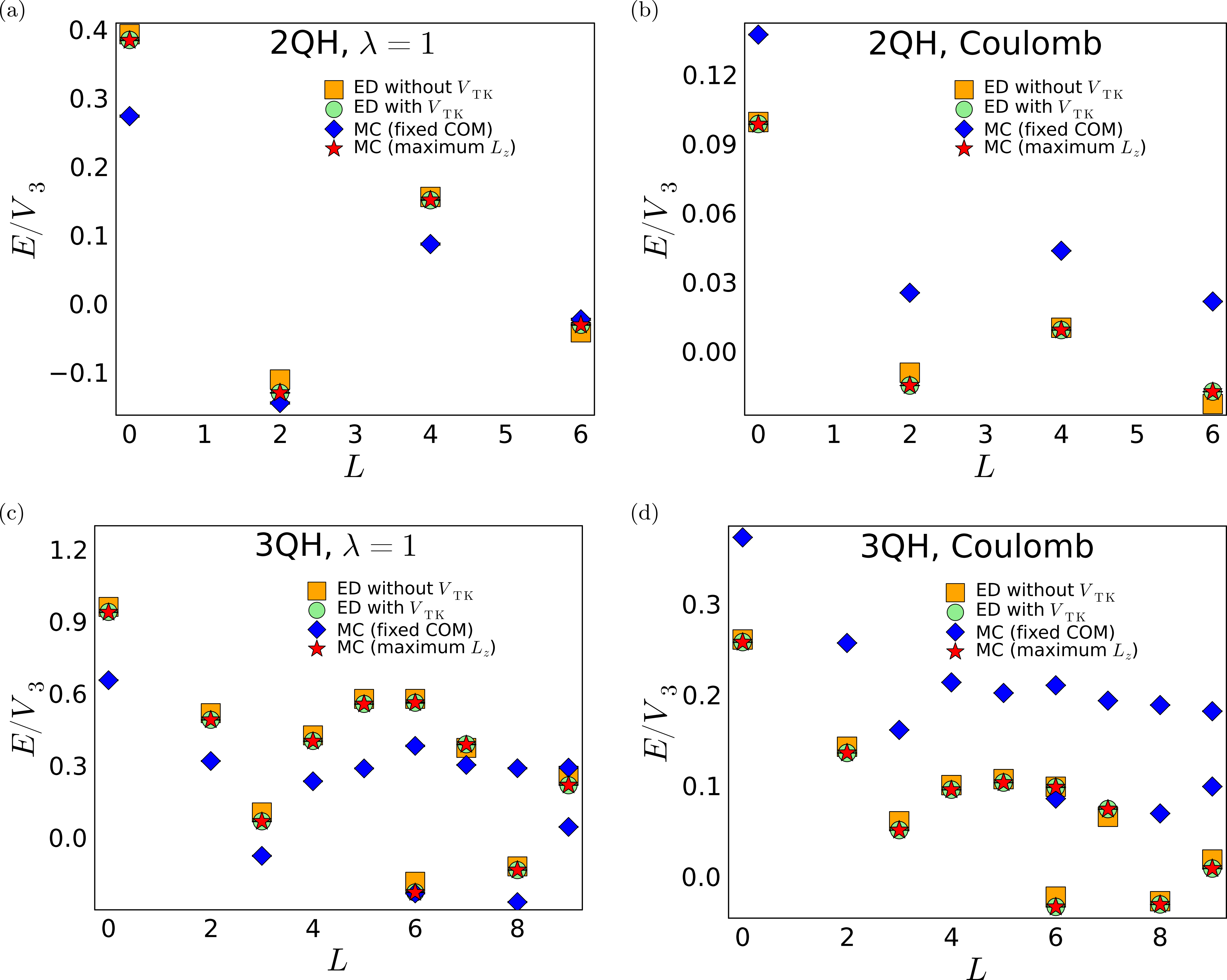}
    \caption{\textbf{Benchmarking Monte Carlo against exact diagonalization:} Energy spectra $E_{2qh} - E_{\rm Laughlin} - 2(E_{1qh} - E_{\rm Laughlin})$ for $Ne=7$ at $\nu = 1/3$, comparing MC and ED results for $N_h = 2$ (a,b) and $N_h = 3$ (c,d), with Yukawa-screened Coulomb interaction at $\lambda = 1$ (a,c) and bare Coulomb interaction (b,d). In each panel, ED results are shown both with and without the Trugman-Kivelson pseudopotential term $\alpha\hat{V}_\mathrm{TK}$ (large $\alpha$), together with MC results for maximum $L_z$ and for the fixed center-of-mass. }
    \label{fig:EDBenchmark}
\end{figure}

Here we develop an alternative approach to the anyon binding problem by working directly in the anyon Hilbert space. The anyon Hamiltonian is constructed by projecting the interaction onto the space of anyon wavefunctions. For any anyons that are zero modes of a Trugman-Kivelson-type pseudopotential, this can be formally achieved by considering an interaction of the form $\alpha \hat{V}_\mathrm{TK} + \hat{V}$, with $\alpha \gg 1$ and $\hat{V}_\mathrm{TK}$ is the
Trugman-Kivelson pseudopotential \cite{Haldane1983_Hierarchy, trugmanExactResultsFractional1985a}, which allows us to project onto the space of zero modes of $\hat{V}_\mathrm{TK}$. For Laughlin quasiholes with short-range interactions with range $\lambda \lesssim \ell_B$, where $\ell_B$ is the magnetic length,
this form is physically natural \footnote{$\hat{V}_\mathrm{TK}
\sim \nabla^2\delta$ while the leading pseudopotential of $\hat{V}$ is $V_3 \sim \nabla^6\delta$,
giving an effective $\alpha \sim (\ell_B/\lambda)^4$} and we show that even away from this limit, our results agree with ED on small systems to remarkable accuracy (cf.~Fig.~\ref{fig:EDBenchmark}). Within the projected quasihole subspace,
we use the formalism of geometric quantization on K\"{a}hler manifolds to construct the
quasihole Hamiltonian exactly in terms of a K\"{a}hler potential
and an effective interaction, describing effective magnetic field felt by the quasiholes and their mutual electrostatic interactions, respectively. Both quantities can be evaluated for very
large system sizes using Monte Carlo, enabling reliable extrapolation to the thermodynamic limit. Our approach effectively enables us to perform exact diagonalization in the Hilbert space of anyons directly, where the exponential scaling is in the number of \emph{anyons} on top of the fractional quantum Hall states, not the number of total electrons. As a result, it allows us to efficiently access the few-anyon problem in a controlled way in the limit where the number of electrons becomes very large. Interestingly, we find that the low-energy spectra for a few quasiholes can be captured to very high accuracy by replacing the complicated anyon Hamiltonian by one of bosons interacting through simple pairwise interactions. 

Applying this framework to the $\nu = 1/3$ Laughlin state in the lowest Landau level (LLL) with Yukawa-screened Coulomb
interactions, we find a surprising tendency of Laughlin quasiholes to form bound states when
the screening length $\lambda \lesssim \ell_B$. This is remarkable because both the bare
electron-electron interaction and the electrostatic quasihole interaction energy are purely
repulsive. In fact, the bound state arises from a combination of electrostatic interactions and Berry phase effects, which lead to oscillations
in the quasihole density profile on the $\ell_B$ scale that are invisible in the electrostatic interaction energy alone. We characterize the full phase diagram as a function of the screening parameter $\lambda$, identifying regimes of free $e/3$ anyons, paired $2e/3$ bound states, three-anyon charge-$e$
clusters, and larger composite objects as $\lambda$ is reduced. We emphasize that, while earlier ED \cite{AnyonClustersYang} found indication of 2 quasihole bound states for screened interactions, it cannot distinguish scenarios where these will be a part of a bigger cluster as more quasiholes are added. The latter scenario corresponds to phase separation and is distinct from the scenario we identify at intermediate $\lambda$ with free $2e/3$ bound states. We also note that Ref.~\cite{AnyonMoleculesJain} used (CF) variational wavefunctions to investigate anyon binding and did not find bound states for unscreened Coulomb, consistent with our finding, but did not study the screened case. At the end, we discuss the implications for experimental detection
via charge imaging and for the phase diagram of FQAH systems.

\emph{Quasihole Hamiltonian and geometric quantization}--- We focus on bound states of quasiholes on top of a Laughlin state in uniform magnetic field, but our approach is generalizable to any state that is a pseudopotential zero mode (satisfying clustering property) \cite{Bernevig2008_Jack,Bernevig2008_clustering} either for LLL or ideal Chern bands \cite{ledwith2020fractional, Wang2021exact, ledwith_vortexability_2023}. As discussed earlier, we consider an interaction of the form $\alpha \hat V_{\rm TK} + \hat V$, with $\alpha \gg 1$, where $\hat V_{\rm TK}$ is the Trugman-Kivelson pseudopotential for the $\nu = 1/q$ Laughlin state. 
Once we dope any number of quasiholes, the spectrum consists of a manifold of multi-quasihole zero modes of $V_{\rm TK}$ separated by a gap of order $\alpha$ from the remaining states. This justifies projecting the remaining interaction $\hat V$ to the manifold of zero modes and expressing the problem purely in terms of quasihole coordinates.

Consider $N_e$ electrons and $N_h$ quasiholes. A zero mode of the TK interaction is a wavefunction $|\xi \rangle$ in the $N_e$ electron coordinates labelled by the complex quasihole coordinates $\xi = (\xi_1, \dots, \xi_{N_h})$ where $\xi_i = \xi_{i,x} + i \xi_{i,y}$. We will use a gauge for which $|\xi \rangle$ depends holomorphically on $\xi$ and is symmetric under exchanging any pair of $\xi$'s. While our general framework is applicable to any geometry, most of our concrete calculations will be done on the sphere, where
\begin{equation} \label{qh-wf}
    \langle z_1,\dots,z_{N_e}|\xi \rangle = \prod_{i=1}^{N_e} \prod_{j=1}^{N_h} (z_i - \xi_j) \prod_{i<j}^{N_e} (z_i - z_j)^q \prod_{i=1}^{N_e} \Gamma(|z_i|)
\end{equation}
where $\Gamma(|z|) = (1 + |z|^2)^{-s}$, $s = \frac{N_\Phi}{2} = \frac{q(N_e-1)+N_h}{2}$, and $N_{\Phi}$ is the number of flux through the sphere. 

The theory of interaction projected onto the space of quasiholes needs to account for both the non-trivial overlaps between quasihole wavefunctions and the interaction between them. While this can be done directly in the Hamiltonian language, we will find it more illustrative to derive this through a path integral (PI) where the quasihole positions become dynamical variables. This makes manifest the connection to conventional anyon phenomenology, including their long-distance physics and their mutual statistics. Concrete calculations will make use of the Hamiltonian, which we will derive later. The PI takes the form 
\begin{gather}
    \langle \xi_F | e^{-\beta \hat V} | \xi_i \rangle = 
    \int \mathcal{D} \mu_\xi \,\, e^{-\int_0^\beta d \tau \mathcal L(\xi, \dot \xi)}\nonumber \\
    \mathcal L = i \sum_l \A_{l}(\bar\xi,\xi) \dot \xi_{l} + U(\bar\xi,\xi), \quad \A_{l}[\bar\xi,\xi] = -i \partial_{\xi_l} \K(\bar \xi, \xi) \nonumber \\
    e^{\K(\bar \xi, \xi)} = \langle \xi|\xi \rangle, \quad U(\bar\xi,\xi) = \frac{\langle \xi|\hat V|\xi \rangle}{\langle \xi|\xi \rangle} = e^{-\K(\bar \xi, \xi)} V(\bar \xi, \xi)
    \label{eq:PathIntegral}
\end{gather}
where $\K(\bar \xi, \xi)$ is the K\"ahler potential in the space of quasiholes and $V(\bar \xi, \xi) := \langle \xi|\hat V|\xi \rangle$ is defined for later convenience. In addition, the path integral has a non-trivial integration measure that yields a resolution of unity $\int d\mu_\xi |\xi \rangle \langle \xi| = \mathbbm{1}$ where the rhs is the identity in the space of quasihole wavefunctions $|\xi \rangle$. Note that we are expressing the path integral in terms of a basis of symmetric functions in $\xi$, so we do not need to include symmetrization in the boundary conditions. This is distinct from worldline PI where we consider trajectories of \emph{distinguishable} particles and include symmetrization in the boundary conditions.

The potential $V(\bar\xi,\xi)$ represents the classical electrostatic potential between quasiholes induced by the interaction. It has the same long distance behavior as the interaction $\hat V$ between particles with charge $1/q$ but may look different at short distances due to smearing of the quasihole charge on $\ell_B$ scale \footnote{When the distance between quasiholes is much larger than the magnetic length $\ell_B$, we can think of them as point particles and replace $V[\bar \xi, \xi] = \sum_{i<j} \Delta V(|\xi_i - \xi_j|)$}. The K\"ahler potential simultaneously encodes information about overlaps between quasihole wavefunctions (quantum metric) and their Berry phase due to the general relation between metric and Berry curvature on K\"ahler manifolds \cite{Kibble1979,Ashtekar1999,Provost1980}. It defines a notion of \emph{many-body quantum geometry} on the space of anyons. The Berry curvatures includes the Berry phase from the background magnetic field and the mutual Berry phase corresponding to the anyonic statistics, which simplifies at long distances due to plasma screening, giving rise to the known expression \cite{Arovas1985}. We emphasize that quasihole binding is a short-distance phenomenon sensitive to the physics on $\ell_B$ scale. Thus, we cannot use the asymptotic values of $U(\bar \xi, \xi)$ and $\K(\bar \xi, \xi)$ at long distances obtained from plasma screening. Instead, we need to know the functions $V(\bar \xi, \xi)$ and $\K(\bar \xi, \xi)$ explicitly at distances $\lesssim \ell_B$. Fortunately, this can be done in a scalable way for very large system sizes using Monte Carlo integration. Our problem then becomes: given $U(\bar \xi, \xi)$ and $\K(\bar \xi, \xi)$, how do we construct a corresponding Hamiltonian?

Let us first note that our Lagrangian is first order in time derivatives, indicating it describes a phase space (or coherent state) PI. 
Due to the holomorphic structure of our PI, reflected in the fact that the Berry phase is derived from a Kahler potential, we can construct the Hilbert space and  Hamiltonian corresponding to our problem using the formalism of geometric quantization on K\"ahler manifold (Berezin-Teoplitz quantization) \cite{Berezin1975_quantization,Rawnsley1990,Nair2020}. Given a K\"ahler potential $\K(\bar \xi, \xi)$, we identify the Hilbert space with that of fully symmetric anti-holomorphic functions $f(\bar \xi)$ with the inner product defined by an integration measure $d\mu_\xi$ satisfying the reproducing Kernel identity
\begin{equation}
    \int d\mu_\xi e^{\K(\bar \omega, \xi)} f(\bar \xi) = f(\bar \omega)
    \label{eq:ReproducingKernel}
\end{equation}
where $\K(\bar \omega, \xi)$ is defined by analytic continuation. This equation defines the measure $d\mu_\xi$. It generalizes the Bergman reproducing Kernel relation of single particle LLL wavefunctions where $\K(\bar \omega,\xi) = \frac{1}{2} \bar \omega \xi$ and $d\mu_\xi = \frac{d \xi d \bar \xi}{2\pi} e^{-\bar \xi \xi/2}$  (with $\ell_B = 1$) \cite{girvin1984FormalismQuantumHall}. We emphasize that both $\K(\bar \omega, \xi)$ and $d\mu_\xi$ in our case depend in a complicated way on the quasihole coordinates and do not factorize in a simple sum/product of terms. Note that this relation is equivalent to the resolution of unity inserted inside $f(\bar \xi) = \langle \xi|f \rangle$. The quantum operator $\hat H_{\rm QH}$ corresponding to the potential $V(\bar \xi, \xi) = \langle \xi|\hat V|\xi \rangle$ is defined by requiring its matrix elements to satisfy
\begin{equation}
    \langle g|\hat H_{\rm QH}|f \rangle = \int d\mu_\xi d\mu_\omega g^*(\bar \omega) V(\bar \omega,\xi) f(\bar \xi)
\end{equation}
This definition ensures that if we derive the PI for the operator $e^{-\beta \hat H_{\rm QH}}$ in the Hilbert space defined by Eq.~\ref{eq:ReproducingKernel}, we reproduce the PI defined in Eq.~\ref{eq:PathIntegral}. Our problem now depends only on $V(\bar \xi, \xi)$ and $\K(\bar \xi, \xi)$, where electron coordinates have been integrated i.e. we are working fully in the anyon Hilbert space.

To make the procedure more concrete, we choose a basis $\phi_\alpha(\xi)$ (not necessarily orthonormal) for the space of symmetric holomorphic functions such that $f(\bar \xi)$ admits the expansion $f(\bar \xi) = \sum_\alpha f_\alpha \bar \phi_\alpha(\xi)$. 
The dimension of such space agrees with the count obtained from the thin torus limit \cite{Bernevig2008_Jack,Bergholtz2006}.
We can expand $e^{\K(\bar \xi,\xi)} = \sum_{\alpha \beta} K_{\alpha,\beta} \bar \phi_\alpha(\xi) \phi_\beta(\xi)$. The reproducing Kernel identity becomes the condition 
\begin{equation}
    \Gamma_{\alpha,\beta} := \int d\mu_\xi \phi_\alpha(\xi)  \bar \phi_\beta(\xi) = (K^{-1})_{\alpha,\beta}
\end{equation}
which defines the measure $d\mu_\xi$ through its moments in the chosen basis. Expanding $V(\bar \xi,\xi) = \sum_{\alpha \beta} V_{\alpha,\beta}  \bar \phi_\alpha(\xi) \phi_\beta(\xi)$, and $f(\bar \xi) = \sum_\alpha f_\alpha \bar \phi_\alpha(\bar \xi)$, the eigenvalue equation $\hat H_{\rm QH}|f \rangle = E |f \rangle$ becomes the matrix equation
\begin{equation}
        H_{\rm QH} f = E f, \qquad H_{\rm QH} = V \Gamma = V K^{-1}
        \label{QHHamiltonian}
\end{equation}
Thus, computing the spectrum of the quasihole Hamiltonian reduces to diagonalizing the matrix $V K^{-1}$ or alternatively solving the generalized eigenvalue problem $V f = E K f$, whose complexity scales only polynomially with the dimension of the quasihole Hilbert space. 
Given an anyon wavefunction $f(\bar \xi)$, we can construct the corresponding electron wavefunction via
\begin{equation}
    |f \rangle = \int d\mu_\xi f(\bar \xi) |\xi \rangle
\end{equation}
The matrix $K$ relates the complicated quasihole Hilbert space structure to the simpler Hilbert space structure in the basis $\phi_\alpha(\xi)$, which can be taken to be a basis of non-interacting boson states. This makes it closely related to `flux attachment' gauge transformations mapping anyons to bosons. In fact, we show in SM, that in the limit where the quasiholes are well-separated, the unitary matrix which diagonalizes $K$ corresponds precisely to the matrix elements of the standard flux attachment transformation \cite{SM}. We emphasize, however, that our approach does not rely on any long-distance assumption and that the matrix $K$ (and the unitary which diagonalizes it), unlike the standard flux attachment transformation, is non-singular at short distances.

\emph{The two quasihole problem}--- Let us begin with the case of two quasiholes, $N_h = 2$, which illustrates our approach. Then we can express the $K$ and $V$ in terms of the center of mass coordinate $\xi_{\rm CM} = (\xi_1 + \xi_2)/2$ and the relative coordinate $\xi_r = \xi_1 - \xi_2$. In the limit of a large system size, we can assume the energy is independent of the center of mass coordinate \footnote{On the plane, there is no full continuous magnetic translation symmetry, but we expect an approximate translation symmetry in the bulk. On the torus, translation symmetry is exact, but we are only allowed to translate by a discrete amount, and rotation symmetry is no longer exact. On the sphere, rotation symmetry is exact, and the analog of translation symmetry is generated by the action of total angular momentum raising/lowering operators, which is distinct from translation for finite systems. For a large enough system, these effects can be neglected, and we can assume both rotation and magnetic translation symmetry regardless of geometry. We verify this effect is negilible for system sizes we choose in SM.}, which we will set to 0. Now expand $K$ and $V$ in terms of $\xi_r$ as
\begin{equation}
    e^{\K(\bar \xi_r, \xi_r)} = \sum_{L = 2\mathbbm{Z} \geq 0} K_{L} \bar \xi_r^{L} \xi_r^{L},\ V(\bar \xi_r, \xi_r) = \sum_{L = 2\mathbbm{Z} \geq 0} V_{L} \bar \xi_r^{L} \xi_r^{L}
\end{equation}
Note that both only depend on $|\xi_r|$ due to rotation symmetry, and the sum is over even (non-negative) integers because our wavefunction basis is symmetric. Eq.~\ref{QHHamiltonian} implies that the Hamiltonian is diagonal in the angular momentum basis with eigenvalue $E_{L} = V_{L}/K_{L}$ and eigenfunctions $f_{L}(\bar \xi_r) = \bar \xi_r^{L}$. 

\begin{figure}
    \centering
    \includegraphics[width=0.95\linewidth]{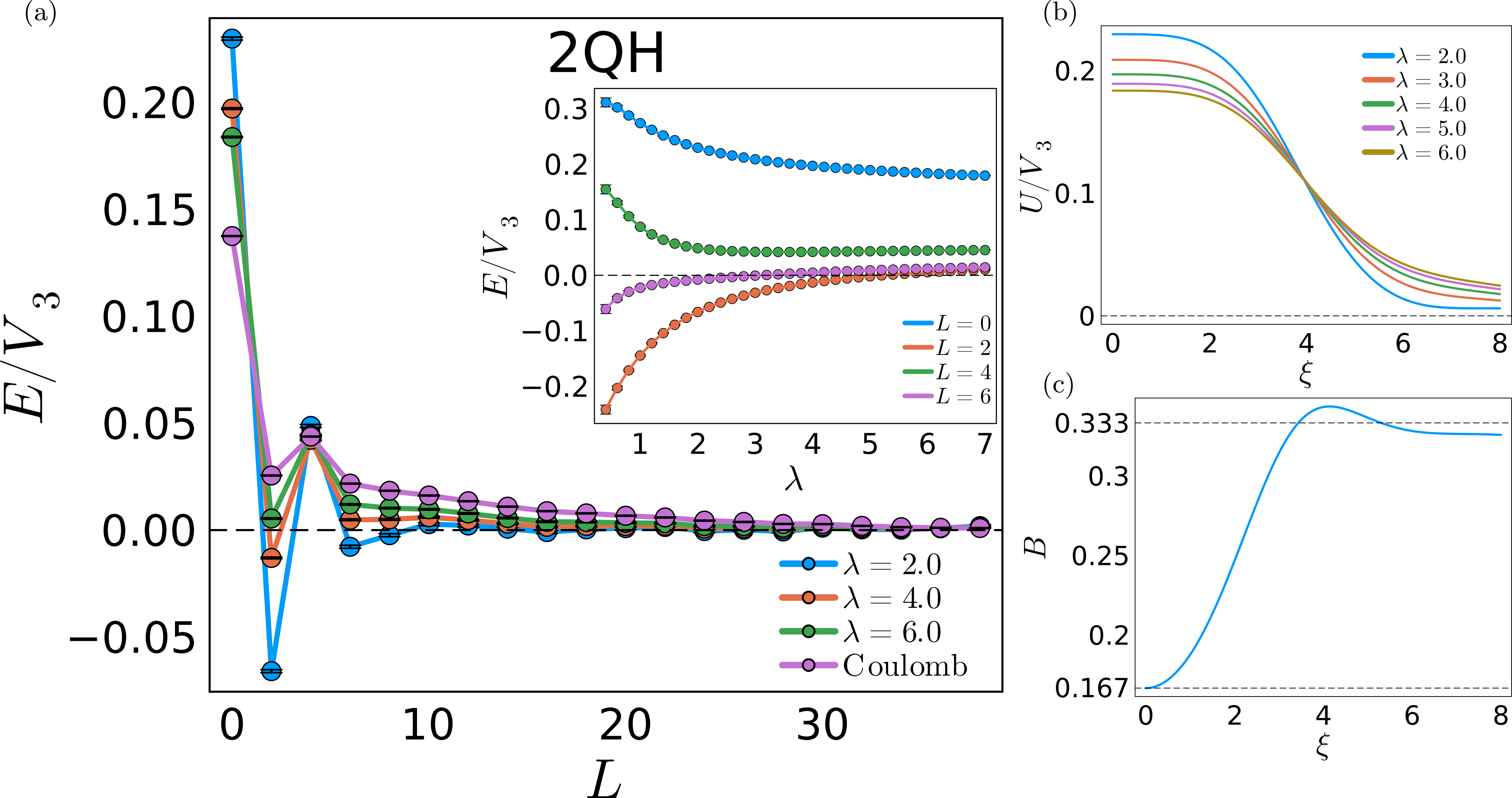}
    \caption{\textbf{Results for two quasiholes}: (a) Energy spectrum of two Laughlin quasiholes, measured in units of the Haldane pseudopotential $V_3$
    relative to well-separated quasiholes, for Yukawa-screened Coulomb interaction $V(r) = \frac{e^{-r/\lambda}}{r}$ for different values of $\lambda$ as a function of the relative angular momentum $L$. (b) Effective electrostatic potential between two quasiholes as a function of the relative position. (c) Effective magnetic field felt by one quasihole as a function of the distance from the other quasihole.}
    \label{fig:2QHBoundState}
\end{figure}

 For definiteness, let us focus on the $1/3$ state ($q = 3$) and take a Coulomb  interaction with Yukawa screening $V(r) = \frac{e^{-r/\lambda}}{r}$. Fig.~\ref{fig:2QHBoundState}a summarizes the corresponding spectra measured in units of the Haldane pseudopotential $V_3 = \int \frac{d^2\bp}{(2\pi)^2} V_\bp e^{-\bp^2} L_3(\bp^2)$\cite{Haldane1983_Hierarchy} where $V_\bp = \int d^2 \br e^{i\bp\cdot\br} V(r)$ and relative to the energy of two well separated holes. Here $L_n$ is the $n$-th Laguerre polynomial. At large $L$, the energy saturates to zero, consistent with well-separated quasiholes. At small $L$, the spectrum shows oscillations as a function of $L$. For $\lambda \lesssim 6$, the lowest-energy state occurs at $L=2$, whose energy is negative, indicating a bound state. There is at least another shallower bound state at $L = 6$ for small $\lambda \lesssim 2$.  We note that a similar tendency for binding was found using ED in \cite{AnyonClustersYang}, but the value of screening for which such binding persists is almost half the value we found here. Results for the $1/5$ state are included in SM \cite{SM}, where a similar trend is observed, but with bigger oscillations and with the bound state persisting to larger screening.

To validate our approach and establish its advantage over small-system ED, we benchmark our
results against ED for $N_e = 7$ with $N_h = 2$ and $N_h = 3$ in Fig.~\ref{fig:EDBenchmark}, for an
intermediate value of $\lambda$ and for the Coulomb case $\lambda \rightarrow \infty$. In
each panel, we show ED results for $\hat{V}$ alone and $\hat{V} + \alpha
\hat{V}_\mathrm{TK}$ with large $\alpha$. The latter is the limit in which our approach is
fully controlled and should reproduce ED exactly; this is confirmed by the MC results for maximum $L_z$, which coincide with the ED$+\hat{V}_\mathrm{TK}$ spectra to within
numerical precision. Notably, the ED results with and without $\hat{V}_\mathrm{TK}$ agree to
remarkable accuracy even for the Coulomb interaction, providing independent justification for
the validity of our approach outside the regime where it is parametrically controlled.
Finally, comparing the maximum-$L_z$ MC results with those obtained for the fixed center-of-mass
reveals substantial finite-size deviations, indicating that finite-size effects are the
dominant source of error in small-system ED. Our approach, which can extrapolate to the
thermodynamic limit, is therefore significantly more reliable than finite-size ED for
extracting the low-energy physics of quasihole bound states.

\emph{Origin of quasihole binding}--- We note a rather puzzling and remarkable feature of the existence of bound states in this problem. Not only is the original electron-electron interaction purely repulsive, but even the projected classical interaction energy between quasiholes -- the potential $U(\bar \xi,\xi)$ defined in Eq.~\ref{eq:PathIntegral} -- is also purely repulsive and monotonically decreasing with the relative separation as shown in Fig.~\ref{fig:2QHBoundState}b. This suggests that the formation of the bound state is not a purely electrostatic effect but arises instead from a combination of Berry phase (quantum) and electrostatic (classical) terms. To see this, note that we can interpret our problem in terms of the relative coordinate as that of a single particle in the LLL with effective magnetic field given by $B_{\rm eff}(\bar \xi_r,\xi_r) = \Delta_{\xi_r} \K(\bar \xi_r,\xi_r)$ and a potential $U(\bar \xi_r,\xi_r)$, shown in Fig.~\ref{fig:2QHBoundState}b,c \footnote{The magnetic field, plotted approaches 1/3 at long distances and has a dip around the origin whose integrated weight is $-1/3$ such that at long distance the magnetic field is $1/3 - 1/3 \delta(\xi_r)$ as expected; the other anyon acts as a flux tube with flux $1/3$}.  If we consider the simpler problem where the magnetic field is replaced by its average value, then the corresponding Hamiltonian is obtained by promoting $\xi_r$ to a guiding center operator with $[\hat \xi_r, \hat \xi_r^\dagger] = 2 \ell_{B,\rm eff}^2$, which promotes $U(\bar \xi_r, \xi_r)$ to an operator \footnote{We need to first normal-order the expression such that $\bar \xi_r$ is to the left of $\xi_r$, then replace $\xi_r$ with the corresponding operator}. It is not difficult to show that the eigenvalues of this operator are given by $E_{2n} = L_n(-q \Delta) U(\bxi)|_{\bxi = 0}$ where $\Delta$ is the 2D Laplacian \cite{SM}. Clearly, $E_n$ can be negative even for positive and monotonically decreasing $U$. In fact, for $L = 2$, we have $E_2 = U + q \Delta U|_{\bxi = 0}$. Since the potential has a local maximum at $0$, the second term is negative and has a large prefactor $q = 3$, explaining the tendency to form a bound state in the $L = 2$ channel. 
Physically, the quantum theory probes oscillations on the scale of $\ell_B$ that are washed out in the LLL projected coherent state basis and thus invisible at the level of the classical potential $U(\bar \xi_r, \xi_r)$.

While our formalism has mapped the 2-quasihole binding problem simply to the spectrum of a single particle in the LLL subject to a radial potential, it is important to point out an important caveat. Here again let us assume a uniform field for simplicity. If we start from the microscopic Lagrangian for a particle in a magnetic field $\L_{\rm micr} = \frac{1}{2} m \dot \bxi^2 + i \bA(\bxi) \cdot \dot \bxi + U_P(\bxi)$, taking the limit $m \rightarrow 0$ in the PI \emph{does not} reproduce our PI due to operator ordering ambiguities. In fact, as discussed earlier, the quantum operator corresponding to $U(\bar \xi,\xi)$ in the coherent state PI (\ref{eq:PathIntegral}) is obtained by placing $\bar \xi$ to the left of $\xi$ then promoting them to operators. On the other hand, for the microscopic Lagrangian in the limit $m \rightarrow 0$, the procedure is the opposite: first place $\xi$ to the left of $\bar \xi$, then promote to operators. The potential $U$ and $U_P$ are related via $U(\bxi) = e^{q \Delta_\bxi} U_p(\bxi)$ which is precisely the LLL projection with $\ell_{\rm eff} = \sqrt{2q}$. The potentials $U$ and $U_P$ correspond to the $Q$ and $P$ symbols studied in quantum optics \cite{GlauberI, GlauberII, Sudarshan, ExperimentPFunction}. It is known that the $P$ symbol can be negative for positive $Q$, and its negativity is used to characterize how nonclassical a given state of light is. In fact, since the unprojected theory corresponds to the Hamiltonian $\frac{1}{2m}\big(\bp - q \bA(\bxi)\big)^2 + U_p(\bxi)$, the energy eigenvalues $E_n$ should be bounded from below by the minimum value of $U_p$. This means that negative $E_n$ necessarily implies negative $U_p$.

An alternative way to understand the bound state formation is to go back to electron coordinates and consider the two-point electron density correlator in a state with definite angular momentum $L$ (see SM for details \cite{SM}). This object can be computed analytically for $q = 1$ (free fermions), and it does not show any oscillations. However, once $q$ deviates from one \cite{Jancovici1981,Can2014}, this object starts to display oscillations. In SM, we analytically show that for $q = 1 + \epsilon$, the first-order correction in $\epsilon$ is always negative for $L = 2$. While this limit is far from the physically relevant case of $\epsilon = 2$, it reveals the emergence of density fluctuations responsible for binding once we deviate from the non-interacting $q = 1$ state. This reveals the origin of binding to be the alignment of the quasihole relative oscillation pattern in certain relative angular momentum channels. Clearly, this effect is only possible for interactions whose range is of the order of $\ell_B$ or smaller.

\emph{Multi-quasiholes}--- The procedure for multiple quasiholes proceeds similarly. We set the center-of-mass coordinate to 0 and work within a total angular momentum sector $L$. The details of the basis choice within each $L$ sector are discussed in SM \cite{SM}. With the chosen basis (which is not generally orthogonal), we introduce the overlap matrix and the interaction matrix within $L$-sector, $K^{(L)}_{m',m} =\langle \psi_{m'}^{(L)} | \psi_{m}^{(L)}\rangle$, $V^{(L)}_{m',m} =\langle \psi_{m'}^{(L)} |\hat V| \psi_{m}^{(L)}\rangle$. The spectrum in each angular-momentum sector is then obtained from the eigenvalue problem $V^{(L)} f = E K^{(L)} f$. The overlap and interaction matrix elements are evaluated numerically using Monte Carlo sampling. The spectrum is measured relative to well-separated quasiholes by shifting the energies as $E \rightarrow E -E_{\textrm{Laughlin}} - N_h(E_{\textrm{qh}} - E_{\textrm{Laughlin}})$. 

Before discussing our results, let us comment on two issues regarding the numerical implementation. First, unlike the two-quasihole case, the overlaps we need to compute now include different wavefunctions indicating the Monte Carlo can encounter sign problem. We note that we can in principle, extract the matrices $K^{(L)}$ and $V^{(L)}$ from the real space overlaps $K(\bar \xi,\xi)$ and $V(\bar \xi,\xi)$, which do not encounter any sign problem, by fitting their expansion coefficients at small $\xi$. However, in practice, we found that expanding in a basis as explained leads to a convergent MC without sign problem. This works because we are considering a small number of quasiholes, where the MC weight is dominated by the part coming from the parent state which is the same for all the basis states entering the overlap.

We also note that our total Hilbert space dimension is $\binom{N_e + N_h}{N_h}$ which scales as $\sim N_e^{N_h}$ which is polynomial in the system size but exponential in the number of holes. By further exploiting rotation symmetry and eliminating center-of-mass degeneracy, we find that the Hilbert space dimension in a given $L$ sector can becomes very small (but generally grows with $L$) \cite{SM}.

\begin{figure*}
    \centering
    \includegraphics[width=0.98\linewidth]{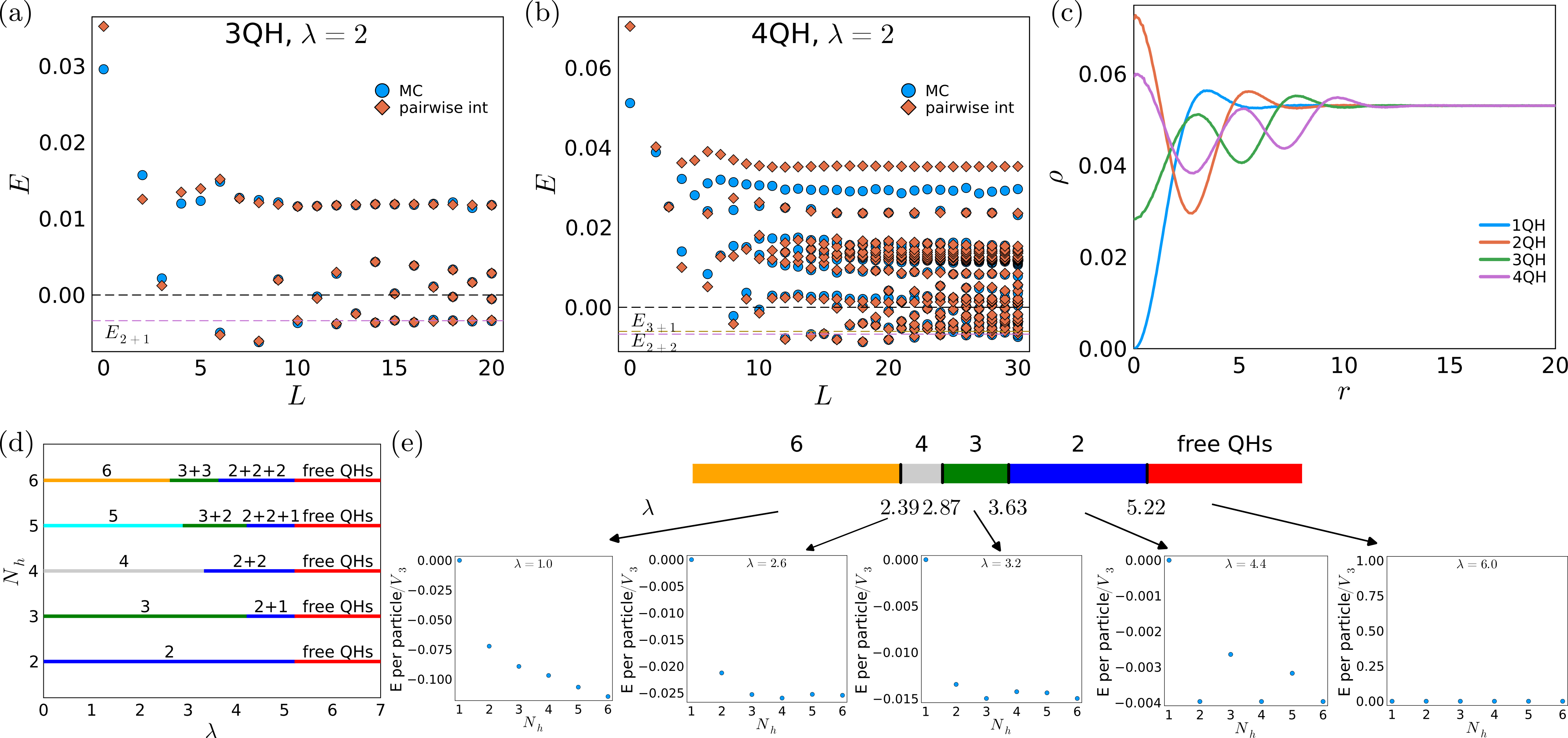}
    \caption{\textbf{Results for multi-quasiholes}: (a,b) Energy spectra of three and four Laughlin quasiholes, respectively, for the Yukawa-screened Coulomb interaction at $\lambda=2$. (c) Density profiles of a single quasihole and multi-quasihole bound states at $L=2,8$ and $18$ corresponding to $N_h=2,3$ and $4$, respectively. (d) Phase diagram for $N_h=2,...,6$ as a function of the screening length $\lambda$. (e) Phase diagram in the thermodynamic limit for different screening lengths $\lambda$, limited to cluster phases with up to six quasiholes. 
    $N_h$=1,2,3,4 results are obtained using the exact interaction evaluated in MC, whereas $N_h=5,6$ results are obtained from pairwise interaction approximation.}
    \label{fig:MultiQH}
\end{figure*}

The spectrum for $N_h=3$ and $N_h=4$ is shown in Fig.~\ref{fig:MultiQH}. For $N_h = 3$, we identify 3 regimes as a function of $\lambda$: (i) for large $\lambda$, the charge exists as 3 well-separated $e/3$ anyons, (ii) for intermediate $\lambda$, we have one $e/3$ anyon + one $2e/3$ bound state, and for (iii) for small $\lambda$, we get a charge $e$ bound state of the three $e/3$ quasiholes, which is \emph{distinct} from the simple hole (see its charge distribution in Fig.~\ref{fig:MultiQH}). Similarly, for $N_h = 4$, we go from a regime with 4 free $e/3$ anyons for large $\lambda$, two free $2e/3$ anyons for intermediate $\lambda$, and one $4e/3$ bound state for small $\lambda$. We do not find any region with $3+1$.

We can understand the emergence of multi-quasihole bound states for small $\lambda$ from the two-body approximation for the interaction as follows. From the two-quasihole spectrum, we see that the two-body interaction is dominated by the small $L$ channels with $L = 0$ strongly repulsive (quasiholes do not want to sit on top of each other) and $L = 2$ attractive but weaker. Thus, for small $\lambda$, each pair of quasiholes avoids being in the $L = 0$ channel, but they otherwise try to stay close to optimize the attractive component. This suggests the simple ansatz $f(\bar \xi) = \prod_{i<j}^{N_h} (\bar \xi_i - \bar \xi_j)^2$, which corresponds to $L = 6$ for $N_h = 3$ and $L = 12$ for $N_h = 4$. This indeed corresponds to the first bound state at small $\lambda$. In SM, we confirm that such a state indeed has a large overlap with the numerically obtained bound states. These quasihole clusters can be understood as precursors to the $2/7$ Haldane-Halperin hierarchy \cite{Haldane1983_Hierarchy,Halperin1984_Hierarchy} state formed when the doped quasiholes are placed in a $\nu=1/2$ bosonic Laughlin state. We also find other bound states corresponding to more complicated wavefunctions with the form $f_L(\bar \xi) = F_{L-N_h(N_h-1)}(\bar \xi) \prod_{i<j} (\bar \xi_i - \bar \xi_j)^2$ for some symmetric polynomial $F$. Indeed, the lowest energy states for 3 and 4 quasiholes occur at $L = 8$ and $L = 18$, respectively. 

To consider a larger number of quasiholes, it is instructive to perform an approximation that we expect describes the low-energy states very well: replace the few-body Hamiltonian in the space of quasiholes by a sum of 2-body terms. This can be justified by noting that the projected quasihole interactions decay over a length scale $\sim \ell_B$ due to plasma screening combined with the fact that the holes avoid sitting right on top of each other to minimize repulsion. These mean that, for low energy states, we should not have more than two holes within a region of size $\sim \ell_B$, justifying the two-body approximation.
Such two-body terms can be easily read off from the solution to the 2-body problem $V_{ij} = \sum_{L = 2\mathbbm{Z} \geq 0} E_L P_L^{ij}$ where $P_L^{ij}$ projects on the relative angular momentum state $L$ of the quasiholes $i$ and $j$.
We verify the accuracy of this approximation by showing the spectrum for 3 and 4 quasiholes using the 2-body approximation compared to the exact answer in Fig.~\ref{fig:MultiQH}(a,b). We observe that this approximation works very well for low-energy states as long as $L$ is not too small. We analyze the error induced by this approximation in more detail in SM \cite{SM}. 

Within the pairwise interaction approximation, we went up to six quasiholes (total charge $2e$). By considering the energy per particle as a function of the number of quasiholes, we identify five distinct regimes as we decrease $\lambda$ (cf. Fig.~\ref{fig:MultiQH}(e)): (i) free $e/3$ quasiholes, (ii) free $2e/3$ bound states, (iii) free $3e/3$ bound states, (iv) free $4e/3$ bound states, (iv) a bound cluster of 6 anyons. The latter is compatible with two possibilities: stable 6-particle clusters with charge $2e$ or phase separation where all doped quasiholes cluster close to each other. We anticipate the latter to be the case for sufficiently small $\lambda$, where the analogy to Type-I superconductors in the Chern-Simons–Landau-Ginzburg formulation suggests a tendency toward phase separation \cite{Parameswaran2012_typeI_FQH,SM}.

\emph{Discussion and relevance to experiment}--- Our results show the surprising tendency of even the simplest anyons, Laughlin quasiholes, to form bound states provided that the long-range part of the Coulomb interaction is screened. The existence of such bound states can be easily detected through the charge profile measured in STM: they correspond to characteristic charge rings around a given point rather than a single dip, as shown in Fig.~\ref{fig:MultiQH}(c). Such charge patterns are expected in the vicinity of impurities whose potential is either sufficiently shallow (depth small compared to the binding energy) or sufficiently wide that it fits the whole bound state. The binding energies computed here set the temperature scale below which bound-state formation becomes relevant for bulk transport. They are likewise expected to provide an important input for a proper treatment of edge transport observables, such as the shot-noise Fano factor \cite{Ghosh2025_anyon_bunching}.

While our results have been restricted to the LLL, we can discuss its implications for Chern bands and possible relevance for anyons superconductivity, leaving a more quantitative discussion for future work. Assuming the Aharonov-Casher limit, we can map a Chern band to the LLL in a non-uniform magnetic field. This generates a periodic single-particle potential $\sum_i v(\bar \xi_i, \xi_i)$ to the Lagrangian in Eq.~\ref{eq:PathIntegral} as discussed in Refs.~\cite{YanAnyonDispersion, SchleithAnyonSphere}. This potential generates a dispersion for the anyons, which is expected to be larger for smaller anyons since for larger anyons, the charge is smeared over a larger region, which averages out the potential. Thus, including weak periodic modulations to the magnetic field (small Berry curvature variations), we expect lighter anyon composites to be favored by kinetic energy, moving the phase boundaries in Fig.~\ref{fig:MultiQH}(d) to smaller $\lambda$. Thus, for small screening, we anticipate the cheapest charge excitations to go from large composites to smaller anyon-bound states and eventually to single $e/3$ quasiholes. In the dilute limit, the main effect of the interaction would be to stabilize certain anyon bound states whose remaining dynamics will be dominated by the kinetic energy, leading to the standard itinerant anyon description \cite{LaughlinAnyonSC, WilczekWittenHalperinAnyonSC, ShiSenthil}.

Let us now compare this to the anyon superconductivity scenario proposed in Refs.~\cite{ShiSenthil,NosovAnyonSCPRL,ShiSenthilAnyonDelocalization} for MoTe${}_2$, where the superconductor (SC) is realized when the lowest charged excitations are (dispersive) charge $2e/3$ anyons, and a re-entrant integer quantum anomalous Hall state is realized when they are charge $e/3$ anyons (in the presence of disorder). In this scenario, increasing the Berry curvature inhomogeneity should induce a transition between SC and IQAH state. This is consistent with the observation of such a transition by increasing the displacement field, which is known to make Berry curvature more inhomogeneous and increase anyon dispersion \cite{TMDAharonocCasher, YanAnyonDispersion}.

\emph{Conclusion}--- In closing, we note that our approach gives access to both the microscopic physics of anyon binding, most naturally described through the quasihole Hamiltonian, and the long-distance topological structure, most naturally expressed through the path integral. The two are unified by the Kähler potential, which simultaneously encodes the anyon wavefunction overlaps, the quantum geometry of the anyon Hilbert space, and their mutual Berry phase. Crucially, the Kähler potential interpolates smoothly between short distances — where it captures the microscopic structure responsible for binding — and long distances, where it reduces to the standard Chern-Simons flux attachment transformation, recovering the known anyonic statistics. Our approach thus provides a microscopic and controlled framework that connects short-distance binding physics to long-distance topological structure without relying on any separation of scales.

Beyond the few-anyon problem studied here, the pairwise approximation yields a simple but accurate quasihole Hamiltonian that is directly amenable to many-body methods such as DMRG or quantum Monte Carlo at finite anyon density. This opens a route to studying anyon phases — including possible anyon superconducting or crystalline phases — at finite doping and finite temperature, in regimes where the separation between short-distance and long-distance physics is no longer assumed.  The explicit construction of the quasihole PI also enables a semiclassical analysis, providing a natural foundation for extending the cooperative ring-exchange theory \cite{Kivelson1987_cooperative_righ_exchange_1,Lee1987_cooperative_righ_exchange_2} to the Haldane-Halperin hierarchy.

Finally, while we have focused on Laughlin quasiholes, the approach extends naturally to broader classes of anyons. For anyons that are exact zero modes of a pseudopotential — including non-Abelian quasiholes of the e.g. Moore-Read state, where three-body pseudopotentials play the role of $\hat{V}_\mathrm{TK}$  \cite{Greiter1991} — the projection is fully controlled, and the formalism applies directly. For anyons that are not exact zero modes, such as quasielectrons, the approach amounts to diagonalizing the Hamiltonian in the projected multi-anyon subspace. While this projection lacks a parametric justification in general, our benchmarking against ED demonstrates that it remains quantitatively accurate even for Coulomb interactions without an explicit pseudopotential term. This empirical robustness suggests the approach will be reliable for quasielectrons as well, opening the door to studying quasihole-quasielectron pairs, anyon-exciton, and anyon-trion complexes in bilayer and moiré systems.\\

\emph{Acknowledgements}--- We are very grateful to Zihan Yan, Mina-low Schleith, Zhaoyu Han, and Tomohiro Soejima for collaboration on related projects. We acknowledge discussions with Ashvin Vishwanath, Bertrand Halperin, and Steven Kivelson. E.~K., Q.~L., and T.~W. are supported by NSF CAREER grant DMR award No. 2441781. The computations in this paper were run on the FASRC Cannon cluster supported by the FAS Division of Science Research Computing Group at Harvard University.

\begingroup
\renewcommand{\addcontentsline}[3]{}
\ifdefined\DeclarePrefChars\DeclarePrefChars{'’-}\else\fi  \newcommand{\noop}[1]{}
\endgroup

\begin{appendix}
\onecolumngrid
\newpage
\makeatletter 
\setcounter{page}{1} 
\begin{center}
\textbf{\large Supplementary material for ``\@title ''}

\vspace{10pt}

\end{center}
\vspace{20pt}

\setcounter{figure}{0}
\setcounter{section}{0}
\setcounter{equation}{0}
\renewcommand{\thefigure}{S\arabic{figure}}
\renewcommand{\theHfigure}{S\arabic{figure}}
\addtocontents{toc}{\protect\setcounter{tocdepth}{2}}     
\tableofcontents

\section{Review of geometric quantization}

In this section, we will present a self-contained derivation of the Hamiltonian operator from the path integral using geometric quantization (Berezin-Teoplitz quantization). This is valid since our path integral is defined on a K\"ahler manifold. A K\"ahler manifold is a $2m$-dimensional manifold with a complex structure, a symplectic structure, and a metric that are compatible with each other. In our problem, the K\"ahler potential is given by $\K(\bar \xi, \xi) = \ln \langle \xi|\xi \rangle$ where $\xi = (\xi_1,\dots,\xi_m)$ describes $N_h$ complex coordinates describing quasihole positions, $|\xi \rangle$ is a set of states spanning the Hilbert space of interest that depends holomorphically on $\xi$. The K\"ahler potential determines the effective magnetic field for the $l$-th quasihole via
\begin{equation}
    {\mathcal B}_l(\bar \xi,\xi) = 2\partial_{\bar \xi_l} \partial_{\xi_l} \K(\bar \xi, \xi)
\end{equation}
The Hilbert space can be identified with anti-holomorphic functions of $\xi$ with the appropriate boundary condition $f(\bar \xi) = \langle \xi|f \rangle$ and inner product that can be constructed from the knowledge of $\K$. This is done by introducing the measure $d\mu_\xi$ defined by requiring that the reproducing Kernel relation 

\begin{equation}
    \int d\mu_\xi \, e^{\K(\bar \omega,\xi)} f(\bar \xi) = f(\bar \omega), 
    \label{RepKernel}
\end{equation}
is satisfied for any anti-holomorphic function $f(\bar \xi)$. Here, $\K(\bar \xi, \omega)$ is obtained from $\K(\bar \xi, \xi)$ by analytic continuation. This identity is equivalent to the existence of a resolution of unity of the form 
\begin{equation}
    \int d \mu_\xi |\xi \rangle \langle \xi| = 1
\end{equation}
which leads to (\ref{RepKernel}) upon the identification $f(\bar \xi) = \langle \xi|f \rangle$. The measure $d\mu_\xi$ defines an inner product on the space of holomorphic functions given by
\begin{equation}
    \langle f|g \rangle = \int d\mu_\xi [f(\bar \xi)]^* g(\bar \xi)
\end{equation}

For a given quantum operator $\hat V$, the coherent state path integral associates a classical function known as the Q-symbol defined via
\begin{equation}
    U(\bar \xi, \xi) = \frac{\langle \xi| \hat V|\xi \rangle}{\langle \xi|\xi \rangle} = e^{-\K(\bar \xi, \xi)} V(\bar \xi, \xi), \qquad V(\bar \xi, \xi) =  \langle \xi| \hat V|\xi \rangle
\end{equation}
One can also define a different classical potential $U_P$, known as the Glauber-Sudarshan P-symbol \cite{GlauberI, GlauberII, Sudarshan} via
\begin{equation}
    \hat V = \int d \mu_\xi U_P(\bar \xi, \xi) |\xi \rangle \langle \xi|
\end{equation}
The two are related via
\begin{equation}
    U(\bar \omega, \omega) = e^{-\K(\bar \omega, \omega)} \int d\mu_\xi e^{\K(\bar \xi, \omega) + \K(\bar \omega, \xi)} U_P(\bar \xi, \xi) 
    \label{PtoQ}
\end{equation}

For the standard harmonic oscillator coherent states, $\K(\bar \xi,\xi) = \frac{1}{2} \bar \xi \xi$ which correspond to  $d\mu_\xi = \frac{d \xi d \bar \xi}{2\pi} e^{-\bar \xi \xi/2}$ and $[\hat \xi, \hat \xi^\dagger] = 2$ \footnote{The way we define $\hat \xi$ is related to the standard raising/lowering operators by a factor of $\sqrt{2}$}. Then the $Q$-symbol corresponds to \emph{normal-ordering} such that all factors of $\hat \xi^\dagger$ are brought to the left of $
\hat \xi$ followed by the replacement $\hat \xi^\dagger \mapsto \bar \xi$ and $\hat \xi \mapsto \xi$. On the other hand, the $P$-symbol corresponds to \emph{anti normal-ordering} where $\hat \xi$ is placed to the left of $\hat \xi^\dagger$ before the replacement with the classical variables. In this case, the mapping between $U$ and $U_P$ in this case is precisely the standard LLL projection (Weierstrass transform)
\begin{align}
    U_Q(\bar \omega, \omega) &= \int \frac{d^2 \v{\xi}}{2\pi} e^{-\frac{|\xi - \omega|^2}{2}} U_P(\bar \xi, \xi) = \int \frac{d^2 \v{\xi}}{2\pi} e^{-\frac{|\xi|^2}{2}} U_P(\bar \xi + \bar \omega, \xi + \omega) \nonumber \\
    &= \int \frac{d^2 \v{\xi}}{2\pi} e^{-\frac{|\xi|^2}{2}} e^{\bar \xi \partial_{\bar \omega} + \xi \partial_{\omega}} U_P(\bar \omega, \omega) = e^{2 \partial_\omega \partial_{\bar \omega}} U_P(\bar \omega, \omega) = e^{\frac{1}{2} \Delta_\omega} U_P(\bar \omega, \omega)
    \label{VQUniform}
\end{align}
We can check this is indeed the case for simple examples. For example, the $Q$ symbol corresponding to the operator $\hat \xi^\dagger \hat \xi$ is $\bar \xi \xi$ while the $P$ symbol is obtained by first anti-normal ordering $\hat \xi^\dagger \hat \xi = \hat \xi \hat \xi^\dagger - 2$ leading to the $P$ symbol $\xi \bar \xi - 2$. Clearly we have $e^{-2 \partial_\xi \partial_{\bar \xi}} \xi \bar \xi = (1 -2 \partial_\xi \partial_{\bar \xi}) \xi \bar \xi = \xi \bar \xi - 2$.

The Berezin-Teoplitz quantization provides a concrete recipe for constructing the Hilbert space operators from the K\"ahler given one of the potentials: $V(\bar \xi, \xi)$, the Q-symbol $U(\bar \xi, \xi)$ or the P-symbol $U_P(\bar \xi, \xi)$ via
\begin{equation}
    [\hat V f](\bar \omega) = \int d \mu_\xi e^{\K(\bar \omega, \xi)} U_P(\bar \xi, \xi) f(\bar \xi) = \int d \mu_z U(\bar \omega, z) e^{\K(\bar \omega, z)} f(\bar z) = \int d \mu_z V(\bar \omega, z) f(\bar z)
    \label{QuantizationV}
\end{equation}

\section{Effective potentials in different path integral representations}
In the main text, we formulate the quasihole dynamics through a path integral built from a resolution of identity of the form  $ \int d\mu_\xi |\psi_\xi\rangle\langle\psi_\xi|$. In the path-integral formulation, the measure $\mu$, Berry connection $\A$, and the potential $V$ are not independent objects. By a suitable redefinition of the quasihole states we can bring $\mu$ and $\A$ to the canonical form of a two-boson problem in the LLL at magnetic field $B=1/q$. All remaining microscopic complexity is encoded in the effective interaction. In the following, we will illustrate this for the case of two quasiholes.

Concretely, we expand the quasihole state in powers of the center-of-mass and relative coordinates, $|\psi_{\xi}\rangle = \sum_{n_0,n_1=0}^{\infty} \xi_{cm}^{n_0} \xi_{r}^{2n_1} | \psi_{n_0,n_1}\rangle$ where only even powers of $\xi_r$ appear due to exchange symmetry. In the thermodynamic limit, the K\"ahler potential takes the form
\begin{align}
    e^{\mathcal{K}(\bar\xi,\xi)} = \exp\bigg(\frac{|\xi_{cm}|^2}{q} + \frac{|\xi_r|^2}{4q} + \mathcal{Q}(\xi_{r})\bigg) = e^{\frac{|\xi_{cm}|^2}{q}}\sum_{n_1=0}^{\infty} b_{n_1}|\xi_r|^{4n_1}.
\end{align}
Matching this expansion to the wavefunction expansion above fixes the normalization $\mathcal{N}_{n_0,n_1} =\langle \psi_{n_0,n_1}| \psi_{n_0,n_1}\rangle = \frac{b_{n_1}}{q^{n_0}n_0!}$.

Next, we introduce a canonical orthonormal basis in the quasihole-coordinate representation, $\varphi_{n_0,n_1}(\xi) =\frac{1}{\sqrt{4^{2n_1}q^{n_0+2n_1} n_0!(2n_1)!}} \xi_{cm}^{n_0} \xi_{r}^{2n_1}$, which is orthonormal with respect to $\langle f,g \rangle_{\xi}:=\int \frac{ d ^2\xi_{cm}  d ^2 \xi_r}{4\pi^2 q^2} e^{-\frac{|\xi_{cm}|^2}{q}- \frac{|\xi_r|^2}{4q}} f^*(\xi) g(\xi)$. We then define normalized coherent states $|\tilde\psi_{\xi}\rangle = \sum_{n_0,n_1} \frac{1}{\sqrt{\mathcal{N}_{n_0,n_1}}} \varphi_{n_0,n_1}(\xi) |\psi_{n_0,n_1}\rangle$. 

With these definitions, the resolution of identity and the inner product take the form
\begin{align}
     \mathds{1} = \int \frac{ d ^2\xi_{cm}  d ^2 \xi_r}{4\pi^2 q^2} e^{-\frac{|\xi_{cm}|^2}{q}- \frac{|\xi_r|^2}{4q}} |\tilde\psi_{\xi} \rangle \langle \tilde\psi_{\xi}|,\qquad \langle \tilde \psi_{\xi'}| \tilde\psi_{\xi} \rangle = \exp\bigg( \frac{\xi_{cm} \bar \xi_{cm}'}{q} \bigg)\cosh\bigg( \frac{\xi_{r} \bar \xi_{r}'}{4q} \bigg)
\end{align}
These expressions are identical to those obtained from the symmetrized two-boson basis in the LLL at external magnetic field $B = 1/q$. The corresponding path integral is
\begin{align}
    \textrm{Tr} e^{-\beta H} = \int_{\substack{\xi(0)=\xi_{i}\\\xi(\beta)=\xi_{f}}} \mathcal{D}(\bar\xi,\xi) e^{- \int_0^\beta  d  \tau \mathcal L_{\rm sym}(\xi,\dot\xi)},
\end{align}
where 
\begin{align}
    \mathcal L_{\rm sym}(\xi,\dot\xi) = \frac{1}{q} \bar\xi_{cm} \dot \xi_{cm} + \frac{1}{4q}\bar\xi_r \dot\xi_r  - \frac{1}{2q}\frac{\bar\xi_r}{e^{\frac{|\xi_r|^2}{2q}}+1}\dot\xi_r + U_{\rm sym}(\bar\xi,\xi),\qquad U_{\rm sym}(\bar\xi,\xi) = \frac{\langle \tilde\psi_\xi|H| \tilde\psi_{\xi}\rangle}{\langle \tilde\psi_\xi| \tilde\psi_{\xi}\rangle}.
\end{align}
With one particle pinned at the origin, the other experiences an effective magnetic field induced by the Berry connection, 
\begin{align}
    B_{\rm sym}(\bar\xi,\xi) = \frac{1}{q}\frac{e^{\frac{|\xi|^2}{2q}}}{e^{\frac{|\xi|^2}{2q}}+1} + \frac{|\xi|^2}{2q^2}\frac{e^{\frac{|\xi|^2}{2q}}}{(e^{\frac{|\xi|^2}{2q}}+1)^2}.
\end{align}

An alternative path integral representation of two bosons in the LLL uses a distinguishable-particle basis,
\begin{align}
     \textrm{Tr} e^{-\beta H} =  \frac{1}{2!} \sum_{\pi} \int_{\substack{\xi_l(0)=\xi_{i,l}\\\xi_l(\beta)=\xi_{f,\pi(l)}}} \mathcal{D}(\bar\xi,\xi) e^{- \int_0^\beta  d  \tau \mathcal L_{\mathrm{dist}}(\xi,\dot\xi)},
     \label{eq:worldline}
\end{align}
where $\mathcal L_{\rm dist} = \frac{1}{2q} \bar\xi_1\dot\xi_1 + \frac{1}{2q} \bar\xi_2\dot\xi_2  + U_{\rm dist}(\bar\xi,\xi)$. 

\begin{figure}[h]
    \centering
    \includegraphics[width=0.65\linewidth]{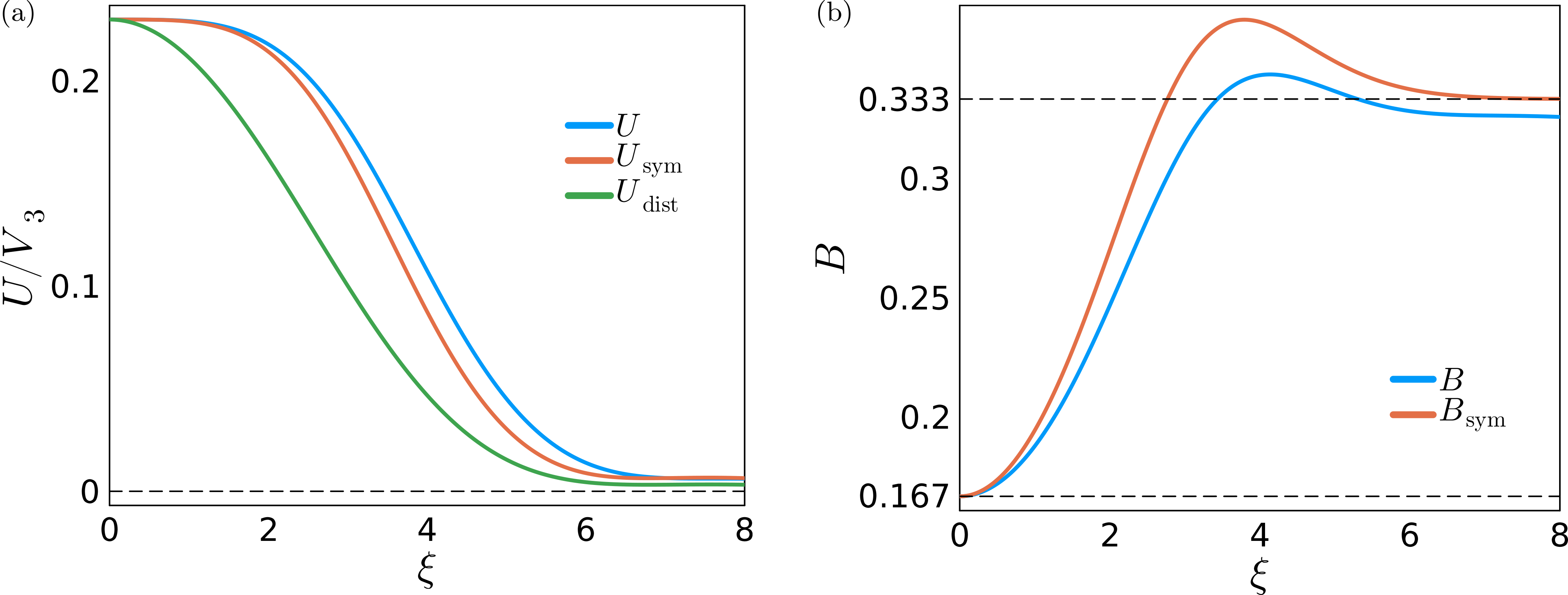}
    \caption{Comparison of effective potentials and Berry-curvature-induced magnetic fields obtained from different path-integral representations. (a) Effective potentials $U$, $U_{\mathrm{sym}}$ and $U_{\mathrm{dist}}$ as functions of the quasihole separation $\xi$ for $\lambda = 2$. (b) Effective magnetic fields $B$ and $B_{\mathrm{sym}}$ as functions of $\xi$. Dashed lines mark $1/2q$ and $1/q$.}
    \label{fig-threeVandB}
\end{figure}
FIG.~\ref{fig-threeVandB} summarizes the effective potentials and Berry-curvature-induced magnetic fields obtained in the original, symmetrized, and distinguishable-particle representations. The effective magnetic fields extracted from the original basis and from the symmetrized basis behave similarly: they start at $1/(2q)$ at $\xi=0$, increase with $|\xi|$, exhibit a mild overshoot, and finally saturate to $1/q$ at large $|\xi|$. They differ from the uniform external field $1/q$ because the symmetrized basis modifies the Berry connection.

The effective potentials reconstructed from the two-boson spectrum can be written as
\begin{align}
    U_{\rm sym}(\bar\xi,\xi) =  \frac{1}{\cosh\big( \frac{\xi \bar \xi}{4q} \big)}\sum_{n=0}^{\infty} \frac{|\xi|^{4n}}{(4q)^{2n} (2n)!} E_{2n},\qquad U_{\rm dist}(\bar\xi,\xi) = e^{-\frac{|\xi|^2}{4q}}\sum_{n} \frac{|\xi|^{4n}}{(4q)^{2n} (2n)!} E_{2n},
    \label{VEexpr}
\end{align}
where $E_{L}$ denotes the energy in the angular-momentum channel $L$. Since the symmetrized two-boson sector contains only even relative angular momentum, we set $E_{2n+1}=0$. The three effective potentials $U(\bar\xi,\xi)$, $U_{\rm sym}(\bar\xi,\xi)$ and $U_{\rm dist}(\bar\xi,\xi)$ are qualitatively very similar: they are monotonically decaying and everywhere repulsive, even though the two-quasihole spectrum contains a bound state. To understand this apparent contradiction, we focus on $U_{\rm dist}(\bar\xi,\xi)$ and emphasize that $U_{\rm dist}(\bar\xi,\xi)$ is not the bare interaction. Instead, it is the interaction projected onto the LLL. In particular, $U_{\rm dist}$ relates to the unprojected interaction $U_{\textrm{int}}$ by the Weierstrass transformation, $U_{\rm dist}(\bar\xi,\xi) = \int \frac{ d ^2 \xi'}{4\pi q} U_{\textrm{int}}(\bar\xi',\xi') e^{-\frac{|\xi-\xi'|^2}{4q}}$, which smears out structure on the scale of $\ell_B$. $U_{\textrm{int}}$ is oscillatory on the scale of $\ell_B$ and becomes negative in certain regions. After LLL projection, these oscillations are washed out, so $U_{\rm dist}$ appears repulsive and decaying.

\section{Bound states from repulsive interaction in the LLL}
In this section, we will provide an explanation for the seemingly paradoxical emergence of a bound state from a purely repulsive interaction. 

\subsection{Bound states in LLL in a uniform field}
To start, we notice that the Lagrangian for the two quasihole problem, when focusing on the relative coordinate, takes the form of the Lagrangian for a particle in the LLL with magnetic field $B_{\rm eff}(\bar \xi_r, \xi_r)$ and potential $U(\bar \xi_r, \xi_r)$. For simplicity, we will first consider the case of a uniform magnetic field. This corresponds to the worldline path integral \ref{eq:worldline} for distinguishable particles. However, below we will not worry about its origin since we are mainly interested in the physics responsible for the appearance of bound states. 

The LLL projection promotes $\xi_r$  to an operator with $[\hat \xi_r, \hat \xi_r^\dagger] = 4q$. This identifies them with the standard harmonic oscillator creation/annihilation operators via
\begin{equation}
    \hat \xi_r = 2\sqrt{q} a, \qquad \hat \xi_r^\dagger = 2\sqrt{q} a^\dagger
\end{equation}
Then the quantum Hamiltonian corresponding to this Lagrangian is obtained by the standard normal ordering procedure placing $\bar \xi_r$ to the left of $\xi_r$ and promoting them to operators. This is conveniently done by introducing the Fourier transform $U(\bar \xi_r, \xi_r) = \int \frac{d^2 \bp}{(2\pi)^2} e^{i \bp \cdot \bxi_r} U(\bp)$ and inserting a resolution of unity in terms of the harmonic oscillator eigenstates $a^\dagger a|n \rangle = n|n \rangle$
\begin{gather}
    \hat H_{\rm QH} = \int \frac{d^2 \bp}{(2\pi)^2}  U(\bp) e^{i p \sqrt{q} \hat a^\dagger} e^{i \bar p \sqrt{q} a} = \sum_{n} E_{n} |n \rangle \langle n|, \\
    E_{n} = \int \frac{d^2 \bp}{(2\pi)^2}  U(\bp) L_n(q \bp^2) = \int \frac{d^2 \bp \, d^2 \bxi}{(2\pi)^2} e^{-i \bp \cdot \bxi} U(\bxi) L_n(q \bp^2)  = L_n(-q \Delta_\bxi) U(\bxi)|_{\bxi = 0}
\end{gather}
This is consistent with the expression in Eq.~\ref{VEexpr} (there, the only non-zero terms correspond to even $n$). Importantly, $E_n$ can be negative even for $U$, which is strictly positive and monotonically decreasing. 

Now the potential $U$ corresponds to the $Q$ symbol corresponding to the expectation value evaluated in the coherent state basis that has a built-in smearing on the scale of $\ell_{B_{\rm eff}} = \sqrt{2q}$. The potential that provides a better indicator for bound states is the $P$-symbol. It is a well-studied object in quantum optics since the pioneering work of Glauber and Sudarshan \cite{GlauberI, GlauberII, Sudarshan}. It is known that the P-symbol can be negative and its negativity diagnoses how quantum a state of light is. Using Eq.~\ref{VQUniform}, we can write the relation between the P and Q symbols as
\begin{equation}
     U(\bxi) = e^{q \Delta_\bxi}U_P(\bxi) 
\end{equation}
While this relation can be formally inverted to write $U_P(\br)$, the inverse is usually not well-defined since it corresponds to 'unsmearing' or 'unprojection' operator. The energy eigenvalues can be expressed in terms of $U_P$ as
\begin{align}
    E_n &= \int \frac{d^2 \bp \, d^2 \bxi}{(2\pi)^2} e^{-i \bp \cdot \bxi} L_n(q \bp^2) e^{q \Delta_\bxi} U_P(\bxi) = \int \frac{d^2 \bp \, d^2 \bxi}{(2\pi)^2} e^{-q \bp^2} e^{-i \bp \cdot \bxi} L_n(q \bp^2) U_P(\bxi) \notag\\
    &= \frac{1}{n!(4q)^{n+1}\pi} \int d^2 \bxi \, \bxi^{2n} U_P(\bxi) e^{-\frac{1}{4q} \bxi^2}
\end{align}
The function integrated again $U_P$ is non-negative and it is peaked around $\xi_n \approx 2 \sqrt{q n}$. This means that a bound state is possible only if $U_P$ is negative somewhere. This is an important consistency check; the unprojected Hamiltonian is given by $\frac{1}{2m}\big(\bp - q \bA(\bxi)\big)^2 + U_p(\bxi)$ whose energy is bounded from below by the minimum of $U_P$. Since the projection cannot lower the energy, any energy eigenvalue is bounded from below by the minimum of $U_P$.

\subsection{Simple example: spin 1 Hamiltonians}
The simplest possible example to illustrate this effect is to consider a spin problem. It is well-known that the Hilbert space of a spin $S$ is equivalent to the LLL on the sphere with monopole charge $2S$. Now consider the spin Hamiltonian $H = 2S_z^2 - 1$ which has eigenvalues $+1, -1, +1$ for $m = -1, 0, 1$. On the other hand, for the coherent state path integral, we have the coherent states 
\begin{equation}
    |\theta,\phi \rangle = \cos^2 \frac{\theta}{2} |1 \rangle + \frac{\sin \theta}{\sqrt{2}} e^{-i \phi}|0 \rangle + \cos^2 \frac{\theta}{2} e^{-2i \phi} |-1 \rangle 
\end{equation}
with
\begin{equation}
    U(\theta) = \langle \theta,\phi|H|\theta,\phi \rangle = 2 \langle \theta,\phi|S_z^2|\theta,\phi\rangle - 1 = 2 - \sin^2 \theta -1 = \cos^2 \theta \geq 0
\end{equation}
This demonstrates that the positivity of the coherent state expectation value does not imply the absence of negative energy eigenstates.

\section{Connection to plasma limit and Chern-Simons gauge transformation}
Consider a basis of single quasihole states $\chi_n(\xi_l)$, where $\xi_l$ is a single complex coordinate (to be contrasted with the multiquasihole basis introduced in the main text $\phi_\alpha(\xi)$). For definiteness, we will consider the sphere geometry where such basis has dimension $N_e + 1$ and can be chosen explicitly to be 
\begin{equation}
    \chi_{m}(\xi_l) = \beta_{N_e,m} \xi_l^m, \qquad \beta_{N_e, m} =  \sqrt{\frac{N_e+1}{4\pi} \binom{N_e}{m}}, \qquad m=0,\dots,N_e
\end{equation}
which is orthonormal under the integration measure on the sphere $d\nu_{\xi_l} =  \frac{4 d^2 \xi_l}{(1 + |\xi_l|^2)^{N_e + 2}}$. A basis for $N_h$ quasihole states can then be labelled by a set of $N_h$ integers $\eta = \{n_l\}$ with $n_1 \leq n_2 \dots n_{N_h}$ such that
\begin{equation}
    \chi_{\eta = \{n_l\}}(\xi) = \mathcal N_\eta \mathcal S \prod_{l=1}^{N_h} \chi_{n_l}(\xi_l)
\end{equation}
where $\mathcal S$ is the symmetrization operator and $\mathcal N_\eta$ is a normalization constant. Concretely, if we define $\mathcal S$ as the sum over permutations (without dividing by $N_h!$), then this factor is the standard bosonic factor $\frac{1}{\sqrt{N! \prod_k n_k}}$ where $n_k$ is the occupation of the $k$-th mode (how many times it appears). This is nothing but the standard symmetrized basis for bosons whose dimension is ${\rm dim}(N_e, N_h) = \binom{N_e + N_h}{N_h}$. Any object of the form $\langle \psi|\xi \rangle$ can be expanded in this basis
\begin{equation}
    \langle \psi|\xi \rangle = \sum_\eta \psi_\eta \chi_\eta(\xi), \qquad \psi_\eta = \int d\nu_\xi \chi^*_\eta(\xi) \langle \psi|\xi \rangle, \qquad d\nu_\xi = \prod_{l=1}^{N_h} d\nu_{\xi_l}
    \label{PsiChiExpansion}
\end{equation}

We can now consider an orthonormal basis of the quasihole wavefunctions expressed in electron coordinates such that
\begin{equation}
    |\xi \rangle = \sum_I \psi_I(\xi) |\psi_I \rangle, \qquad \langle \psi_I|\psi_J \rangle = \delta_{I,J}
\end{equation}
where $I = 1, \dots, {\rm dim}(N_e, N_h)$. Using Eq.~\ref{PsiChiExpansion}, we can expand $\psi_I(\xi)$ in terms of $\chi_\eta(\xi)$:
\begin{equation}
    \psi_I(\xi) = \sum_\eta \Gamma_{I,\eta} \chi_\eta(\xi)
\end{equation}
Here, $\Gamma$ is a square matrix that relates two kinds of bases: one is an orthonormal basis of bosons and the other is the orthonormal basis in the space of quasiholes with the proper measure defined through wavefunction overlap. The matrix $\Gamma$ plays a role similar to a 'flux attachment' gauge transformation which turns anyons to bosons. In the following, we will make this picture more precise. 

In the limit where the distance between quasiholes is much larger than $\ell_B$, we can use plasma screening to simplify the K\"ahler potential (here we have also assumed $N_e \gg N_h$ so effects from the curvature of the sphere can be neglected).
\begin{equation}
    \K(\bar \xi, \xi) = -\frac{1}{q} \sum_{i < j} \ln (\xi_i - \xi_j) -\frac{1}{q} \sum_{i < j} \ln (\bar \xi_i - \bar \xi_j) + \frac{1}{2q} \sum_l \bar \xi_l \xi_l
\end{equation}
In the limit of large $N_e$ where the curvature of the sphere can be neglected, the product measure on the sphere simply reduces to the product measure on the plane $d\nu_{\xi} \approx \prod_l d^2 \xi_l e^{-\frac{1}{2q} |\xi_l|^2}$. Furthermore, the simple form of the K\"ahler potential implies the integration measure $d\mu_\xi$ is
\begin{equation}
    d\mu_\xi = |\Delta(\xi)|^{2/q} \prod_l d^2 \xi_l e^{-\frac{1}{2q} |\xi_l|^2} = |\Delta(\xi)|^{2/q} d\nu_\xi, \qquad \Delta(\xi) = \prod_{i<j} (\xi_i - \xi_j)
\end{equation}
because 
\begin{align}
    \int d\mu_\xi e^{\K(\bar\omega,\xi)} f(\bar\xi) = \bar\Delta^{-\frac{1}{q}}(\omega) \int \prod_ld^2\xi_l  e^{-\frac{\sum_{l}\bar\xi_l\xi_l}{2q}+\frac{\sum_l \bar\omega_l\xi_l}{2q}} \bar\Delta^{\frac{1}{q}}(\xi) f(\bar\xi) = f(\bar\omega)
\end{align}
where we restrict to the regime $|\xi_i-\xi_j|\gg \ell_B$ so that no branch cut of $\Delta^{1/q}$ is encountered in the region of interest.
We can now write
\begin{equation}
    \delta_{I,J} = \int d\mu_\xi \psi^*_I(\xi) \psi_J(\xi) = \int d\nu_\xi |\Delta(\xi)|^{2/q} \psi^*_I(\xi) \psi_J(\xi)
\end{equation}
However, since the functions $\chi_\nu(\xi)$ are orthonormal with respect to the integration measure $d\nu_\xi$, we can make the identification
\begin{equation}
    \psi_{I = \nu}(\xi) = \Delta(\xi)^{-1/q} \chi_\nu(\xi)
\end{equation}
which tells us that the square matrix $\Gamma_{I,\eta}$ can be expressed as
\begin{equation}
    \Gamma_{\eta',\eta} = \int d\nu_\xi \chi^*_\nu(\xi) \Delta(\xi)^{-1/q} \chi_{\nu'}(\xi)
\end{equation}
That is, $\Gamma$ is the matrix representation of the Chern-Simons flux attachment transformation. Note that because we are working in terms of holomorphic functions, this differs from the conventional form. In particular, it is not unitary. This will be fixed if we define $\psi$ and $\chi$ to include the integration measure as is usual in LLL physics
\begin{equation}
    \tilde \psi_I(\xi) =  \psi_I(\xi) e^{-\frac{1}{2} \K(\bar \xi,\xi)}, \qquad \tilde \chi_\eta(\xi) = \chi_\eta(\xi) e^{-\frac{1}{4q} \sum_l \bar \xi_l \xi_l}
\end{equation}
Then, the relation between the two becomes
\begin{equation}
    \tilde \psi_\nu(\xi) = \left(\frac{\Delta(\xi)}{|\Delta(\xi)|}\right)^{-1/q} \tilde \chi_\nu(\xi)
\end{equation}
with the matrix relating them becoming
\begin{equation}
    \tilde \Gamma_{\eta',\eta} = \int d^2 \xi \, \tilde \chi^*_\nu(\xi) \left(\frac{\Delta(\xi)}{|\Delta(\xi)|}\right)^{-1/q} \tilde \chi_{\nu'}(\xi)
\end{equation}
which reduces to the standard form of the flux attachment transformation. Note that
\begin{equation}
    e^{\K(\bar \xi,\xi)} = \langle \xi|\xi \rangle = \sum_{I} \psi_I^*(\xi) \psi_I(\xi) = \sum_{\eta,\eta'} K_{\eta,\eta'} \chi^*_\eta(\xi) \chi_{\eta'}(\xi)
\end{equation}
Thus, $K = \Gamma^\dagger \Gamma$ and $\tilde \Gamma$ is associated with the singular value decomposition of $\Gamma = V D \tilde \Gamma$. Thus, $\tilde \Gamma$ can be identified with the unitary that diagonalizes $K$.

\section{Basis construction for quasihole Hilbert space}
A basis for quasihole Hilbert space spanned by $\{|\xi\rangle\}$ can be constructed from the coefficients of the polynomial expansion of $|\xi\rangle$ in the quasihole coordinates $\xi$. Since $\hat V$ preserves the rotational symmetry, it is convenient to remove the center-of-mass component by fixing $\xi_{cm} = 0$, and to work within the subspace of definite angular momentum. Label sectors by the angular-momentum deficit, $L = L_{z,\textrm{max}} - L_z$. A basis for the $L$-sector is
\begin{equation}
    \psi^{(L)}_{m} = m_{\lambda_m}(\{z^{-1}\}) \prod_{i=1}^{N_e} z_i^{N_h} \prod_{i<j}^{N_e} (z_i - z_j)^q \prod_{i=1}^{N_e} \Gamma(|z_i|).
\end{equation}
where $m_{\lambda_m}(\{x\})$ denotes the monomial symmetric polynomial. $m=(m_2,m_3,\dots,m_{N_h})$, $m_i \in \mathbb{Z}_{\ge 0}$ specifies the partition $\lambda_m=(2^{m_2},3^{m_3},\dots,N_h^{m_{N_h}})$ and the admissible basis states in a given sector are subject to the constraint $\sum_{a=2}^{N_h} a m_a = L$.
With this non-orthonormal basis, we introduce the overlap matrix and the interaction matrix within the $L$-sector, $K^{(L)}_{m',m} =\langle \psi_{m'}^{(L)} | \psi_{m}^{(L)}\rangle$, $V^{(L)}_{m',m} =\langle \psi_{m'}^{(L)} |\hat V| \psi_{m}^{(L)}\rangle$. The spectrum in each angular-momentum sector is then obtained from the eigenvalue problem $V^{(L)} f = E K^{(L)} f$. The overlap and interaction matrix elements are evaluated numerically using Monte Carlo sampling. We do not encounter a significant sign/phase problem in the Monte Carlo computation. Within a fixed $L$-sector, the basis states differ only by the factor $m_{\lambda_m}(\{z^{-1}\})$. When $L$ is not large, this factor has low degree and does not induce rapidly oscillating phases in the Monte Carlo integrand. As an example, we list admissible $\{m\}$ for three quasiholes and $L \leq 10$ in TABLE~\ref{tab:Nh3_solutions}.

\begin{table}[h]
\centering
\begin{tabular}{c @{\hspace{1.5em}} l}
\hline
$L$ & Solutions $(m_2,m_3)$  \\
\hline
0  & $(0,0)$ \\
1  & no solution \\
2  & $(1,0)$ \\
3  & $(0,1)$ \\
4  & $(2,0)$ \\
5  & $(1,1)$ \\
6  & $(3,0)$, $(0,2)$ \\
7  & $(2,1)$ \\
8  & $(4,0)$, $(1,2)$ \\
9  & $(3,1)$, $(0,3)$ \\
10 & $(5,0)$, $(2,2)$ \\
\hline
\end{tabular}
\caption{Solutions $(m_2,m_3)$ to $2m_2+3m_3=L$ for three quasiholes ($N_h=3$) and $L\le 10$.}
\label{tab:Nh3_solutions}
\end{table}

For the two-quasihole problem, the effective potential $U$ between two quasiholes as a function of the relative position is
\begin{align}
    U(\bar\xi_r,\xi_r) = \frac{\sum_{L = 2\mathbbm{Z} \geq 0} V_{L} \bar \xi_r^{L} \xi_r^{L}}{\sum_{L = 2\mathbbm{Z} \geq 0} K_{L} \bar \xi_r^{L} \xi_r^{L}},
\end{align}
and the effective magnetic field $B$ felt by one quasihole as a function of the distance from the other quasihole is
\begin{align}
    B(\bar\xi_r,\xi_r) = \frac{1}{2q} + 2 \frac{\sum_{L = 2\mathbbm{Z} \geq 0} L^2 K_{L} \bar \xi_r^{L-1} \xi_r^{L-1}}{\sum_{L = 2\mathbbm{Z} \geq 0} K_{L} \bar \xi_r^{L} \xi_r^{L}} - 2 |\xi_r|^2 \bigg( \frac{\sum_{L = 2\mathbbm{Z} \geq 0} L K_{L} \bar \xi_r^{L-1} \xi_r^{L-1}}{\sum_{L = 2\mathbbm{Z} \geq 0} K_{L} \bar \xi_r^{L} \xi_r^{L}} \bigg)^2.
\end{align}
We evaluate the ratios \(K_{L+1}/K_L\) and the sector energies \(E_L \equiv V_L/K_L\) via Monte Carlo sampling, and reconstruct \(K_L\) and \(V_L\) up to \(L<40\). The quantities \(U(\bar\xi_r,\xi_r)\) and \(B(\bar\xi_r,\xi_r)\) then follow from the above series expressions.

\section{Additional numerical results}
In this section, we present additional numerical results supporting the main text.

\subsection{Monte Carlo for the fixed center-of-mass and for a fixed angular momentum}
\begin{figure}[h]
    \centering
    \includegraphics[width=0.65\linewidth]{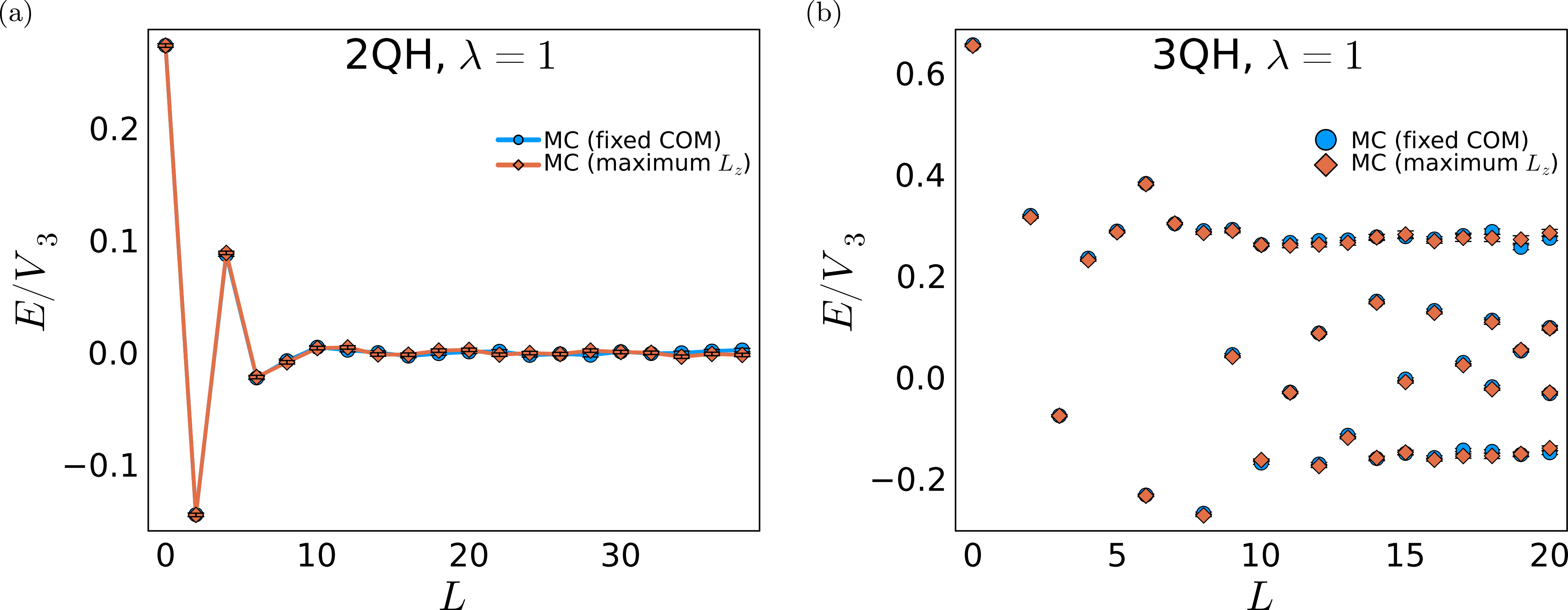}
    \caption{MC spectra on the sphere at $\nu=1/3$ for $N_e=100$ with Yukawa-screened Coulomb interaction with $\lambda=1$.(a,b) Spectra for $N_h=2$ and $3$. Each panel shows MC results for the fixed center-of-mass and MC results for maximum $L_z$.}
    \label{fig-MC}
\end{figure}

All results in the main text except those obtained at small systems for ED benchmarking were obtained for large system size $N_e = 100$. In this case, we argued in the main text that there is an emergent translation symmetry that allows us to simply set $\xi_{\rm CM} = 0$. Here, we test this claim by comparing MC done by fixing the center-of-mass and MC done by fixing the total $L_z$, which we choose to be the maximum value i.e. we impose $L_+ |\psi\rangle> = 0$. For small finite system size, there are discrepancies between these two but these discrepancies disappear as we increase the system size. Fig.~\ref{fig-MC} shows a comparison between the two for $N_e = 100$.

\subsection{Quasihole spectra at 1/5 filling}
\begin{figure}[h]
    \centering
    \includegraphics[width=0.65\linewidth]{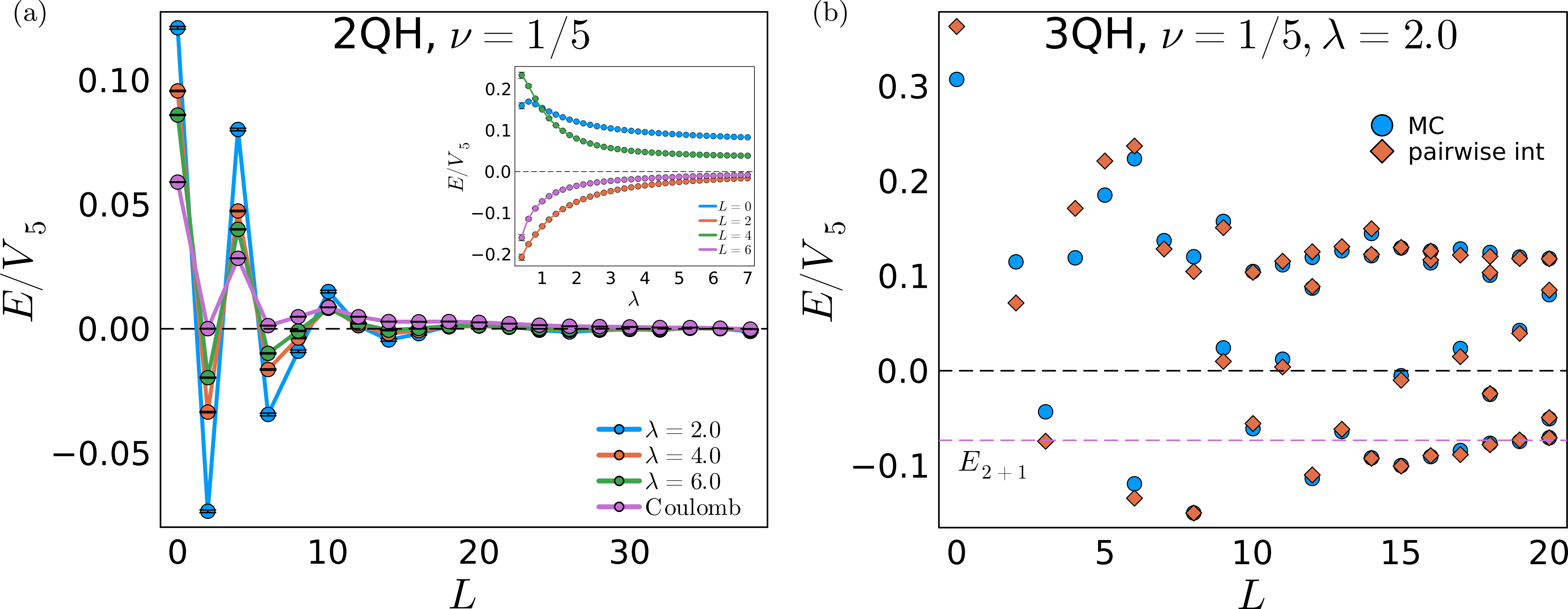}
    \caption{Energy spectra on the sphere for $N_e=100$ and $N_h=2,3$ at $\nu=1/5$. (a,b) Two- and three-quasihole spectra relative to well-separated quasiholes, in which energies are measured in units of the Haldane pseudopotential $V_5$ component.} 
    \label{fig-spectra-q5}
\end{figure}
The main text focuses on $\nu=1/3$. Here we extend our analysis to $\nu=1/5$ and study the two- and three-quasihole cases. The quasihole spectra shown in FIG.~\ref{fig-spectra-q5}(a,b) exhibit both similarities and differences compared with $\nu=1/3$. In both fillings, they display oscillatory features. The energy approaches that of separated quasihole clusters at large $L$, and the lowest-energy state occurs at $L=2$ for $N_h=2$ and $L=8$ for $N_h=3$. Moreover, for the three-quasihole problem, the pairwise-interaction approximation remains accurate for the low-energy levels. A key difference is that the oscillations at $\nu=1/5$ are stronger. This can be explained by the lower electron density at $\nu=1/5$, which enhances the tendency toward Wigner crystallization.

\subsection{Quasihole spectra for double-gate screened Coulomb interaction}
\begin{figure}[h]
    \centering
    \includegraphics[width=0.65\linewidth]{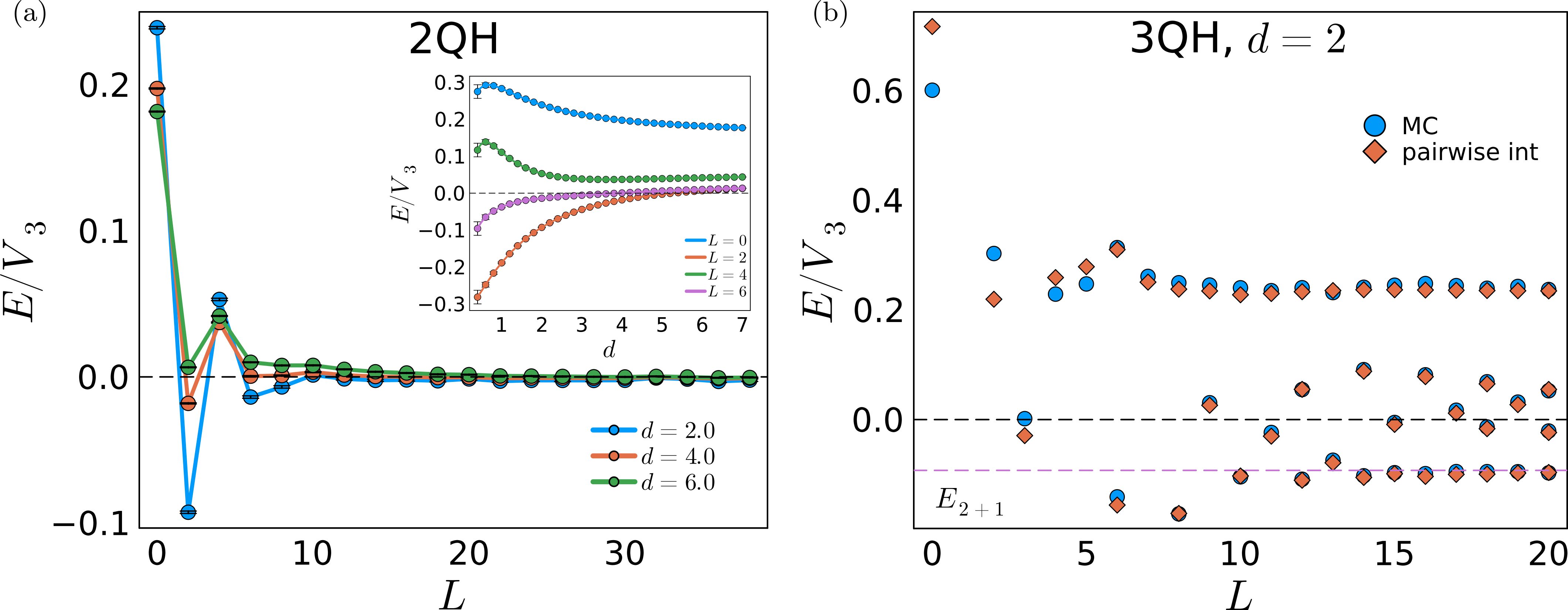}
    \caption{Energy spectra on the sphere for $N_e=100$ and $N_h=2,3$. (a,b) Two- and three-quasihole spectra, relative to well-separated quasiholes, for the double-gate screened Coulomb interaction.} 
    \label{fig-spectra-sgc}
\end{figure}
We present the quasihole spectra for $N_h=2$ and $N_h=3$ obtained for the double-gate screened Coulomb interaction whose Fourier transform is 
\begin{align}
    V(q) = \frac{2\pi}{q} \tanh(qd),
\end{align}
where $d$ denotes the gate distance. We find that the resulting spectra shown in FIG.~\ref{fig-spectra-sgc}(a,b) are qualitatively similar to the Yukawa spectra shown in the main text. This suggests that our results do not rely on the specific form of the screened interaction.

\subsection{Error analysis}
In this subsection, we quantitatively assess the Monte Carlo statistical uncertainty $\delta E_{\rm MC}$ and the deviation introduced by the pairwise-interaction approximation.

\begin{figure}[h]
    \centering
    \includegraphics[width=0.65\linewidth]{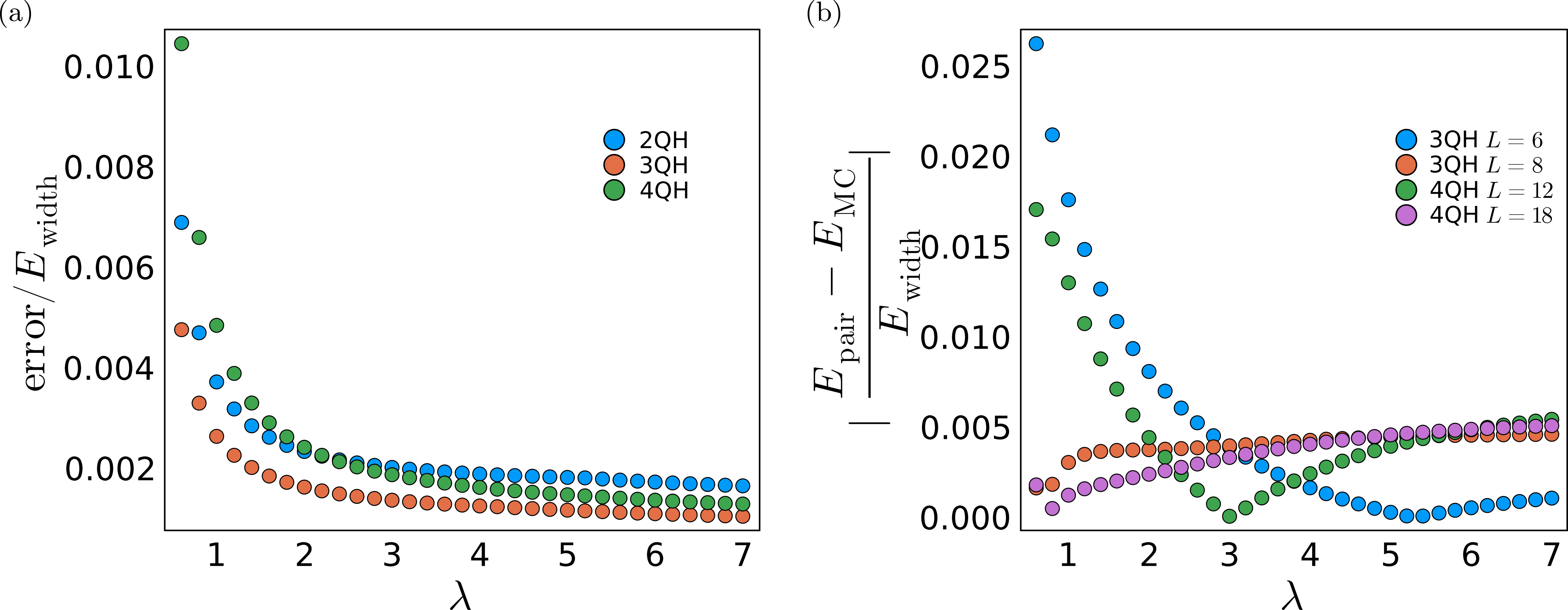}
    \caption{(a) Relative MC uncertainty of the ground-state energy $|\delta E_{\rm MC}/E_{\rm width}|$, as a function of the screening length $\lambda$ for $N_h=2,3,4$. (b) Relative deviation of the pairwise-interaction approximation from the MC results as a function of the screening length $\lambda$ for $N_h=3,4$.}
    \label{fig-error}
\end{figure}

FIG.~\ref{fig-error}(a) shows the relative error $\delta E_{\rm MC}/E_{\rm width}$ as a function of $\lambda$, where $E_{\rm width}$ denotes the spectral width. We find that the relative MC uncertainty is small ($\lesssim 1\%$) for $N_h=2,3,4$. We therefore conclude that MC statistical uncertainty does not affect our conclusions regarding the existence of bound states.

FIG.~\ref{fig-error}(b) compares the energy of low-energy states at $L=6$ and $8$ for $N_h=3$, and $L=12$ and $18$ for $N_h=4$, obtained from MC with those obtained from the pairwise interaction approximation. The discrepancy is small relative to the spectral width, indicating that the pairwise approximation is reliable for the low-energy states of interest.

\subsection{Finite-size scaling analysis}
We consider two finite-size extrapolations: the effective magnetic field and the ground-state energy.

\begin{figure}[h]
    \centering
    \includegraphics[width=0.65\linewidth]{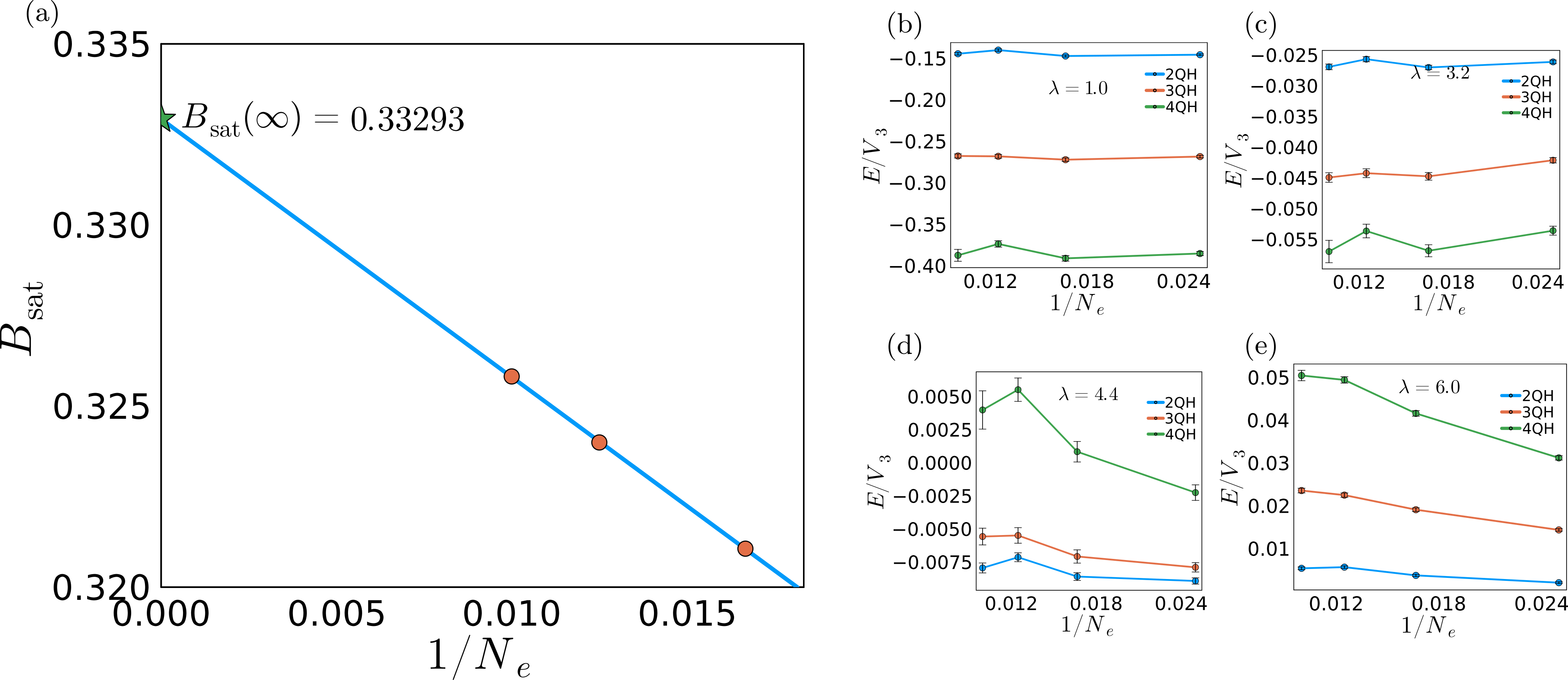}
    \caption{Finite-size scaling of the saturated effective magnetic field and bound-state energies. (a) Saturated effective field $B_{\mathrm{sat}}$ at large quasihole separation plotted versus $1/N_e$. The solid line is a linear fit yielding $B_0=0.33293$. (b-e) Bound-state energies as functions of $1/N_e$ for $N_h=2,3,4$, evaluated in the angular-momentum sectors where the bound states occur ($L=2,8,18$, respectively). Results are shown for Yukawa interactions with $\lambda=1.0,3.2,4.4,6.0$.}
    \label{fig-finitesize}
\end{figure}

In the limit where the two quasiholes are well separated, the effective magnetic field is expected to approach the external field, $B_{\rm ext}=1/q$. In the main text we observe that the saturated effective field at finite $N_e$ does not reach $1/3$ exactly. To understand this deviation, we extract the saturated value $B_{\rm{sat}}(N_e)$ for several electron numbers $N_e$ and fit it to $B_{\rm{sat}}(N_e)=B_0+\frac{c}{N_e}$. Fig.~\ref{fig-finitesize}(a) shows the linear fit. Extrapolating to $N_e\rightarrow\infty$ yields
\begin{align*}
    B_0=0.33293,
\end{align*}
with $R^2 = 0.999913$. This result is consistent with the expected thermodynamic-limit result $B_0=1/3$.

We analyze finite-size effects in the ground-state energy by computing the quasihole bound-state energy in the angular-momentum sector where the bound state occurs: $L=2$ for $N_h=2$, $L=8$ for $N_h=3$, and $L=18$ for $N_h=4$. FIG.~\ref{fig-finitesize}(b-e) displays the finite-size scaling of these bound-state energy for $\lambda=1.0,3.2,4.4,6.0$. For short-range interactions ($\lambda=1,3.2$), the bound state remains compact, and the finite-size drift is smaller than or comparable to our numerical uncertainty for all system sizes considered, except for a mild deviation at $N_e=40$ when $\lambda=3.2$, which disappears for $N_e\ge 60$. For longer-range interactions ($\lambda=4.4,6$), we observe a more pronounced size dependence for $N_h=4$ at $\lambda=4.4$ and all at $\lambda=6$; however, in this regime these quasihole clusters tend to separate into spatially distinct clusters. We therefore conclude that finite-size effects do not affect our conclusions.

\subsection{Ground-state quasihole wavefunctions: an ansatz}
Let $|\Psi\rangle$ denote the ground state in the $N_h$-quasihole sector (in numerics we choose $\lambda=2$). The corresponding wavefunction in the quasihole-coordinate representation is defined by $\Psi(\bar \xi) = \langle \xi | \Psi \rangle$. In our construction, the quasihole coordinates $\xi$ enter only through the factor $\prod_{j=1}^{N_h} (z-\xi_j) = \sum_{r=0}^{N_h} (-1)^r e_r(\xi) z^{N_h-r}$ where $e_r(\xi)$ are the elementary symmetric polynomials. Therefore, $\Psi(\bar\xi)$ can be expressed as a polynomial in $e_r(\bar\xi)$. Fixing the center-of-mass coordinate $\xi_{cm}=0$ removes the dependence on $e_1$, so the wavefunction can be constructed from monomials of the form $e_2^{n_2} e_3^{n_3}\cdots e_{N_h}^{n_{N_h}}$. 

We propose the following ansatz for the ground-state wavefunction: in the quasihole-coordinate representation it contains a universal Jastrow factor $\Delta(\bar\xi):=\prod_{i<j}(\bar \xi_i-\bar \xi_j)^2$, motivated by the fact that configurations in which two quasiholes coincide are energetically unfavorable, and are therefore suppressed in the ground state.

The $N_h=2$ ground state occurs at $L=2$. In the quasihole-coordinate representation, its wavefunction takes the form $(\bar\xi_1-\bar\xi_2)^2$, in agreement with the ansatz.

The $N_h=3$ ground state occurs at $L=8$. For three quasiholes, the Jastrow factor can be expressed as $\Delta = -27e_3^2 - 4e_2^3$. Using this, the $L=8$ ansatz wavefunction takes the form $\Psi_{\rm ansatz}(\bar\xi) = -27e_2e_3^2 - 4 e_2^4$. Equivalently, in a translationally invariant form, it can be written as 
\begin{align}
    \Psi_{\rm ansatz}(\bar\xi) = \sum_{i<j=1}^{3}(\bar\xi_i-\bar\xi_j)^2 \Delta(\bar\xi).
\end{align}
Let $E_{\rm ansatz}$, $E$ and $E_{\rm bind}$ denote the energies of the ansatz state, the lowest-energy state and the binding energy, respectively. For our choice of $\lambda=2$, $E_{\rm bind}$ is the energy gain relative to the corresponding asymptotically separated cluster configuration: for $N_h=3$, it is the energy difference between the ground state at $L=8$ and the $2+1$ cluster, and for $N_h=4$, it is the energy difference between the ground state at $L=18$ and the $2+2$ cluster. We quantify the quality of the ansatz by the overlap $O = \frac{|\langle \Psi_{\rm ansatz}|\Psi\rangle|}{||\Psi_{\rm ansatz}||\cdot ||\Psi||} = 0.999997$, and by the relative energy error $\delta = \frac{|E_{\rm ansatz} - E|}{|E_{\rm bind}|} = 4\times 10^{-5}$. This indicates excellent agreement with the exact ground state. 

The $N_h=4$ ground state occurs at $L=18$. We take the ansatz
\begin{align}
    \Psi_{\rm ansatz}(\bar\xi) = F_6(\bar\xi) \Delta(\bar\xi),
\end{align}
where $F_6(\bar\xi) = x_1 e_2^3 + x_2 e_3^2 + x_3 e_2e_4$ and the coefficients $x_i$ are determined variationally by minimizing the energy. The overlap and the relative energy error are $O = 0.990885$ and $\delta = 0.118$, respectively. The overlap is high and $\delta$ is moderate, indicating that the ansatz provides an excellent description of the wave function and captures the energy reasonably well, though not with high precision.

We further test whether $\Delta(\bar\xi)$ alone captures other low-energy states. For $L=6$ ($N_h=3$) we obtain $O=0.976574$ and $\delta=0.326$, and for $L=12$ ($N_h=4$) we find $O=0.942771$ and $\delta=0.748$. In both cases, the overlap remains relatively high, suggesting that the ansatz captures the dominant structure of the wave functions; however, the large $\delta$ values indicate that the energies are not reproduced quantitatively.

\section{Expansion of the binding energy near integer filling}
In this section, we use the two-dimensional one-component plasma description of Laughlin-like states, in which the parameter $q$ (and hence $\nu=1/q$) can be treated as a continuous variable. We focus on the $N_h=2$ case, and define the binding energy as the energy of the $L=2$ two-quasihole state measured relative to the well-separated two-quasihole configuration,
\begin{align}
    E_{\mathrm{bind}}(q) = E_{L=2}(q) - E_{\rm Laughlin}(q) - 2\Big[ E_{\rm qh}(q) - E_{\rm Laughlin}(q)\Big].
\end{align}
Here $E_{\rm Laughlin}(q)$, $E_{\rm qh}(q)$ and $E_{L=2}(q)$ denote the energies of the Laughlin state, the single-quasihole state and the $L=2$ two-quasihole state, respectively. In the thermodynamic limit, it is convenient to work with unnormalized wavefunctions $\tilde \psi$, which can be written in the unified form,
\begin{align}
    \tilde\psi(z;q) = w(z) \prod_{i<j} (z_i-z_j) ^q e^{-\frac{\sum_i|z_i|^2}{4}},\qquad Z(q) = \int \prod_i  d ^2 z_i |w(z)|^2  \prod_{i<j} |z_i-z_j| ^{2q} e^{-\frac{\sum_i|z_i|^2}{2}}.
\end{align} 
The normalized state is $|\psi(q)\rangle=\frac{1}{\sqrt{Z(q)}}|\tilde\psi(q)\rangle$.
Specifically, (1) $w(z)=1$ for the Laughlin state $\psi_{\mathrm{Laughlin}}$, (2) $w(z)=\prod_i z_i$ for the single-quasihole state $\psi_{\mathrm{qh}}$, and (3) $w(z) = e_1(z^{-2})\prod_i z_i^2$ for the $L=2$ state $\psi_{L=2}$ where $e_1$ is the elementary symmetric polynomial. 

For a translationally invariant electron--electron interaction $V(z_i,z_j) = V(z_i-z_j)$, the energy of the state $\psi(q)$ can be written as
\begin{align}
    E(q) = \langle \psi(q) | \hat V |\psi(q)\rangle = \frac{1}{2}\int d ^2 z d ^2 z'\; V(z,z') \rho^{(2)}(z,z';q).
\end{align}
Here $\rho^{(2)}(z,z';q)$ is the 2-point correlation,
\begin{align}
    \rho^{(2)}(z,z';q) = \langle \psi(q)|\hat\rho_2(z,z')|\psi(q)\rangle,\qquad \hat\rho_2(z,z') = \sum_{i\neq j} \delta(z,\hat z_i)\delta(z',\hat z_j).
\end{align}
Define the relative-coordinate pair density by
\begin{align}
    n^{(2)}(r;q) = \int  d ^2 R\; \rho^{(2)}\Big(R+\frac{r}{2},R-\frac{r}{2};q\Big),
\end{align}
and the binding energy can be expressed as
\begin{align}
    E_{\rm bind}(q) = \frac{1}{2} \int  d ^2 r\; V(r) n_{\rm bind}^{(2)}(r;q),\qquad  n_{\rm bind}^{(2)}(r;q) = n^{(2)}_{L=2}(r;q) + n^{(2)}_{\rm Laughlin}(r;q) - 2 n^{(2)}_{\rm qh}(r;q).
\end{align}

\begin{figure}[h]
    \centering
    \includegraphics[width=0.65\linewidth]{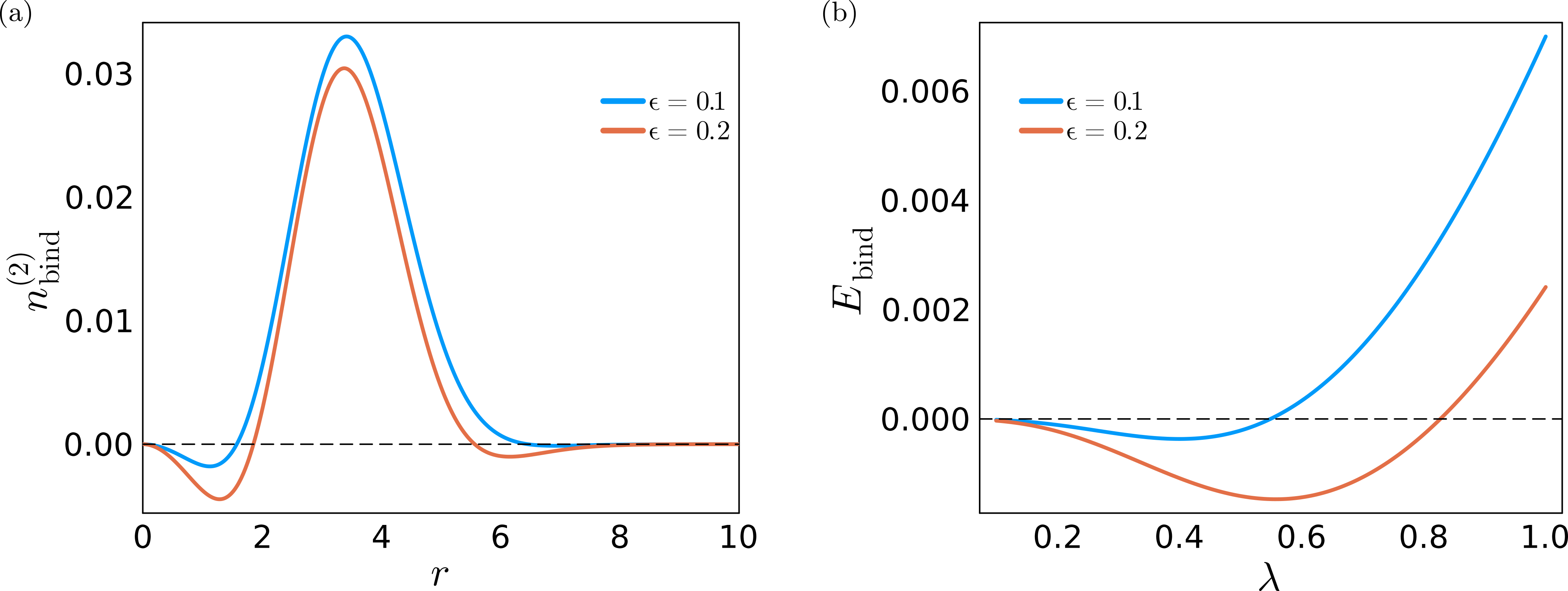}
    \caption{Perturbative expansion of the two-quasihole binding near integer filling $q=1$. (a) The relative-coordinate pair density $n^{(2)}_{\rm bind}(r;1+\epsilon)$ evaluated to first order in $\epsilon$. (b) Corresponding binding energy $E_{\rm bind}(1+\epsilon)$, evaluated to first order in $\epsilon$, as a function of the Yukawa screening length $\lambda$.}
    \label{fig-expansion}
\end{figure}

We study the expansion near integer filling $q=1$ by setting $q = 1 + \epsilon$, because the system at $q=1$ is solvable. The binding energy and the two-point correlation can be expanded as (throughout this subsection, quantities without an explicit $q$ are evaluated at $q=1$)
\begin{align}
    E_{\rm bind}(q) = E_{\rm bind} + \epsilon \delta E_{\rm bind} + O(\epsilon^2),\qquad    \rho^{(2)}(z,z';q) = \rho^{(2)}(z,z') + \epsilon \delta \rho^{(2)}(z,z') + O(\epsilon^2).
\end{align}
Using the definition of $\rho^{(2)}$, we can write $ \delta \rho^{(2)}$ as 
\begin{align}
    \delta \rho^{(2)}(z,z') = & \int  d ^2 w d ^2 w' U(w,w') \Big[ \rho^{(4)}(z,z',w,w') - \rho^{(2)}(z,z') \rho^{(2)}(w,w') \Big] + 2 \int  d ^2 w U(z,w) \rho^{(3)}(z,z',w)\notag\\
    &+ 2\int d ^2 w U(z',w)\rho^{(3)} (z,z',w) + 2U(z,z')\rho^{(2)}(z,z'),
\end{align}
where $U(z,z') = U(|z-z'|) = \ln|z-z'|$, and $\rho^{(3)}$ and $\rho^{(4)}$ are the three-point and four-point correlation. We can also expand $n^{(2)}_{\rm bind}$ to obtain $n^{(2)}_{\rm bind}(r;q) = n^{(2)}_{\rm bind}(r) + \epsilon\delta n^{(2)}_{\rm bind} + O(\epsilon^2)$.

At $q=1$, the Laughlin state reduces to a filled lowest Landau level. Let $\phi_m(z)$ be the orthonormal basis in LLL,
\begin{align*}
    \phi_m(z) = \frac{1}{\sqrt{h_m}}z^m e^{-\frac{|z|^2}{4}},\qquad h_m = 2^{m+1}\pi m!,
\end{align*}
and let $c_m$ annihilate an electron in orbital $m$. We introduce $\hat\psi(z) = \sum_{m} \phi_m(z) c_m$.

The $k$-point correlation for $|\psi\rangle$ can be written as 
\begin{align}
    \rho^{(k)}(z_1,\dots,z_k) = \langle \psi| \hat\psi^\dagger(z_1)\cdots\hat\psi^\dagger(z_k) \hat\psi(z_k)\cdots\hat\psi(z_1)|\psi\rangle.
\end{align}
The Laughlin state at $q=1$ occupies all orbitals, so the $k$-point correlation can be written as 
\begin{align}
    \rho_{\rm Laughlin}^{(k)}(z_1,\dots,z_k) = \det[K(z_i,z_j)]_{i,j=1}^k,\qquad K(z,z') = \frac{1}{2\pi} e^{-\frac{|z|^2}{4} - \frac{|z'|^2}{4}+\frac{z\bar{z}'}{2}}.
\end{align}
The single-quasihole state at $q=1$ occupies all orbitals except $m=0$, so the $k$-point correlation takes the form 
\begin{align}
    \rho_{\rm qh}^{(k)}(z_1,\dots,z_k) = \det[K_{\rm qh}(z_i,z_j)]_{i,j=1}^k,\qquad K_{\rm qh}(z,z') = \frac{1}{2\pi} e^{-\frac{|z|^2}{4} - \frac{|z'|^2}{4}} \Big(e^{\frac{z\bar{z}'}{2}} - 1\Big).
\end{align}
The $L=2$ two-quasihole state at $q=1$  can be represented as a linear combination of two orthogonal Slater determinants. Let $\tilde\psi_{A}(z)$ and $\tilde\psi_{B}(z)$ be the wavefunctions of electrons occupying $\mathcal{S}_A=\{0,3,4,\cdots,N+1\}$ and $\mathcal{S}_B=\{1,2,4,\cdots,N+1\}$. The $L=2$ two-quasihole state at $q=1$ can be written as $\psi_1(z) = \frac{\sqrt{3}}{2}\psi_{A}(z) - \frac 12\psi_{B}(z)$. The $k$-point correlation is
\begin{align}
    \rho^{(k)}_{L=2}(z_1,\dots,z_k) = \frac{3}{4} \rho^{(k)}_{A}(z_1,\cdots,z_k) + \frac{1}{4}\rho^{(k)}_{B}(z_1,\cdots,z_k)-\frac{\sqrt 3}{4}\Big[T^{(k)}(z_1,\dots,z_k) +{T^{(k)}}^*(z_1,\dots,z_k)\Big],
\end{align}
where $\rho^{(k)}_{A/B}(z_1,\cdots,z_k) =\det [K_{A/B}(z_i,z_j)]_{i,j=1}^k$ and $T^{(k)}(z_1,\dots,z_k) =\det M(z_1,\cdots,z_k)$ with 
\begin{align}
    K_{A}(z_i,z_j) &= K(z_i,z_j) - \phi_1(z_i)\phi_1^*(z_j) - \phi_2(z_i)\phi_2^*(z_j),\notag\\
    K_{B}(z_i,z_j) &= K(z_i,z_j) - \phi_0(z_i)\phi_0^*(z_j) - \phi_3(z_i)\phi_3^*(z_j),\notag\\
    M(z_1,\dots,z_k) &=
    \begin{pmatrix}
        0 & 0 & \phi_1(z_1) & \phi_1(z_2) & \cdots & \phi_1(z_k)\\
        0 & 0 & \phi_2(z_1) & \phi_2(z_2) & \cdots & \phi_2(z_k)\\
        \phi_0^*(z_1) & \phi_3^*(z_1) & K(z_1,z_1) & K(z_2,z_1) & \cdots & K(z_k,z_1)\\
        \phi_0^*(z_2) & \phi_3^*(z_2) & K(z_1,z_2) & K(z_2,z_2) & \cdots & K(z_k,z_2)\\
        \vdots & \vdots & \vdots & \vdots & \ddots & \vdots\\
        \phi_0^*(z_k) & \phi_3^*(z_k) & K(z_1,z_k) & K(z_2,z_k) & \cdots & K(z_k,z_k)
    \end{pmatrix}.
\end{align}

Using the correlation functions obtained above, we find $n^{(2)}_{\rm bind}(r)$ and $\delta n^{(2)}_{\rm bind}(r)$ are
\begin{align}
    n^{(2)}_{\rm bind}(r) = \frac{r^6}{2^8\times 3 \pi}e^{-\frac{r^2}{4}}
\end{align} 
and 
\begin{align}
    \delta n^{(2)}_{\rm bind}(r) = \frac{1}{373248\pi}\Bigg[&-16e^{-x/3}\bigl(-51552-300x+601x^2\bigr)+16e^{-x/6}\bigl(-51552+300x+601x^2\bigr)\notag\\
&-746496\Bigl(\mathrm{Ei}(-x/4)-\mathrm{Ei}(-x/6)\Bigr)\notag\\
&+81e^{-x/4}\Bigl(13824+768x-192x^2+(11+6\gamma)x^3\notag\\
&\qquad\qquad\qquad+6x^3\Bigl[4\mathrm{Ei}(-x/4)-2\bigl(\mathrm{Ei}(-x/12)+\mathrm{Ei}(x/12)\bigr)+\ln(x/16)\Bigr]
\Bigr)\notag\\
&+7776e^{-x/2}\Bigl(x^2-4x-96\mathrm{Ei}(x/6)+96\ln x-48\bigl(3-2\gamma+4\ln 2\bigr)\Bigr)\Bigg].
\end{align}
where we have defined $x=r^2$, $\gamma$ is the Euler constant, and $\mathrm{Ei}$ is the exponential integral. Fig.~\ref{fig-expansion}(a) shows the function $ n^{(2)}_{\rm bind}(r) + \epsilon \delta  n^{(2)}_{\rm bind}(r)$ for $\epsilon=0.1,0.2$. The small-$r$ expansion of $\delta n^{(2)}_{\rm bind}(r)$ is 
\begin{align*}
   \delta n^{(2)}_{\rm bind}(r)\approx \frac{1}{\pi}\Bigl(\ln\frac{2}{3}+\frac{35}{108}\Bigr)r^2 +\frac{1}{\pi}\Bigl(\frac14\ln\frac{3}{2}-\frac{35}{432}\Bigr)r^4 + \frac{r^6}{384\pi}\ln r + O(r^6).
\end{align*}
Since $\ln\frac{2}{3}+\frac{35}{108}<0$, the leading correction is negative and scales as $r^2$. Therefore, $ n^{(2)}_{\rm bind}(r;1+\epsilon)$ becomes negative at small $r$, supporting a bound state for short-range repulsive interactions. In FIG.~\ref{fig-expansion}(b), the perturbative result for the Yukawa interaction shows that $E_{\rm bind}(1+\epsilon)$ is negative for sufficiently small $\lambda$, indicating a two-quasihole bound state for short-range repulsion. As $\lambda$ increases, $E_{\rm bind}(1+\epsilon)$ crosses zero and becomes positive, signaling the disappearance of the bound state in the long-range regime. This trend is consistent with the main text results.

\section{Relation to the  Chern-Simons Landau-Ginzburg theory near the Bogomol’nyi point}

A useful way to place our microscopic construction in a broader context is through its relation to the Chern–Simons Landau–Ginzburg (CSLG) description near a Bogomol’nyi, or self-dual, point. In the conventional Landau–Ginzburg theory, the Bogomol’nyi point is the special choice of couplings separating Type-I and Type-II superconductors, at which vortices are non-interacting: the second-order equations of motion reduce to first-order self-duality equations, and the energy depends only on the total vorticity, not on the spatial arrangement of the vortices. In the presence of a Chern–Simons term, an analogous Bogomol’nyi point exists, but with an important difference: because time-reversal symmetry is broken, vortices and anti-vortices are no longer on the same footing, and only vortices remain non-interacting, while the anti-vortices are generically coupled (or vice versa) \cite{Parameswaran2012_typeI_FQH}. Microscopically, the same structure is realized in our setup with the Trugman–Kivelson pseudopotential, where the Laughlin state, together with all multi-quasihole states, forms a degenerate manifold of zero modes, whereas quasielectrons do not. In this sense, the pseudopotential problem may be viewed as a microscopic realization of the CSLG Bogomol’nyi point, with Laughlin quasiholes identified as vortices. It is therefore natural to follow Refs. \cite{Parameswaran2011_Pfaffian,Parameswaran2012_typeI_FQH} and study perturbations away from this point, supplemented by a repulsive Yukawa interaction $V(r)=e^{-r/\lambda}/r$. This procedure illustrates that the sequence of bound states as a function of the screening length found in our exact microscopic treatment could be phenomenologically interpreted as a qualitative analog of the transition from the type-II behavior to finite-cluster formation, and then to the type-I behavior (phase separation) within the CSLG theory.

Concretely, the bosonic Chern-Simons--Landau-Ginzburg theory near $\nu=1/q$ (written in the Coulomb gauge) is
\begin{equation}
\mathcal{L}
=
\bar\phi D_\tau \phi
+
\frac{|D_i\phi|^2}{2m}
+
u\bigl(|\phi|^2-\rho\bigr)^2
+
i a_0\,\frac{\nabla\times a}{2\pi q},
\qquad
D_\mu = \partial_\mu - i(a_\mu +A_\mu).
\label{eq:CSLG}
\end{equation}
Here $A_\mu$ is the external gauge field, $B=\nabla\times A=2\pi q\rho$, $\rho$ is the uniform density, and the statistical gauge field $a_0$ imposes the usual flux-attachment constraint $b \equiv \nabla\times a = -2\pi q |\phi|^2 $. For static configurations, the equations of motion are
\begin{equation}
-\frac{D^2\phi}{2m}
+
2u\bigl(|\phi|^2-\rho\bigr)\phi
=0,
\qquad
e\equiv-\nabla a_0 = 2\pi q\,\hat z\times j,
\qquad
j_i=\frac{1}{m}\,\mathrm{Im}\!\left(\bar\phi\,D_i\phi\right).
\label{eq:EOM}
\end{equation}
Using the Bogomol'nyi identity
\begin{equation}
|D\phi|^2
=
|D_\pm\phi|^2
\mp (b+B)|\phi|^2
\pm m \nabla\times j,
\qquad
D_\pm \equiv D_x \pm i D_y ,
\label{eq:Bogoid}
\end{equation}
together with the flux attachment constraint, the static energy can be written as
\begin{equation}
E
=
\int d^2r\,
\left[
\frac{|D_-\phi|^2}{2m}
+
(u-u_B)\bigl(|\phi|^2-\rho\bigr)^2
\right]
+ E_{\rm top},
\qquad
u_B \equiv \frac{\pi q}{m},
\label{eq:EBPS}
\end{equation}
where $E_{\rm top}$ depends only on the total vorticity (equivalently, total quasihole charge), and not on the vortex positions. Exactly at the Bogomol'nyi point $u=u_B$, the quartic term cancels, and the minimum is obtained from the first-order equations
\begin{equation}
D_-\phi = 0,
\qquad
b=-2\pi q |\phi|^2 .
\label{eq:selfdual}
\end{equation}
Therefore, for fixed total vorticity, all vortex configurations have the same energy: quasiholes are exactly non-interacting at the Bogomol'nyi point. We now perturb slightly to the type-I side,
\begin{equation}
u = u_B - \delta u,
\qquad
\delta u > 0,
\label{eq:deltau}
\end{equation}
and add an explicit repulsive density-density interaction of Yukawa form,
\begin{equation}
E_V
=
\frac12
\int d^2r\,d^2r'\,
(|\phi(\mathbf r)|^2-\rho )\,
V(|\mathbf r-\mathbf r'|)\,
(|\phi(\mathbf r')|^2-\rho )\;,
\qquad
V(r)=\frac{e^{-r/\lambda}}{r}\;.
\label{eq:Yukawa}
\end{equation}
The full energy functional is then
\begin{equation}
E_{\rm pert}
=
E_{\rm top}
+
\int d^2r\,
\left[
\frac{|D_-\phi|^2}{2m}
-
\delta u\,(|\phi(\mathbf r)|^2-\rho )^2
\right]
+
E_V .
\label{eq:Epert}
\end{equation}
Since $\delta u>0$, overlapping density deficits lower the local energy: this is the short-distance type-I attraction. The Yukawa term frustrates unlimited clustering at distances $r\gtrsim \lambda$.

To estimate the competition between these two tendencies, we assume a simple variational ansatz where $N$ quasiholes (vortices)  form a bound state (droplet) of the radius $R_N\approx c l_B \sqrt{N}$, where $c$ is a constant of the order of one. The quasihole charge density in the droplet is assumed to be uniform
\begin{equation}
    h_N(r)\equiv \rho-|\phi_N(r)|^2=\frac{(N/q)}{\pi R_N^2}\theta(R_N-r)\;, \quad h_N(k)=2(N/q)\frac{J_1(kR_N)}{k R_N}\;,
\end{equation}
and $J_1$ is the Bessel function, and $ h_N(r)$ integrates to the correct total missing charge $\int d^2r  h_N(r) = N/q$. In order to determine the optimal binding, we should minimize the energy of such a droplet $h_N(r)$ over $N$. For the Yukawa repulsion given by Eq.~\eqref{eq:Yukawa}, we find
\begin{equation}
    E_V= \frac{1}{2}\int \frac{d^2k}{(2\pi)^2}V(k)h_N^2(k)=\frac{2(N/q)^2}{R_N}\mathcal{I}\left(\frac{R_N}{\lambda}\right),\quad \quad \mathcal{I}(x)=\int_0^{\infty}\frac{dt}{t} \frac{J_1^2(t)}{\sqrt{t^2+x^2}}\;.
\end{equation}
where we used $V(k)=2\pi/\sqrt{k^2+\lambda^{-2}}$. The second term in Eq.~\eqref{eq:Epert} yields a negative contribution $\approx -\sigma N$, where $\sigma>0$ is the effective scale determining the bulk energy gain associated with $\sim \delta u /(q^2 l_B^2)$  (to leading order away from the Bogomol’nyi point, the perturbation is evaluated on the Bogomol’nyi moduli space, where $D_-\phi=0$, and therefore only the $-\delta u$ and Yukawa terms contribute).  Thus, we obtain
\begin{equation}
    \Delta E_{\rm pert}\equiv E_{\rm pert}
-E_{\rm top} \approx -\sigma N +\frac{2N^{3/2}}{c q^2 l_B }\mathcal{I}\left(\frac{c l_B \sqrt{N}}{\lambda}\right)\;.
\end{equation}
When the droplet is much larger than the Yukawa screening length, $\lambda \ll R_N$, then we can approximate $\mathcal{I}(c l_B \sqrt{N}/\lambda)\approx \lambda/(2 c l_B \sqrt{N})$, and thus $\Delta E_{\rm pert} \approx [-\delta u/\pi  +\lambda]N/ (c q l_B)^2$. This implies that for sufficiently small $\lambda$, the energy gain for making a large droplet dominates and the system phase-separates. This is the analog of type-I superconductivity. On the other hand, when the droplet fits inside the Yukawa range, $ R_N \ll \lambda$, we can approximate $\mathcal{I}(x)\approx 4/3\pi$, and so $\Delta E_{\rm pert} \approx -\sigma N +8N^{3/2}/(3\pi cq^2 l_B)$. The second term has a different scaling with $N$, and so the energy acquires a minimum at finite $N_*\sim (c q^2 l_B\sigma )^2\sim \delta u^2/l_B^2$. If $\delta u$ is sufficiently small, then $N_*$ eventually becomes of the order of $1$ for large $\lambda$, and the system transitions into the type-II regime without any clustering.

Thus, we find that for short screening length, the repulsion is effectively local and the energy stays extensive in the number of quasiholes, so phase separation wins; for long screening length, the repulsion becomes superextensive in $N$, which cuts off unlimited growth and gives a finite optimal cluster size.

\end{appendix}


\begin{thebibliography}{72}%
\makeatletter
\providecommand \@ifxundefined [1]{%
 \@ifx{#1\undefined}
}%
\providecommand \@ifnum [1]{%
 \ifnum #1\expandafter \@firstoftwo
 \else \expandafter \@secondoftwo
 \fi
}%
\providecommand \@ifx [1]{%
 \ifx #1\expandafter \@firstoftwo
 \else \expandafter \@secondoftwo
 \fi
}%
\providecommand \natexlab [1]{#1}%
\providecommand \enquote  [1]{``#1''}%
\providecommand \bibnamefont  [1]{#1}%
\providecommand \bibfnamefont [1]{#1}%
\providecommand \citenamefont [1]{#1}%
\providecommand \href@noop [0]{\@secondoftwo}%
\providecommand \href [0]{\begingroup \@sanitize@url \@href}%
\providecommand \@href[1]{\@@startlink{#1}\@@href}%
\providecommand \@@href[1]{\endgroup#1\@@endlink}%
\providecommand \@sanitize@url [0]{\catcode `\\12\catcode `\$12\catcode `\&12\catcode `\#12\catcode `\^12\catcode `\_12\catcode `\%12\relax}%
\providecommand \@@startlink[1]{}%
\providecommand \@@endlink[0]{}%
\providecommand \url  [0]{\begingroup\@sanitize@url \@url }%
\providecommand \@url [1]{\endgroup\@href {#1}{\urlprefix }}%
\providecommand \urlprefix  [0]{URL }%
\providecommand \Eprint [0]{\href }%
\providecommand \doibase [0]{https://doi.org/}%
\providecommand \selectlanguage [0]{\@gobble}%
\providecommand \bibinfo  [0]{\@secondoftwo}%
\providecommand \bibfield  [0]{\@secondoftwo}%
\providecommand \translation [1]{[#1]}%
\providecommand \BibitemOpen [0]{}%
\providecommand \bibitemStop [0]{}%
\providecommand \bibitemNoStop [0]{.\EOS\space}%
\providecommand \EOS [0]{\spacefactor3000\relax}%
\providecommand \BibitemShut  [1]{\csname bibitem#1\endcsname}%
\let\auto@bib@innerbib\@empty
\bibitem [{\citenamefont {Tafelmayer}(1993)}]{Tafelmayer1993topological}%
  \BibitemOpen
  \bibfield  {author} {\bibinfo {author} {\bibfnamefont {R.}~\bibnamefont {Tafelmayer}},\ }\href@noop {} {\bibfield  {journal} {\bibinfo  {journal} {Nuclear Physics B}\ }\textbf {\bibinfo {volume} {396}},\ \bibinfo {pages} {386} (\bibinfo {year} {1993})}\BibitemShut {NoStop}%
\bibitem [{\citenamefont {Gattu}\ and\ \citenamefont {Jain}(2025)}]{AnyonMoleculesJain}%
  \BibitemOpen
  \bibfield  {author} {\bibinfo {author} {\bibfnamefont {M.}~\bibnamefont {Gattu}}\ and\ \bibinfo {author} {\bibfnamefont {J.~K.}\ \bibnamefont {Jain}},\ }\href {https://doi.org/10.1103/scl5-8pv6} {\bibfield  {journal} {\bibinfo  {journal} {Phys. Rev. Lett.}\ }\textbf {\bibinfo {volume} {135}},\ \bibinfo {pages} {236601} (\bibinfo {year} {2025})}\BibitemShut {NoStop}%
\bibitem [{\citenamefont {Xu}\ \emph {et~al.}(2025)\citenamefont {Xu}, \citenamefont {Ji}, \citenamefont {Wang}, \citenamefont {Trung},\ and\ \citenamefont {Yang}}]{AnyonClustersYang}%
  \BibitemOpen
  \bibfield  {author} {\bibinfo {author} {\bibfnamefont {Q.}~\bibnamefont {Xu}}, \bibinfo {author} {\bibfnamefont {G.}~\bibnamefont {Ji}}, \bibinfo {author} {\bibfnamefont {Y.}~\bibnamefont {Wang}}, \bibinfo {author} {\bibfnamefont {H.~Q.}\ \bibnamefont {Trung}},\ and\ \bibinfo {author} {\bibfnamefont {B.}~\bibnamefont {Yang}},\ }\href {https://doi.org/10.1103/vgz6-z98r} {\bibfield  {journal} {\bibinfo  {journal} {Phys. Rev. B}\ }\textbf {\bibinfo {volume} {112}},\ \bibinfo {pages} {235112} (\bibinfo {year} {2025})}\BibitemShut {NoStop}%
\bibitem [{\citenamefont {Gonçalves}\ \emph {et~al.}(2025)\citenamefont {Gonçalves}, \citenamefont {Mendez-Valderrama}, \citenamefont {Herzog-Arbeitman}, \citenamefont {Yu}, \citenamefont {Xu}, \citenamefont {Xiao}, \citenamefont {Bernevig},\ and\ \citenamefont {Regnault}}]{bernevig2025ED}%
  \BibitemOpen
  \bibfield  {author} {\bibinfo {author} {\bibfnamefont {M.}~\bibnamefont {Gonçalves}}, \bibinfo {author} {\bibfnamefont {J.~F.}\ \bibnamefont {Mendez-Valderrama}}, \bibinfo {author} {\bibfnamefont {J.}~\bibnamefont {Herzog-Arbeitman}}, \bibinfo {author} {\bibfnamefont {J.}~\bibnamefont {Yu}}, \bibinfo {author} {\bibfnamefont {X.}~\bibnamefont {Xu}}, \bibinfo {author} {\bibfnamefont {D.}~\bibnamefont {Xiao}}, \bibinfo {author} {\bibfnamefont {B.~A.}\ \bibnamefont {Bernevig}},\ and\ \bibinfo {author} {\bibfnamefont {N.}~\bibnamefont {Regnault}},\ }\href {https://arxiv.org/abs/2506.05330} {\bibinfo {title} {Spinless and spinful charge excitations in moir\'e fractional chern insulators}} (\bibinfo {year} {2025}),\ \Eprint {https://arxiv.org/abs/2506.05330} {arXiv:2506.05330 [cond-mat.str-el]} \BibitemShut {NoStop}%
\bibitem [{\citenamefont {Hu}\ \emph {et~al.}(2025)\citenamefont {Hu}, \citenamefont {Tsui}, \citenamefont {He}, \citenamefont {Kamber}, \citenamefont {Wang}, \citenamefont {Mohammadi}, \citenamefont {Watanabe}, \citenamefont {Taniguchi}, \citenamefont {Papi{\'c}}, \citenamefont {Zaletel},\ and\ \citenamefont {Yazdani}}]{Yazdani2025_STM}%
  \BibitemOpen
  \bibfield  {author} {\bibinfo {author} {\bibfnamefont {Y.}~\bibnamefont {Hu}}, \bibinfo {author} {\bibfnamefont {Y.-C.}\ \bibnamefont {Tsui}}, \bibinfo {author} {\bibfnamefont {M.}~\bibnamefont {He}}, \bibinfo {author} {\bibfnamefont {U.}~\bibnamefont {Kamber}}, \bibinfo {author} {\bibfnamefont {T.}~\bibnamefont {Wang}}, \bibinfo {author} {\bibfnamefont {A.~S.}\ \bibnamefont {Mohammadi}}, \bibinfo {author} {\bibfnamefont {K.}~\bibnamefont {Watanabe}}, \bibinfo {author} {\bibfnamefont {T.}~\bibnamefont {Taniguchi}}, \bibinfo {author} {\bibfnamefont {Z.}~\bibnamefont {Papi{\'c}}}, \bibinfo {author} {\bibfnamefont {M.~P.}\ \bibnamefont {Zaletel}},\ and\ \bibinfo {author} {\bibfnamefont {A.}~\bibnamefont {Yazdani}},\ }\href {https://doi.org/10.1038/s41567-025-02830-y} {\bibfield  {journal} {\bibinfo  {journal} {Nature Physics}\ }\textbf {\bibinfo {volume} {21}},\ \bibinfo {pages} {716} (\bibinfo {year} {2025})}\BibitemShut {NoStop}%
\bibitem [{\citenamefont {Chiu}\ \emph {et~al.}(2025)\citenamefont {Chiu}, \citenamefont {Wang}, \citenamefont {Fan}, \citenamefont {Watanabe}, \citenamefont {Taniguchi}, \citenamefont {Liu}, \citenamefont {Zaletel},\ and\ \citenamefont {Yazdani}}]{Yazdani2025_PNAS}%
  \BibitemOpen
  \bibfield  {author} {\bibinfo {author} {\bibfnamefont {C.-L.}\ \bibnamefont {Chiu}}, \bibinfo {author} {\bibfnamefont {T.}~\bibnamefont {Wang}}, \bibinfo {author} {\bibfnamefont {R.}~\bibnamefont {Fan}}, \bibinfo {author} {\bibfnamefont {K.}~\bibnamefont {Watanabe}}, \bibinfo {author} {\bibfnamefont {T.}~\bibnamefont {Taniguchi}}, \bibinfo {author} {\bibfnamefont {X.}~\bibnamefont {Liu}}, \bibinfo {author} {\bibfnamefont {M.~P.}\ \bibnamefont {Zaletel}},\ and\ \bibinfo {author} {\bibfnamefont {A.}~\bibnamefont {Yazdani}},\ }\href {https://doi.org/10.1073/pnas.2424781122} {\bibfield  {journal} {\bibinfo  {journal} {Proceedings of the National Academy of Sciences}\ }\textbf {\bibinfo {volume} {122}},\ \bibinfo {pages} {e2424781122} (\bibinfo {year} {2025})},\ \Eprint {https://arxiv.org/abs/https://www.pnas.org/doi/pdf/10.1073/pnas.2424781122} {https://www.pnas.org/doi/pdf/10.1073/pnas.2424781122} \BibitemShut {NoStop}%
\bibitem [{\citenamefont {Ghosh}\ \emph {et~al.}(2025)\citenamefont {Ghosh}, \citenamefont {Labendik}, \citenamefont {Umansky}, \citenamefont {Heiblum},\ and\ \citenamefont {Mross}}]{Ghosh2025_anyon_bunching}%
  \BibitemOpen
  \bibfield  {author} {\bibinfo {author} {\bibfnamefont {B.}~\bibnamefont {Ghosh}}, \bibinfo {author} {\bibfnamefont {M.}~\bibnamefont {Labendik}}, \bibinfo {author} {\bibfnamefont {V.}~\bibnamefont {Umansky}}, \bibinfo {author} {\bibfnamefont {M.}~\bibnamefont {Heiblum}},\ and\ \bibinfo {author} {\bibfnamefont {D.~F.}\ \bibnamefont {Mross}},\ }\href {https://doi.org/10.1038/s41586-025-09143-3} {\bibfield  {journal} {\bibinfo  {journal} {Nature}\ }\textbf {\bibinfo {volume} {642}},\ \bibinfo {pages} {922} (\bibinfo {year} {2025})}\BibitemShut {NoStop}%
\bibitem [{\citenamefont {Bid}\ \emph {et~al.}(2009)\citenamefont {Bid}, \citenamefont {Ofek}, \citenamefont {Heiblum}, \citenamefont {Umansky},\ and\ \citenamefont {Mahalu}}]{Bid2009_shot_noise_2/3}%
  \BibitemOpen
  \bibfield  {author} {\bibinfo {author} {\bibfnamefont {A.}~\bibnamefont {Bid}}, \bibinfo {author} {\bibfnamefont {N.}~\bibnamefont {Ofek}}, \bibinfo {author} {\bibfnamefont {M.}~\bibnamefont {Heiblum}}, \bibinfo {author} {\bibfnamefont {V.}~\bibnamefont {Umansky}},\ and\ \bibinfo {author} {\bibfnamefont {D.}~\bibnamefont {Mahalu}},\ }\href {https://doi.org/10.1103/PhysRevLett.103.236802} {\bibfield  {journal} {\bibinfo  {journal} {Phys. Rev. Lett.}\ }\textbf {\bibinfo {volume} {103}},\ \bibinfo {pages} {236802} (\bibinfo {year} {2009})}\BibitemShut {NoStop}%
\bibitem [{\citenamefont {Chung}\ \emph {et~al.}(2003)\citenamefont {Chung}, \citenamefont {Heiblum},\ and\ \citenamefont {Umansky}}]{Chung2003_Bunched_anyons}%
  \BibitemOpen
  \bibfield  {author} {\bibinfo {author} {\bibfnamefont {Y.~C.}\ \bibnamefont {Chung}}, \bibinfo {author} {\bibfnamefont {M.}~\bibnamefont {Heiblum}},\ and\ \bibinfo {author} {\bibfnamefont {V.}~\bibnamefont {Umansky}},\ }\href {https://doi.org/10.1103/PhysRevLett.91.216804} {\bibfield  {journal} {\bibinfo  {journal} {Phys. Rev. Lett.}\ }\textbf {\bibinfo {volume} {91}},\ \bibinfo {pages} {216804} (\bibinfo {year} {2003})}\BibitemShut {NoStop}%
\bibitem [{\citenamefont {Xu}\ \emph {et~al.}(2023)\citenamefont {Xu}, \citenamefont {Sun}, \citenamefont {Jia}, \citenamefont {Liu}, \citenamefont {Xu}, \citenamefont {Li}, \citenamefont {Gu}, \citenamefont {Watanabe}, \citenamefont {Taniguchi}, \citenamefont {Tong}, \citenamefont {Jia}, \citenamefont {Shi}, \citenamefont {Jiang}, \citenamefont {Zhang}, \citenamefont {Liu},\ and\ \citenamefont {Li}}]{xu2023observation}%
  \BibitemOpen
  \bibfield  {author} {\bibinfo {author} {\bibfnamefont {F.}~\bibnamefont {Xu}}, \bibinfo {author} {\bibfnamefont {Z.}~\bibnamefont {Sun}}, \bibinfo {author} {\bibfnamefont {T.}~\bibnamefont {Jia}}, \bibinfo {author} {\bibfnamefont {C.}~\bibnamefont {Liu}}, \bibinfo {author} {\bibfnamefont {C.}~\bibnamefont {Xu}}, \bibinfo {author} {\bibfnamefont {C.}~\bibnamefont {Li}}, \bibinfo {author} {\bibfnamefont {Y.}~\bibnamefont {Gu}}, \bibinfo {author} {\bibfnamefont {K.}~\bibnamefont {Watanabe}}, \bibinfo {author} {\bibfnamefont {T.}~\bibnamefont {Taniguchi}}, \bibinfo {author} {\bibfnamefont {B.}~\bibnamefont {Tong}}, \bibinfo {author} {\bibfnamefont {J.}~\bibnamefont {Jia}}, \bibinfo {author} {\bibfnamefont {Z.}~\bibnamefont {Shi}}, \bibinfo {author} {\bibfnamefont {S.}~\bibnamefont {Jiang}}, \bibinfo {author} {\bibfnamefont {Y.}~\bibnamefont {Zhang}}, \bibinfo {author} {\bibfnamefont {X.}~\bibnamefont {Liu}},\ and\ \bibinfo {author} {\bibfnamefont {T.}~\bibnamefont {Li}},\ }\href {https://doi.org/10.1103/PhysRevX.13.031037} {\bibfield  {journal} {\bibinfo  {journal} {Phys. Rev. X}\ }\textbf {\bibinfo {volume} {13}},\ \bibinfo {pages} {031037} (\bibinfo {year} {2023})}\BibitemShut {NoStop}%
\bibitem [{\citenamefont {Zeng}\ \emph {et~al.}(2023)\citenamefont {Zeng}, \citenamefont {Xia}, \citenamefont {Kang}, \citenamefont {Zhu}, \citenamefont {Kn{\"u}ppel}, \citenamefont {Vaswani}, \citenamefont {Watanabe}, \citenamefont {Taniguchi}, \citenamefont {Mak},\ and\ \citenamefont {Shan}}]{zeng2023thermodynamic}%
  \BibitemOpen
  \bibfield  {author} {\bibinfo {author} {\bibfnamefont {Y.}~\bibnamefont {Zeng}}, \bibinfo {author} {\bibfnamefont {Z.}~\bibnamefont {Xia}}, \bibinfo {author} {\bibfnamefont {K.}~\bibnamefont {Kang}}, \bibinfo {author} {\bibfnamefont {J.}~\bibnamefont {Zhu}}, \bibinfo {author} {\bibfnamefont {P.}~\bibnamefont {Kn{\"u}ppel}}, \bibinfo {author} {\bibfnamefont {C.}~\bibnamefont {Vaswani}}, \bibinfo {author} {\bibfnamefont {K.}~\bibnamefont {Watanabe}}, \bibinfo {author} {\bibfnamefont {T.}~\bibnamefont {Taniguchi}}, \bibinfo {author} {\bibfnamefont {K.~F.}\ \bibnamefont {Mak}},\ and\ \bibinfo {author} {\bibfnamefont {J.}~\bibnamefont {Shan}},\ }\href {https://doi.org/10.1038/s41586-023-06452-3} {\bibfield  {journal} {\bibinfo  {journal} {Nature}\ }\textbf {\bibinfo {volume} {622}},\ \bibinfo {pages} {69} (\bibinfo {year} {2023})}\BibitemShut {NoStop}%
\bibitem [{\citenamefont {Cai}\ \emph {et~al.}(2023)\citenamefont {Cai}, \citenamefont {Anderson}, \citenamefont {Wang}, \citenamefont {Zhang}, \citenamefont {Liu}, \citenamefont {Holtzmann}, \citenamefont {Zhang}, \citenamefont {Fan}, \citenamefont {Taniguchi}, \citenamefont {Watanabe}, \citenamefont {Ran}, \citenamefont {Cao}, \citenamefont {Fu}, \citenamefont {Xiao}, \citenamefont {Yao},\ and\ \citenamefont {Xu}}]{cai2023signatures}%
  \BibitemOpen
  \bibfield  {author} {\bibinfo {author} {\bibfnamefont {J.}~\bibnamefont {Cai}}, \bibinfo {author} {\bibfnamefont {E.}~\bibnamefont {Anderson}}, \bibinfo {author} {\bibfnamefont {C.}~\bibnamefont {Wang}}, \bibinfo {author} {\bibfnamefont {X.}~\bibnamefont {Zhang}}, \bibinfo {author} {\bibfnamefont {X.}~\bibnamefont {Liu}}, \bibinfo {author} {\bibfnamefont {W.}~\bibnamefont {Holtzmann}}, \bibinfo {author} {\bibfnamefont {Y.}~\bibnamefont {Zhang}}, \bibinfo {author} {\bibfnamefont {F.}~\bibnamefont {Fan}}, \bibinfo {author} {\bibfnamefont {T.}~\bibnamefont {Taniguchi}}, \bibinfo {author} {\bibfnamefont {K.}~\bibnamefont {Watanabe}}, \bibinfo {author} {\bibfnamefont {Y.}~\bibnamefont {Ran}}, \bibinfo {author} {\bibfnamefont {T.}~\bibnamefont {Cao}}, \bibinfo {author} {\bibfnamefont {L.}~\bibnamefont {Fu}}, \bibinfo {author} {\bibfnamefont {D.}~\bibnamefont {Xiao}}, \bibinfo {author} {\bibfnamefont {W.}~\bibnamefont {Yao}},\ and\ \bibinfo {author} {\bibfnamefont {X.}~\bibnamefont {Xu}},\ }\href {https://doi.org/10.1038/s41586-023-06289-w} {\bibfield  {journal} {\bibinfo  {journal} {Nature}\ }\textbf {\bibinfo {volume} {622}},\ \bibinfo {pages} {63} (\bibinfo {year} {2023})}\BibitemShut {NoStop}%
\bibitem [{\citenamefont {Park}\ \emph {et~al.}(2023)\citenamefont {Park}, \citenamefont {Cai}, \citenamefont {Anderson}, \citenamefont {Zhang}, \citenamefont {Zhu}, \citenamefont {Liu}, \citenamefont {Wang}, \citenamefont {Holtzmann}, \citenamefont {Hu}, \citenamefont {Liu}, \citenamefont {Taniguchi}, \citenamefont {Watanabe}, \citenamefont {Chu}, \citenamefont {Cao}, \citenamefont {Fu}, \citenamefont {Yao}, \citenamefont {Chang}, \citenamefont {Cobden}, \citenamefont {Xiao},\ and\ \citenamefont {Xu}}]{park2023observation}%
  \BibitemOpen
  \bibfield  {author} {\bibinfo {author} {\bibfnamefont {H.}~\bibnamefont {Park}}, \bibinfo {author} {\bibfnamefont {J.}~\bibnamefont {Cai}}, \bibinfo {author} {\bibfnamefont {E.}~\bibnamefont {Anderson}}, \bibinfo {author} {\bibfnamefont {Y.}~\bibnamefont {Zhang}}, \bibinfo {author} {\bibfnamefont {J.}~\bibnamefont {Zhu}}, \bibinfo {author} {\bibfnamefont {X.}~\bibnamefont {Liu}}, \bibinfo {author} {\bibfnamefont {C.}~\bibnamefont {Wang}}, \bibinfo {author} {\bibfnamefont {W.}~\bibnamefont {Holtzmann}}, \bibinfo {author} {\bibfnamefont {C.}~\bibnamefont {Hu}}, \bibinfo {author} {\bibfnamefont {Z.}~\bibnamefont {Liu}}, \bibinfo {author} {\bibfnamefont {T.}~\bibnamefont {Taniguchi}}, \bibinfo {author} {\bibfnamefont {K.}~\bibnamefont {Watanabe}}, \bibinfo {author} {\bibfnamefont {J.-H.}\ \bibnamefont {Chu}}, \bibinfo {author} {\bibfnamefont {T.}~\bibnamefont {Cao}}, \bibinfo {author} {\bibfnamefont {L.}~\bibnamefont {Fu}}, \bibinfo {author} {\bibfnamefont {W.}~\bibnamefont {Yao}}, \bibinfo {author} {\bibfnamefont {C.-Z.}\ \bibnamefont {Chang}}, \bibinfo {author} {\bibfnamefont {D.}~\bibnamefont {Cobden}}, \bibinfo {author} {\bibfnamefont {D.}~\bibnamefont {Xiao}},\ and\ \bibinfo {author} {\bibfnamefont {X.}~\bibnamefont {Xu}},\ }\href {https://doi.org/10.1038/s41586-023-06536-0} {\bibfield  {journal} {\bibinfo  {journal} {Nature}\ }\textbf {\bibinfo {volume} {622}},\ \bibinfo {pages} {74} (\bibinfo {year} {2023})}\BibitemShut {NoStop}%
\bibitem [{\citenamefont {Lu}\ \emph {et~al.}(2024)\citenamefont {Lu}, \citenamefont {Han}, \citenamefont {Yao}, \citenamefont {Reddy}, \citenamefont {Yang}, \citenamefont {Seo}, \citenamefont {Watanabe}, \citenamefont {Taniguchi}, \citenamefont {Fu},\ and\ \citenamefont {Ju}}]{Lu2024}%
  \BibitemOpen
  \bibfield  {author} {\bibinfo {author} {\bibfnamefont {Z.}~\bibnamefont {Lu}}, \bibinfo {author} {\bibfnamefont {T.}~\bibnamefont {Han}}, \bibinfo {author} {\bibfnamefont {Y.}~\bibnamefont {Yao}}, \bibinfo {author} {\bibfnamefont {A.~P.}\ \bibnamefont {Reddy}}, \bibinfo {author} {\bibfnamefont {J.}~\bibnamefont {Yang}}, \bibinfo {author} {\bibfnamefont {J.}~\bibnamefont {Seo}}, \bibinfo {author} {\bibfnamefont {K.}~\bibnamefont {Watanabe}}, \bibinfo {author} {\bibfnamefont {T.}~\bibnamefont {Taniguchi}}, \bibinfo {author} {\bibfnamefont {L.}~\bibnamefont {Fu}},\ and\ \bibinfo {author} {\bibfnamefont {L.}~\bibnamefont {Ju}},\ }\href {https://doi.org/10.1038/s41586-023-07010-7} {\bibfield  {journal} {\bibinfo  {journal} {Nature}\ }\textbf {\bibinfo {volume} {626}},\ \bibinfo {pages} {759} (\bibinfo {year} {2024})}\BibitemShut {NoStop}%
\bibitem [{\citenamefont {Laughlin}(1988{\natexlab{a}})}]{LaughlinAnyonSC}%
  \BibitemOpen
  \bibfield  {author} {\bibinfo {author} {\bibfnamefont {R.~B.}\ \bibnamefont {Laughlin}},\ }\href {https://doi.org/10.1103/PhysRevLett.60.2677} {\bibfield  {journal} {\bibinfo  {journal} {Phys. Rev. Lett.}\ }\textbf {\bibinfo {volume} {60}},\ \bibinfo {pages} {2677} (\bibinfo {year} {1988}{\natexlab{a}})}\BibitemShut {NoStop}%
\bibitem [{\citenamefont {Fetter}\ \emph {et~al.}(1989)\citenamefont {Fetter}, \citenamefont {Hanna},\ and\ \citenamefont {Laughlin}}]{FetterHannaLaughlin}%
  \BibitemOpen
  \bibfield  {author} {\bibinfo {author} {\bibfnamefont {A.}~\bibnamefont {Fetter}}, \bibinfo {author} {\bibfnamefont {C.}~\bibnamefont {Hanna}},\ and\ \bibinfo {author} {\bibfnamefont {R.}~\bibnamefont {Laughlin}},\ }\href@noop {} {\bibfield  {journal} {\bibinfo  {journal} {Physical Review B}\ }\textbf {\bibinfo {volume} {39}},\ \bibinfo {pages} {9679} (\bibinfo {year} {1989})}\BibitemShut {NoStop}%
\bibitem [{\citenamefont {Laughlin}(1988{\natexlab{b}})}]{laughlin1988}%
  \BibitemOpen
  \bibfield  {author} {\bibinfo {author} {\bibfnamefont {R.~B.}\ \bibnamefont {Laughlin}},\ }\href {https://doi.org/10.1126/science.242.4878.525} {\bibfield  {journal} {\bibinfo  {journal} {Science}\ }\textbf {\bibinfo {volume} {242}},\ \bibinfo {pages} {525} (\bibinfo {year} {1988}{\natexlab{b}})}\BibitemShut {NoStop}%
\bibitem [{\citenamefont {Chen}\ \emph {et~al.}(1989)\citenamefont {Chen}, \citenamefont {Wilczek}, \citenamefont {Witten},\ and\ \citenamefont {Halperin}}]{WilczekWittenHalperinAnyonSC}%
  \BibitemOpen
  \bibfield  {author} {\bibinfo {author} {\bibfnamefont {Y.-H.}\ \bibnamefont {Chen}}, \bibinfo {author} {\bibfnamefont {F.}~\bibnamefont {Wilczek}}, \bibinfo {author} {\bibfnamefont {E.}~\bibnamefont {Witten}},\ and\ \bibinfo {author} {\bibfnamefont {B.~I.}\ \bibnamefont {Halperin}},\ }\href@noop {} {\bibfield  {journal} {\bibinfo  {journal} {International Journal of Modern Physics B}\ }\textbf {\bibinfo {volume} {3}},\ \bibinfo {pages} {1001} (\bibinfo {year} {1989})}\BibitemShut {NoStop}%
\bibitem [{\citenamefont {Lee}\ and\ \citenamefont {Fisher}(1989)}]{PhysRevLett.63.903}%
  \BibitemOpen
  \bibfield  {author} {\bibinfo {author} {\bibfnamefont {D.-H.}\ \bibnamefont {Lee}}\ and\ \bibinfo {author} {\bibfnamefont {M.~P.~A.}\ \bibnamefont {Fisher}},\ }\href {https://doi.org/10.1103/PhysRevLett.63.903} {\bibfield  {journal} {\bibinfo  {journal} {Phys. Rev. Lett.}\ }\textbf {\bibinfo {volume} {63}},\ \bibinfo {pages} {903} (\bibinfo {year} {1989})}\BibitemShut {NoStop}%
\bibitem [{\citenamefont {Wen}\ \emph {et~al.}(1989)\citenamefont {Wen}, \citenamefont {Wilczek},\ and\ \citenamefont {Zee}}]{PhysRevB.39.11413}%
  \BibitemOpen
  \bibfield  {author} {\bibinfo {author} {\bibfnamefont {X.~G.}\ \bibnamefont {Wen}}, \bibinfo {author} {\bibfnamefont {F.}~\bibnamefont {Wilczek}},\ and\ \bibinfo {author} {\bibfnamefont {A.}~\bibnamefont {Zee}},\ }\href {https://doi.org/10.1103/PhysRevB.39.11413} {\bibfield  {journal} {\bibinfo  {journal} {Phys. Rev. B}\ }\textbf {\bibinfo {volume} {39}},\ \bibinfo {pages} {11413} (\bibinfo {year} {1989})}\BibitemShut {NoStop}%
\bibitem [{\citenamefont {Hosotani}\ and\ \citenamefont {Chakravarty}(1990)}]{PhysRevB.42.342}%
  \BibitemOpen
  \bibfield  {author} {\bibinfo {author} {\bibfnamefont {Y.}~\bibnamefont {Hosotani}}\ and\ \bibinfo {author} {\bibfnamefont {S.}~\bibnamefont {Chakravarty}},\ }\href {https://doi.org/10.1103/PhysRevB.42.342} {\bibfield  {journal} {\bibinfo  {journal} {Phys. Rev. B}\ }\textbf {\bibinfo {volume} {42}},\ \bibinfo {pages} {342} (\bibinfo {year} {1990})}\BibitemShut {NoStop}%
\bibitem [{\citenamefont {Wen}\ and\ \citenamefont {Zee}(1990)}]{PhysRevB.41.240}%
  \BibitemOpen
  \bibfield  {author} {\bibinfo {author} {\bibfnamefont {X.~G.}\ \bibnamefont {Wen}}\ and\ \bibinfo {author} {\bibfnamefont {A.}~\bibnamefont {Zee}},\ }\href {https://doi.org/10.1103/PhysRevB.41.240} {\bibfield  {journal} {\bibinfo  {journal} {Phys. Rev. B}\ }\textbf {\bibinfo {volume} {41}},\ \bibinfo {pages} {240} (\bibinfo {year} {1990})}\BibitemShut {NoStop}%
\bibitem [{\citenamefont {Lee}(1991)}]{doi:10.1142/S0217979291001607}%
  \BibitemOpen
  \bibfield  {author} {\bibinfo {author} {\bibfnamefont {D.-H.}\ \bibnamefont {Lee}},\ }\href {https://doi.org/10.1142/S0217979291001607} {\bibfield  {journal} {\bibinfo  {journal} {International Journal of Modern Physics B}\ }\textbf {\bibinfo {volume} {05}},\ \bibinfo {pages} {1695} (\bibinfo {year} {1991})},\ \Eprint {https://arxiv.org/abs/https://doi.org/10.1142/S0217979291001607} {https://doi.org/10.1142/S0217979291001607} \BibitemShut {NoStop}%
\bibitem [{\citenamefont {Kitazawa}\ and\ \citenamefont {Murayama}(1990)}]{PhysRevB.41.11101}%
  \BibitemOpen
  \bibfield  {author} {\bibinfo {author} {\bibfnamefont {Y.}~\bibnamefont {Kitazawa}}\ and\ \bibinfo {author} {\bibfnamefont {H.}~\bibnamefont {Murayama}},\ }\href {https://doi.org/10.1103/PhysRevB.41.11101} {\bibfield  {journal} {\bibinfo  {journal} {Phys. Rev. B}\ }\textbf {\bibinfo {volume} {41}},\ \bibinfo {pages} {11101} (\bibinfo {year} {1990})}\BibitemShut {NoStop}%
\bibitem [{\citenamefont {Shi}\ and\ \citenamefont {Senthil}(2025{\natexlab{a}})}]{shi2025doping}%
  \BibitemOpen
  \bibfield  {author} {\bibinfo {author} {\bibfnamefont {Z.~D.}\ \bibnamefont {Shi}}\ and\ \bibinfo {author} {\bibfnamefont {T.}~\bibnamefont {Senthil}},\ }\href@noop {} {\bibfield  {journal} {\bibinfo  {journal} {Physical Review X}\ }\textbf {\bibinfo {volume} {15}},\ \bibinfo {pages} {031069} (\bibinfo {year} {2025}{\natexlab{a}})}\BibitemShut {NoStop}%
\bibitem [{\citenamefont {Divic}\ \emph {et~al.}(2024)\citenamefont {Divic}, \citenamefont {Cr{\'e}pel}, \citenamefont {Soejima}, \citenamefont {Song}, \citenamefont {Millis}, \citenamefont {Zaletel},\ and\ \citenamefont {Vishwanath}}]{divic2024anyon}%
  \BibitemOpen
  \bibfield  {author} {\bibinfo {author} {\bibfnamefont {S.}~\bibnamefont {Divic}}, \bibinfo {author} {\bibfnamefont {V.}~\bibnamefont {Cr{\'e}pel}}, \bibinfo {author} {\bibfnamefont {T.}~\bibnamefont {Soejima}}, \bibinfo {author} {\bibfnamefont {X.-Y.}\ \bibnamefont {Song}}, \bibinfo {author} {\bibfnamefont {A.}~\bibnamefont {Millis}}, \bibinfo {author} {\bibfnamefont {M.~P.}\ \bibnamefont {Zaletel}},\ and\ \bibinfo {author} {\bibfnamefont {A.}~\bibnamefont {Vishwanath}},\ }\href@noop {} {\bibfield  {journal} {\bibinfo  {journal} {arXiv preprint arXiv:2410.18175}\ } (\bibinfo {year} {2024})}\BibitemShut {NoStop}%
\bibitem [{\citenamefont {Nosov}\ \emph {et~al.}(2026)\citenamefont {Nosov}, \citenamefont {Han},\ and\ \citenamefont {Khalaf}}]{NosovAnyonSCPRL}%
  \BibitemOpen
  \bibfield  {author} {\bibinfo {author} {\bibfnamefont {P.~A.}\ \bibnamefont {Nosov}}, \bibinfo {author} {\bibfnamefont {Z.}~\bibnamefont {Han}},\ and\ \bibinfo {author} {\bibfnamefont {E.}~\bibnamefont {Khalaf}},\ }\href {https://doi.org/10.1103/6bgj-bfdn} {\bibfield  {journal} {\bibinfo  {journal} {Phys. Rev. Lett.}\ }\textbf {\bibinfo {volume} {136}},\ \bibinfo {pages} {106501} (\bibinfo {year} {2026})}\BibitemShut {NoStop}%
\bibitem [{\citenamefont {Shi}\ and\ \citenamefont {Senthil}(2025{\natexlab{b}})}]{ShiSenthilAnyonDelocalization}%
  \BibitemOpen
  \bibfield  {author} {\bibinfo {author} {\bibfnamefont {Z.~D.}\ \bibnamefont {Shi}}\ and\ \bibinfo {author} {\bibfnamefont {T.}~\bibnamefont {Senthil}},\ }\href {https://doi.org/10.1073/pnas.2520608122} {\bibfield  {journal} {\bibinfo  {journal} {Proceedings of the National Academy of Sciences}\ }\textbf {\bibinfo {volume} {122}},\ \bibinfo {pages} {e2520608122} (\bibinfo {year} {2025}{\natexlab{b}})},\ \Eprint {https://arxiv.org/abs/https://www.pnas.org/doi/pdf/10.1073/pnas.2520608122} {https://www.pnas.org/doi/pdf/10.1073/pnas.2520608122} \BibitemShut {NoStop}%
\bibitem [{\citenamefont {Pichler}\ \emph {et~al.}(2025{\natexlab{a}})\citenamefont {Pichler}, \citenamefont {Kuhlenkamp}, \citenamefont {Knap},\ and\ \citenamefont {Vishwanath}}]{PICHLERanyonSC}%
  \BibitemOpen
  \bibfield  {author} {\bibinfo {author} {\bibfnamefont {F.}~\bibnamefont {Pichler}}, \bibinfo {author} {\bibfnamefont {C.}~\bibnamefont {Kuhlenkamp}}, \bibinfo {author} {\bibfnamefont {M.}~\bibnamefont {Knap}},\ and\ \bibinfo {author} {\bibfnamefont {A.}~\bibnamefont {Vishwanath}},\ }\href {https://doi.org/https://doi.org/10.1016/j.newton.2025.100340} {\bibfield  {journal} {\bibinfo  {journal} {Newton}\ ,\ \bibinfo {pages} {100340}} (\bibinfo {year} {2025}{\natexlab{a}})}\BibitemShut {NoStop}%
\bibitem [{\citenamefont {Han}\ \emph {et~al.}(2025)\citenamefont {Han}, \citenamefont {Wang}, \citenamefont {Dong}, \citenamefont {Zaletel},\ and\ \citenamefont {Vishwanath}}]{han2025anyon}%
  \BibitemOpen
  \bibfield  {author} {\bibinfo {author} {\bibfnamefont {Z.}~\bibnamefont {Han}}, \bibinfo {author} {\bibfnamefont {T.}~\bibnamefont {Wang}}, \bibinfo {author} {\bibfnamefont {Z.}~\bibnamefont {Dong}}, \bibinfo {author} {\bibfnamefont {M.~P.}\ \bibnamefont {Zaletel}},\ and\ \bibinfo {author} {\bibfnamefont {A.}~\bibnamefont {Vishwanath}},\ }\href@noop {} {\bibfield  {journal} {\bibinfo  {journal} {arXiv preprint arXiv:2508.14894}\ } (\bibinfo {year} {2025})}\BibitemShut {NoStop}%
\bibitem [{\citenamefont {Pichler}\ \emph {et~al.}(2025{\natexlab{b}})\citenamefont {Pichler}, \citenamefont {Kadow}, \citenamefont {Kuhlenkamp},\ and\ \citenamefont {Knap}}]{Pichlerspectral}%
  \BibitemOpen
  \bibfield  {author} {\bibinfo {author} {\bibfnamefont {F.}~\bibnamefont {Pichler}}, \bibinfo {author} {\bibfnamefont {W.}~\bibnamefont {Kadow}}, \bibinfo {author} {\bibfnamefont {C.}~\bibnamefont {Kuhlenkamp}},\ and\ \bibinfo {author} {\bibfnamefont {M.}~\bibnamefont {Knap}},\ }\href {https://doi.org/10.1103/PhysRevB.111.075108} {\bibfield  {journal} {\bibinfo  {journal} {Phys. Rev. B}\ }\textbf {\bibinfo {volume} {111}},\ \bibinfo {pages} {075108} (\bibinfo {year} {2025}{\natexlab{b}})}\BibitemShut {NoStop}%
\bibitem [{\citenamefont {Jain}\ and\ \citenamefont {Kamilla}(1997)}]{CFED1997}%
  \BibitemOpen
  \bibfield  {author} {\bibinfo {author} {\bibfnamefont {J.~K.}\ \bibnamefont {Jain}}\ and\ \bibinfo {author} {\bibfnamefont {R.~K.}\ \bibnamefont {Kamilla}},\ }\href {https://doi.org/10.1103/PhysRevB.55.R4895} {\bibfield  {journal} {\bibinfo  {journal} {Phys. Rev. B}\ }\textbf {\bibinfo {volume} {55}},\ \bibinfo {pages} {R4895} (\bibinfo {year} {1997})}\BibitemShut {NoStop}%
\bibitem [{\citenamefont {Balram}\ \emph {et~al.}(2013)\citenamefont {Balram}, \citenamefont {W\'ojs},\ and\ \citenamefont {Jain}}]{CFED2013}%
  \BibitemOpen
  \bibfield  {author} {\bibinfo {author} {\bibfnamefont {A.~C.}\ \bibnamefont {Balram}}, \bibinfo {author} {\bibfnamefont {A.}~\bibnamefont {W\'ojs}},\ and\ \bibinfo {author} {\bibfnamefont {J.~K.}\ \bibnamefont {Jain}},\ }\href {https://doi.org/10.1103/PhysRevB.88.205312} {\bibfield  {journal} {\bibinfo  {journal} {Phys. Rev. B}\ }\textbf {\bibinfo {volume} {88}},\ \bibinfo {pages} {205312} (\bibinfo {year} {2013})}\BibitemShut {NoStop}%
\bibitem [{\citenamefont {Haldane}(1983)}]{Haldane1983_Hierarchy}%
  \BibitemOpen
  \bibfield  {author} {\bibinfo {author} {\bibfnamefont {F.~D.~M.}\ \bibnamefont {Haldane}},\ }\href {https://doi.org/10.1103/PhysRevLett.51.605} {\bibfield  {journal} {\bibinfo  {journal} {Phys. Rev. Lett.}\ }\textbf {\bibinfo {volume} {51}},\ \bibinfo {pages} {605} (\bibinfo {year} {1983})}\BibitemShut {NoStop}%
\bibitem [{\citenamefont {Trugman}\ and\ \citenamefont {Kivelson}(1985)}]{trugmanExactResultsFractional1985a}%
  \BibitemOpen
  \bibfield  {author} {\bibinfo {author} {\bibfnamefont {S.~A.}\ \bibnamefont {Trugman}}\ and\ \bibinfo {author} {\bibfnamefont {S.}~\bibnamefont {Kivelson}},\ }\href {https://doi.org/10.1103/PhysRevB.31.5280} {\bibfield  {journal} {\bibinfo  {journal} {Phys. Rev. B}\ }\textbf {\bibinfo {volume} {31}},\ \bibinfo {pages} {5280} (\bibinfo {year} {1985})}\BibitemShut {NoStop}%
\bibitem [{Note1()}]{Note1}%
  \BibitemOpen
  \bibinfo {note} {$\protect \hat {V}_\protect \mathrm {TK} \sim \nabla ^2\delta $ while the leading pseudopotential of $\protect \hat {V}$ is $V_3 \sim \nabla ^6\delta $, giving an effective $\alpha \sim (\ell _B/\lambda )^4$}\BibitemShut {NoStop}%
\bibitem [{\citenamefont {Bernevig}\ and\ \citenamefont {Haldane}(2008{\natexlab{a}})}]{Bernevig2008_Jack}%
  \BibitemOpen
  \bibfield  {author} {\bibinfo {author} {\bibfnamefont {B.~A.}\ \bibnamefont {Bernevig}}\ and\ \bibinfo {author} {\bibfnamefont {F.~D.~M.}\ \bibnamefont {Haldane}},\ }\href {https://doi.org/10.1103/PhysRevLett.100.246802} {\bibfield  {journal} {\bibinfo  {journal} {Phys. Rev. Lett.}\ }\textbf {\bibinfo {volume} {100}},\ \bibinfo {pages} {246802} (\bibinfo {year} {2008}{\natexlab{a}})}\BibitemShut {NoStop}%
\bibitem [{\citenamefont {Bernevig}\ and\ \citenamefont {Haldane}(2008{\natexlab{b}})}]{Bernevig2008_clustering}%
  \BibitemOpen
  \bibfield  {author} {\bibinfo {author} {\bibfnamefont {B.~A.}\ \bibnamefont {Bernevig}}\ and\ \bibinfo {author} {\bibfnamefont {F.~D.~M.}\ \bibnamefont {Haldane}},\ }\href {https://doi.org/10.1103/PhysRevB.77.184502} {\bibfield  {journal} {\bibinfo  {journal} {Phys. Rev. B}\ }\textbf {\bibinfo {volume} {77}},\ \bibinfo {pages} {184502} (\bibinfo {year} {2008}{\natexlab{b}})}\BibitemShut {NoStop}%
\bibitem [{\citenamefont {Ledwith}\ \emph {et~al.}(2020)\citenamefont {Ledwith}, \citenamefont {Tarnopolsky}, \citenamefont {Khalaf},\ and\ \citenamefont {Vishwanath}}]{ledwith2020fractional}%
  \BibitemOpen
  \bibfield  {author} {\bibinfo {author} {\bibfnamefont {P.~J.}\ \bibnamefont {Ledwith}}, \bibinfo {author} {\bibfnamefont {G.}~\bibnamefont {Tarnopolsky}}, \bibinfo {author} {\bibfnamefont {E.}~\bibnamefont {Khalaf}},\ and\ \bibinfo {author} {\bibfnamefont {A.}~\bibnamefont {Vishwanath}},\ }\href {https://doi.org/10.1103/PhysRevResearch.2.023237} {\bibfield  {journal} {\bibinfo  {journal} {Phys. Rev. Research}\ }\textbf {\bibinfo {volume} {2}},\ \bibinfo {pages} {023237} (\bibinfo {year} {2020})}\BibitemShut {NoStop}%
\bibitem [{\citenamefont {Wang}\ \emph {et~al.}(2021)\citenamefont {Wang}, \citenamefont {Cano}, \citenamefont {Millis}, \citenamefont {Liu},\ and\ \citenamefont {Yang}}]{Wang2021exact}%
  \BibitemOpen
  \bibfield  {author} {\bibinfo {author} {\bibfnamefont {J.}~\bibnamefont {Wang}}, \bibinfo {author} {\bibfnamefont {J.}~\bibnamefont {Cano}}, \bibinfo {author} {\bibfnamefont {A.~J.}\ \bibnamefont {Millis}}, \bibinfo {author} {\bibfnamefont {Z.}~\bibnamefont {Liu}},\ and\ \bibinfo {author} {\bibfnamefont {B.}~\bibnamefont {Yang}},\ }\href {https://doi.org/10.1103/PhysRevLett.127.246403} {\bibfield  {journal} {\bibinfo  {journal} {Phys. Rev. Lett.}\ }\textbf {\bibinfo {volume} {127}},\ \bibinfo {pages} {246403} (\bibinfo {year} {2021})}\BibitemShut {NoStop}%
\bibitem [{\citenamefont {Ledwith}\ \emph {et~al.}(2023)\citenamefont {Ledwith}, \citenamefont {Vishwanath},\ and\ \citenamefont {Parker}}]{ledwith_vortexability_2023}%
  \BibitemOpen
  \bibfield  {author} {\bibinfo {author} {\bibfnamefont {P.~J.}\ \bibnamefont {Ledwith}}, \bibinfo {author} {\bibfnamefont {A.}~\bibnamefont {Vishwanath}},\ and\ \bibinfo {author} {\bibfnamefont {D.~E.}\ \bibnamefont {Parker}},\ }\href {https://doi.org/10.1103/PhysRevB.108.205144} {\bibfield  {journal} {\bibinfo  {journal} {Physical Review B}\ }\textbf {\bibinfo {volume} {108}},\ \bibinfo {pages} {205144} (\bibinfo {year} {2023})},\ \bibinfo {note} {arXiv:2209.15023 [cond-mat]}\BibitemShut {NoStop}%
\bibitem [{Note2()}]{Note2}%
  \BibitemOpen
  \bibinfo {note} {When the distance between quasiholes is much larger than the magnetic length $\ell _B$, we can think of them as point particles and replace $V[\protect \bar \xi , \xi ] = \DOTSB \sum@ \slimits@ _{i<j} \Delta V(|\xi _i - \xi _j|)$}\BibitemShut {NoStop}%
\bibitem [{\citenamefont {Kibble}(1979)}]{Kibble1979}%
  \BibitemOpen
  \bibfield  {author} {\bibinfo {author} {\bibfnamefont {T.~W.~B.}\ \bibnamefont {Kibble}},\ }\href@noop {} {\bibfield  {journal} {\bibinfo  {journal} {Communications in Mathematical Physics}\ }\textbf {\bibinfo {volume} {65}},\ \bibinfo {pages} {189 } (\bibinfo {year} {1979})}\BibitemShut {NoStop}%
\bibitem [{\citenamefont {Ashtekar}\ and\ \citenamefont {Schilling}(1999)}]{Ashtekar1999}%
  \BibitemOpen
  \bibfield  {author} {\bibinfo {author} {\bibfnamefont {A.}~\bibnamefont {Ashtekar}}\ and\ \bibinfo {author} {\bibfnamefont {T.~A.}\ \bibnamefont {Schilling}},\ }\bibinfo {title} {Geometrical formulation of quantum mechanics},\ in\ \href {https://doi.org/10.1007/978-1-4612-1422-9_3} {\emph {\bibinfo {booktitle} {On Einstein's Path: Essays in Honor of Engelbert Schucking}}},\ \bibinfo {editor} {edited by\ \bibinfo {editor} {\bibfnamefont {A.}~\bibnamefont {Harvey}}}\ (\bibinfo  {publisher} {Springer New York},\ \bibinfo {address} {New York, NY},\ \bibinfo {year} {1999})\ pp.\ \bibinfo {pages} {23--65}\BibitemShut {NoStop}%
\bibitem [{\citenamefont {Provost}\ and\ \citenamefont {Vallee}(1980)}]{Provost1980}%
  \BibitemOpen
  \bibfield  {author} {\bibinfo {author} {\bibfnamefont {J.~P.}\ \bibnamefont {Provost}}\ and\ \bibinfo {author} {\bibfnamefont {G.}~\bibnamefont {Vallee}},\ }\href {https://doi.org/10.1007/BF02193559} {\bibfield  {journal} {\bibinfo  {journal} {Communications in Mathematical Physics}\ }\textbf {\bibinfo {volume} {76}},\ \bibinfo {pages} {289} (\bibinfo {year} {1980})}\BibitemShut {NoStop}%
\bibitem [{\citenamefont {Arovas}\ \emph {et~al.}(1985)\citenamefont {Arovas}, \citenamefont {Schrieffer}, \citenamefont {Wilczek},\ and\ \citenamefont {Zee}}]{Arovas1985}%
  \BibitemOpen
  \bibfield  {author} {\bibinfo {author} {\bibfnamefont {D.~P.}\ \bibnamefont {Arovas}}, \bibinfo {author} {\bibfnamefont {R.}~\bibnamefont {Schrieffer}}, \bibinfo {author} {\bibfnamefont {F.}~\bibnamefont {Wilczek}},\ and\ \bibinfo {author} {\bibfnamefont {A.}~\bibnamefont {Zee}},\ }\href {https://doi.org/https://doi.org/10.1016/0550-3213(85)90252-4} {\bibfield  {journal} {\bibinfo  {journal} {Nuclear Physics B}\ }\textbf {\bibinfo {volume} {251}},\ \bibinfo {pages} {117} (\bibinfo {year} {1985})}\BibitemShut {NoStop}%
\bibitem [{\citenamefont {Berezin}(1975)}]{Berezin1975_quantization}%
  \BibitemOpen
  \bibfield  {author} {\bibinfo {author} {\bibfnamefont {F.~A.}\ \bibnamefont {Berezin}},\ }\href {https://doi.org/10.1007/BF01609397} {\bibfield  {journal} {\bibinfo  {journal} {Communications in Mathematical Physics}\ }\textbf {\bibinfo {volume} {40}},\ \bibinfo {pages} {153} (\bibinfo {year} {1975})}\BibitemShut {NoStop}%
\bibitem [{\citenamefont {Rawnsley}\ \emph {et~al.}(1990)\citenamefont {Rawnsley}, \citenamefont {Cahen},\ and\ \citenamefont {Gutt}}]{Rawnsley1990}%
  \BibitemOpen
  \bibfield  {author} {\bibinfo {author} {\bibfnamefont {J.}~\bibnamefont {Rawnsley}}, \bibinfo {author} {\bibfnamefont {M.}~\bibnamefont {Cahen}},\ and\ \bibinfo {author} {\bibfnamefont {S.}~\bibnamefont {Gutt}},\ }\href {https://doi.org/https://doi.org/10.1016/0393-0440(90)90019-Y} {\bibfield  {journal} {\bibinfo  {journal} {Journal of Geometry and Physics}\ }\textbf {\bibinfo {volume} {7}},\ \bibinfo {pages} {45} (\bibinfo {year} {1990})}\BibitemShut {NoStop}%
\bibitem [{\citenamefont {Nair}(2020)}]{Nair2020}%
  \BibitemOpen
  \bibfield  {author} {\bibinfo {author} {\bibfnamefont {V.~P.}\ \bibnamefont {Nair}},\ }\href {https://doi.org/10.1103/PhysRevD.102.025015} {\bibfield  {journal} {\bibinfo  {journal} {Phys. Rev. D}\ }\textbf {\bibinfo {volume} {102}},\ \bibinfo {pages} {025015} (\bibinfo {year} {2020})}\BibitemShut {NoStop}%
\bibitem [{\citenamefont {Girvin}\ and\ \citenamefont {Jach}(1984)}]{girvin1984FormalismQuantumHall}%
  \BibitemOpen
  \bibfield  {author} {\bibinfo {author} {\bibfnamefont {S.~M.}\ \bibnamefont {Girvin}}\ and\ \bibinfo {author} {\bibfnamefont {T.}~\bibnamefont {Jach}},\ }\href {https://doi.org/10.1103/PhysRevB.29.5617} {\bibfield  {journal} {\bibinfo  {journal} {Physical Review B}\ }\textbf {\bibinfo {volume} {29}},\ \bibinfo {pages} {5617} (\bibinfo {year} {1984})}\BibitemShut {NoStop}%
\bibitem [{\citenamefont {Bergholtz}\ and\ \citenamefont {Karlhede}(2006)}]{Bergholtz2006}%
  \BibitemOpen
  \bibfield  {author} {\bibinfo {author} {\bibfnamefont {E.~J.}\ \bibnamefont {Bergholtz}}\ and\ \bibinfo {author} {\bibfnamefont {A.}~\bibnamefont {Karlhede}},\ }\href {https://doi.org/10.1088/1742-5468/2006/04/L04001} {\bibfield  {journal} {\bibinfo  {journal} {Journal of Statistical Mechanics: Theory and Experiment}\ }\textbf {\bibinfo {volume} {2006}},\ \bibinfo {pages} {L04001} (\bibinfo {year} {2006})}\BibitemShut {NoStop}%
\bibitem [{SM()}]{SM}%
  \BibitemOpen
  \bibinfo {note} {See Supplemental Material}\BibitemShut {NoStop}%
\bibitem [{Note3()}]{Note3}%
  \BibitemOpen
  \bibinfo {note} {On the plane, there is no full continuous magnetic translation symmetry, but we expect an approximate translation symmetry in the bulk. On the torus, translation symmetry is exact, but we are only allowed to translate by a discrete amount, and rotation symmetry is no longer exact. On the sphere, rotation symmetry is exact, and the analog of translation symmetry is generated by the action of total angular momentum raising/lowering operators, which is distinct from translation for finite systems. For a large enough system, these effects can be neglected, and we can assume both rotation and magnetic translation symmetry regardless of geometry. We verify this effect is negilible for system sizes we choose in SM.}\BibitemShut {Stop}%
\bibitem [{Note4()}]{Note4}%
  \BibitemOpen
  \bibinfo {note} {The magnetic field, plotted approaches 1/3 at long distances and has a dip around the origin whose integrated weight is $-1/3$ such that at long distance the magnetic field is $1/3 - 1/3 \delta (\xi _r)$ as expected; the other anyon acts as a flux tube with flux $1/3$}\BibitemShut {NoStop}%
\bibitem [{Note5()}]{Note5}%
  \BibitemOpen
  \bibinfo {note} {We need to first normal-order the expression such that $\protect \bar \xi _r$ is to the left of $\xi _r$, then replace $\xi _r$ with the corresponding operator}\BibitemShut {NoStop}%
\bibitem [{\citenamefont {Glauber}(1963{\natexlab{a}})}]{GlauberI}%
  \BibitemOpen
  \bibfield  {author} {\bibinfo {author} {\bibfnamefont {R.~J.}\ \bibnamefont {Glauber}},\ }\href {https://doi.org/10.1103/PhysRev.131.2766} {\bibfield  {journal} {\bibinfo  {journal} {Phys. Rev.}\ }\textbf {\bibinfo {volume} {131}},\ \bibinfo {pages} {2766} (\bibinfo {year} {1963}{\natexlab{a}})}\BibitemShut {NoStop}%
\bibitem [{\citenamefont {Glauber}(1963{\natexlab{b}})}]{GlauberII}%
  \BibitemOpen
  \bibfield  {author} {\bibinfo {author} {\bibfnamefont {R.~J.}\ \bibnamefont {Glauber}},\ }\href {https://doi.org/10.1103/PhysRev.130.2529} {\bibfield  {journal} {\bibinfo  {journal} {Phys. Rev.}\ }\textbf {\bibinfo {volume} {130}},\ \bibinfo {pages} {2529} (\bibinfo {year} {1963}{\natexlab{b}})}\BibitemShut {NoStop}%
\bibitem [{\citenamefont {Sudarshan}(1963)}]{Sudarshan}%
  \BibitemOpen
  \bibfield  {author} {\bibinfo {author} {\bibfnamefont {E.~C.~G.}\ \bibnamefont {Sudarshan}},\ }\href {https://doi.org/10.1103/PhysRevLett.10.277} {\bibfield  {journal} {\bibinfo  {journal} {Phys. Rev. Lett.}\ }\textbf {\bibinfo {volume} {10}},\ \bibinfo {pages} {277} (\bibinfo {year} {1963})}\BibitemShut {NoStop}%
\bibitem [{\citenamefont {Kiesel}\ \emph {et~al.}(2008)\citenamefont {Kiesel}, \citenamefont {Vogel}, \citenamefont {Parigi}, \citenamefont {Zavatta},\ and\ \citenamefont {Bellini}}]{ExperimentPFunction}%
  \BibitemOpen
  \bibfield  {author} {\bibinfo {author} {\bibfnamefont {T.}~\bibnamefont {Kiesel}}, \bibinfo {author} {\bibfnamefont {W.}~\bibnamefont {Vogel}}, \bibinfo {author} {\bibfnamefont {V.}~\bibnamefont {Parigi}}, \bibinfo {author} {\bibfnamefont {A.}~\bibnamefont {Zavatta}},\ and\ \bibinfo {author} {\bibfnamefont {M.}~\bibnamefont {Bellini}},\ }\href {https://doi.org/10.1103/PhysRevA.78.021804} {\bibfield  {journal} {\bibinfo  {journal} {Phys. Rev. A}\ }\textbf {\bibinfo {volume} {78}},\ \bibinfo {pages} {021804} (\bibinfo {year} {2008})}\BibitemShut {NoStop}%
\bibitem [{\citenamefont {Jancovici}(1981)}]{Jancovici1981}%
  \BibitemOpen
  \bibfield  {author} {\bibinfo {author} {\bibfnamefont {B.}~\bibnamefont {Jancovici}},\ }\href {https://doi.org/10.1103/PhysRevLett.46.386} {\bibfield  {journal} {\bibinfo  {journal} {Phys. Rev. Lett.}\ }\textbf {\bibinfo {volume} {46}},\ \bibinfo {pages} {386} (\bibinfo {year} {1981})}\BibitemShut {NoStop}%
\bibitem [{\citenamefont {Can}\ \emph {et~al.}(2014)\citenamefont {Can}, \citenamefont {Forrester}, \citenamefont {T\'ellez},\ and\ \citenamefont {Wiegmann}}]{Can2014}%
  \BibitemOpen
  \bibfield  {author} {\bibinfo {author} {\bibfnamefont {T.}~\bibnamefont {Can}}, \bibinfo {author} {\bibfnamefont {P.~J.}\ \bibnamefont {Forrester}}, \bibinfo {author} {\bibfnamefont {G.}~\bibnamefont {T\'ellez}},\ and\ \bibinfo {author} {\bibfnamefont {P.}~\bibnamefont {Wiegmann}},\ }\href {https://doi.org/10.1103/PhysRevB.89.235137} {\bibfield  {journal} {\bibinfo  {journal} {Phys. Rev. B}\ }\textbf {\bibinfo {volume} {89}},\ \bibinfo {pages} {235137} (\bibinfo {year} {2014})}\BibitemShut {NoStop}%
\bibitem [{\citenamefont {Halperin}(1984)}]{Halperin1984_Hierarchy}%
  \BibitemOpen
  \bibfield  {author} {\bibinfo {author} {\bibfnamefont {B.~I.}\ \bibnamefont {Halperin}},\ }\href {https://doi.org/10.1103/PhysRevLett.52.1583} {\bibfield  {journal} {\bibinfo  {journal} {Phys. Rev. Lett.}\ }\textbf {\bibinfo {volume} {52}},\ \bibinfo {pages} {1583} (\bibinfo {year} {1984})}\BibitemShut {NoStop}%
\bibitem [{\citenamefont {Parameswaran}\ \emph {et~al.}(2012)\citenamefont {Parameswaran}, \citenamefont {Kivelson}, \citenamefont {Rezayi}, \citenamefont {Simon}, \citenamefont {Sondhi},\ and\ \citenamefont {Spivak}}]{Parameswaran2012_typeI_FQH}%
  \BibitemOpen
  \bibfield  {author} {\bibinfo {author} {\bibfnamefont {S.~A.}\ \bibnamefont {Parameswaran}}, \bibinfo {author} {\bibfnamefont {S.~A.}\ \bibnamefont {Kivelson}}, \bibinfo {author} {\bibfnamefont {E.~H.}\ \bibnamefont {Rezayi}}, \bibinfo {author} {\bibfnamefont {S.~H.}\ \bibnamefont {Simon}}, \bibinfo {author} {\bibfnamefont {S.~L.}\ \bibnamefont {Sondhi}},\ and\ \bibinfo {author} {\bibfnamefont {B.~Z.}\ \bibnamefont {Spivak}},\ }\href {https://doi.org/10.1103/PhysRevB.85.241307} {\bibfield  {journal} {\bibinfo  {journal} {Phys. Rev. B}\ }\textbf {\bibinfo {volume} {85}},\ \bibinfo {pages} {241307} (\bibinfo {year} {2012})}\BibitemShut {NoStop}%
\bibitem [{\citenamefont {Yan}\ \emph {et~al.}(2025)\citenamefont {Yan}, \citenamefont {Li}, \citenamefont {Soejima},\ and\ \citenamefont {Khalaf}}]{YanAnyonDispersion}%
  \BibitemOpen
  \bibfield  {author} {\bibinfo {author} {\bibfnamefont {Z.}~\bibnamefont {Yan}}, \bibinfo {author} {\bibfnamefont {Q.}~\bibnamefont {Li}}, \bibinfo {author} {\bibfnamefont {T.}~\bibnamefont {Soejima}},\ and\ \bibinfo {author} {\bibfnamefont {E.}~\bibnamefont {Khalaf}},\ }\href@noop {} {\bibfield  {journal} {\bibinfo  {journal} {arXiv preprint arXiv:2512.15863}\ } (\bibinfo {year} {2025})}\BibitemShut {NoStop}%
\bibitem [{\citenamefont {Schleith}\ \emph {et~al.}(2025)\citenamefont {Schleith}, \citenamefont {Soejima},\ and\ \citenamefont {Khalaf}}]{SchleithAnyonSphere}%
  \BibitemOpen
  \bibfield  {author} {\bibinfo {author} {\bibfnamefont {M.-L.}\ \bibnamefont {Schleith}}, \bibinfo {author} {\bibfnamefont {T.}~\bibnamefont {Soejima}},\ and\ \bibinfo {author} {\bibfnamefont {E.}~\bibnamefont {Khalaf}},\ }\href@noop {} {\bibfield  {journal} {\bibinfo  {journal} {arXiv preprint arXiv:2506.11211}\ } (\bibinfo {year} {2025})}\BibitemShut {NoStop}%
\bibitem [{\citenamefont {Shi}\ and\ \citenamefont {Senthil}(2024)}]{ShiSenthil}%
  \BibitemOpen
  \bibfield  {author} {\bibinfo {author} {\bibfnamefont {Z.~D.}\ \bibnamefont {Shi}}\ and\ \bibinfo {author} {\bibfnamefont {T.}~\bibnamefont {Senthil}},\ }\href@noop {} {\bibfield  {journal} {\bibinfo  {journal} {arXiv preprint arXiv:2409.20567}\ } (\bibinfo {year} {2024})}\BibitemShut {NoStop}%
\bibitem [{\citenamefont {Shi}\ \emph {et~al.}(2024)\citenamefont {Shi}, \citenamefont {Morales-Dur\'an}, \citenamefont {Khalaf},\ and\ \citenamefont {MacDonald}}]{TMDAharonocCasher}%
  \BibitemOpen
  \bibfield  {author} {\bibinfo {author} {\bibfnamefont {J.}~\bibnamefont {Shi}}, \bibinfo {author} {\bibfnamefont {N.}~\bibnamefont {Morales-Dur\'an}}, \bibinfo {author} {\bibfnamefont {E.}~\bibnamefont {Khalaf}},\ and\ \bibinfo {author} {\bibfnamefont {A.~H.}\ \bibnamefont {MacDonald}},\ }\href {https://doi.org/10.1103/PhysRevB.110.035130} {\bibfield  {journal} {\bibinfo  {journal} {Phys. Rev. B}\ }\textbf {\bibinfo {volume} {110}},\ \bibinfo {pages} {035130} (\bibinfo {year} {2024})}\BibitemShut {NoStop}%
\bibitem [{\citenamefont {Kivelson}\ \emph {et~al.}(1987)\citenamefont {Kivelson}, \citenamefont {Kallin}, \citenamefont {Arovas},\ and\ \citenamefont {Schrieffer}}]{Kivelson1987_cooperative_righ_exchange_1}%
  \BibitemOpen
  \bibfield  {author} {\bibinfo {author} {\bibfnamefont {S.}~\bibnamefont {Kivelson}}, \bibinfo {author} {\bibfnamefont {C.}~\bibnamefont {Kallin}}, \bibinfo {author} {\bibfnamefont {D.~P.}\ \bibnamefont {Arovas}},\ and\ \bibinfo {author} {\bibfnamefont {J.~R.}\ \bibnamefont {Schrieffer}},\ }\href {https://doi.org/10.1103/PhysRevB.36.1620} {\bibfield  {journal} {\bibinfo  {journal} {Phys. Rev. B}\ }\textbf {\bibinfo {volume} {36}},\ \bibinfo {pages} {1620} (\bibinfo {year} {1987})}\BibitemShut {NoStop}%
\bibitem [{\citenamefont {Lee}\ \emph {et~al.}(1987)\citenamefont {Lee}, \citenamefont {Baskaran},\ and\ \citenamefont {Kivelson}}]{Lee1987_cooperative_righ_exchange_2}%
  \BibitemOpen
  \bibfield  {author} {\bibinfo {author} {\bibfnamefont {D.~H.}\ \bibnamefont {Lee}}, \bibinfo {author} {\bibfnamefont {G.}~\bibnamefont {Baskaran}},\ and\ \bibinfo {author} {\bibfnamefont {S.}~\bibnamefont {Kivelson}},\ }\href {https://doi.org/10.1103/PhysRevLett.59.2467} {\bibfield  {journal} {\bibinfo  {journal} {Phys. Rev. Lett.}\ }\textbf {\bibinfo {volume} {59}},\ \bibinfo {pages} {2467} (\bibinfo {year} {1987})}\BibitemShut {NoStop}%
\bibitem [{\citenamefont {Greiter}\ \emph {et~al.}(1991)\citenamefont {Greiter}, \citenamefont {Wen},\ and\ \citenamefont {Wilczek}}]{Greiter1991}%
  \BibitemOpen
  \bibfield  {author} {\bibinfo {author} {\bibfnamefont {M.}~\bibnamefont {Greiter}}, \bibinfo {author} {\bibfnamefont {X.-G.}\ \bibnamefont {Wen}},\ and\ \bibinfo {author} {\bibfnamefont {F.}~\bibnamefont {Wilczek}},\ }\href {https://doi.org/10.1103/PhysRevLett.66.3205} {\bibfield  {journal} {\bibinfo  {journal} {Phys. Rev. Lett.}\ }\textbf {\bibinfo {volume} {66}},\ \bibinfo {pages} {3205} (\bibinfo {year} {1991})}\BibitemShut {NoStop}%
\bibitem [{Note6()}]{Note6}%
  \BibitemOpen
  \bibinfo {note} {The way we define $\protect \hat \xi $ is related to the standard raising/lowering operators by a factor of $\protect \sqrt {2}$}\BibitemShut {NoStop}%
\bibitem [{\citenamefont {Parameswaran}\ \emph {et~al.}(2011)\citenamefont {Parameswaran}, \citenamefont {Kivelson}, \citenamefont {Sondhi},\ and\ \citenamefont {Spivak}}]{Parameswaran2011_Pfaffian}%
  \BibitemOpen
  \bibfield  {author} {\bibinfo {author} {\bibfnamefont {S.~A.}\ \bibnamefont {Parameswaran}}, \bibinfo {author} {\bibfnamefont {S.~A.}\ \bibnamefont {Kivelson}}, \bibinfo {author} {\bibfnamefont {S.~L.}\ \bibnamefont {Sondhi}},\ and\ \bibinfo {author} {\bibfnamefont {B.~Z.}\ \bibnamefont {Spivak}},\ }\href {https://doi.org/10.1103/PhysRevLett.106.236801} {\bibfield  {journal} {\bibinfo  {journal} {Phys. Rev. Lett.}\ }\textbf {\bibinfo {volume} {106}},\ \bibinfo {pages} {236801} (\bibinfo {year} {2011})}\BibitemShut {NoStop}%
\end{thebibliography}
\end{document}